\documentclass[preprint2]{aastex}

\usepackage{aas_macros}
\usepackage{color}
\usepackage{graphicx}
\usepackage{amssymb}

\newcommand{\fei}{Fe\,\textsc{i}}
\newcommand{\feii}{Fe\,\textsc{ii}}
\newcommand{\teff}{T_\mathrm{eff}}
\newcommand{\logg}{\log g}
\newcommand{\feh}{\mathrm{[Fe/H]}}

\newcommand{\ms}{\ensuremath{\mathrm{m\ s}^{-1}}}

\newcommand{\mass}{\mathcal{M}}

\newcommand{\msun}{\ensuremath{\mass_\sun}}

\newcommand{\msini}{\ensuremath{\mass\sin i}}
\newcommand{\mjup}{\ensuremath{\mass_{J}}}

\newcommand{\pdb}{\mbox{$\psi^1$\,Dra\,B}}
\newcommand{\pdbb}{\mbox{$\psi^1$\,Dra\,Bb}}

\newcommand{\pd}{\mbox{$\psi^1$\,Draconis}}

\shorttitle{Two New Jupiter Analogs}
\shortauthors{Endl et al.}

\begin{document}

\title{Two New Long-Period Giant Planets from the McDonald Observatory Planet Search\\
    and Two Stars with Long-Period Radial Velocity Signals Related to Stellar Activity Cycles}

\author{Michael Endl\altaffilmark{1}, Erik J. Brugamyer, William D. Cochran, 
Phillip J. MacQueen, Paul Robertson\altaffilmark{2,3}, Stefano Meschiari, Ivan 
Ramirez, 
Matthew Shetrone, Kevin Gullikson, Marshall C. Johnson, }
\affil{\textit{McDonald Observatory and Department of Astronomy, University of Texas at Austin, Austin, TX 78712, USA}}

\author{Robert Wittenmyer, }
\affil{\textit{School of Physics and Australian Centre for Astrobiology, UNSW Australia, Sydney 2052, Australia}}

\author{Jonathan Horner, }
\affil{\textit{
Computational Engineering and Science Research Centre, 
University of Southern Queensland, Toowoomba, Queensland 4350}}
\affil{\textit{Australian Centre for Astrobiology, UNSW Australia, Sydney 2052, Australia}}

\author{David R. Ciardi, }
\affil{\textit{NASA Exoplanet Science Institute \& Infrared Processing and Analysis Center,
California Institute of Technology, USA}}

\author{Elliott Horch, }
\affil{\textit{Department of Physics, Southern Connecticut State University, 501 Crescent St., New Haven, CT 06515, USA}}

\author{Attila E. Simon, }
\affil{\textit{Physikalisches Institut, Center for Space and Habitability, University of Bern, CH-3012 Bern, Switzerland}}

\author{Steve B. Howell, }
\affil{\textit{NASA Ames Research Center, Moffett Field, CA 94035 USA}}

\author{Mark Everett, }
\affil{\textit{National Optical Astronomy Observatory, 950 North Cherry Avenue, Tucson, AZ 85719 USA}}

\author{Caroline Caldwell, }
\affil{\textit{Astrophysics Research Institute, Liverpool John Moores University, 146 Brownlow Hill, Liverpool, L3 5RF, UK}}

\and

\author{Barbara G. Castanheira}
\affil{\textit{McDonald Observatory, University of Texas at Austin, Austin, TX 78712, USA}}

\altaffiltext{1}{e-mail: mike@astro.as.utexas.edu}
\altaffiltext{2}{\textit{Center for Exoplanets and Habitable Worlds, Department of Astronomy and Astrophysics, The Pennsylvania State University}}
\altaffiltext{3}{\textit{NASA Sagan fellow}}

\begin{abstract}
We report the detection of two new long-period giant planets orbiting the stars HD~95872 and HD~162004 (\pdb) by the McDonald Observatory
planet search. The planet HD~95872b has a minimum mass of 4.6~$M_{\rm Jup}$ and an orbital semi-major axis of 5.2~AU. The giant
planet \pdbb \,has a minimum mass of 1.5~$M_{\rm Jup}$ and an orbital semi-major axis of 4.4~AU. Both of these planets qualify as Jupiter analogs.  
These results are based on over one and a half decades of precise radial velocity measurements collected by our program using the
McDonald Observatory Tull Coude spectrograph at the 2.7\,m Harlan J. Smith telescope.
In the case of \pdb \,we also detect a long-term non-linear trend in our data that indicates the presence of an additional giant planet, similar
to the Jupiter-Saturn pair. The primary of the binary star system, $\psi^{1}$~Dra~A, exhibits a very large amplitude radial velocity variation due to another stellar companion.
We detect this additional member using speckle imaging.   
We also report two cases -- HD~10086 and HD~102870 ($\beta$ Virginis) -- of significant radial velocity variation consistent with the presence of a planet, 
but that are probably caused by stellar activity, rather than reflexive Keplerian motion. These two cases stress the importance of
monitoring the magnetic activity level of a target star, as long-term activity cycles can mimic the presence of a Jupiter-analog planet.  
\end{abstract}

\keywords{planetary systems, stars: individual (HD~10086, HD~95872, HD~102870, HD~162003, HD 162004), stars: abundances, stars: activity, techniques: radial velocities, techniques: 
spectroscopic}

\section{Introduction} \label{intro}

``How common are Solar System analogs?''  Until relatively recently, this fundamental question had little in the way of observational answers.  
Although the \textit{Kepler} mission (Borucki et al.~2010)
has provided first constraints on the answer to the related question of ``how common are Earth analogs?'', 
until our instruments and techniques improve to the point that we are capable of detecting planets across a range of masses and orbits analogous to those of the planets in
our Solar System, a definitive answer is currently beyond our 
reach.  However, as a next step we might instead ask: ``how common are Jupiter analogs?'', gas giant planets that have either not significantly migrated inward 
from the location of their formation beyond the ice-line in the protoplanetary disk, or migrated inwards very early, followed by an episode of 
outward migration (the ``Grand Tack'' model; Walsh et al.~2011).  
As the time baseline of radial velocity searches grows, we are becoming better equipped to answer this last question.

The radial velocity (RV) technique has been used to detect/discover $\sim$600 of the $\sim$2000 known, confirmed exoplanets.  
Since the technique is heavily biased towards massive planets in short-period orbits, the majority of these are gas giants in orbits of less than one Earth-year.  
Only about 25 RV detected planets can be considered ``Jupiter analogs'', 
which we define as within a factor of a few Jupiter-masses and in orbits longer than 8 years (about 3000 days).  
Although the \textit{Kepler} mission -- utilizing the planet transit method -- has delivered 
$\sim$1000 planets and nearly 5000 candidates, none of these can be classified as ``long-period'', 
due to the limited time baseline of the mission data.

To answer the question of the uniqueness of our solar system, it is probably more important to find and characterize long-period Jovian planets than to find small-radius terrestrial 
planets. Other studies (Howard et al.~2012, Wittenmyer et al.~2011b, Fressin et al.~2013, Petigura et al.~2013) have shown that terrestrial-size planets are quite common around other 
stars, but the data concerning Jupiter analogs are quite incomplete due to the need for a time baseline of over 10-15 years.
A handful of RV surveys have ``outgrown'' this time baseline selection bias: the Lick Observatory planet search from 1987 to 2011 (Fischer, Marcy \& 
Spronck~2014), our ongoing McDonald Observatory Planet Search, the Anglo-Australian Planet Search (e.g. Wittenmyer et al.~2014a), the Keck/HIRES RV survey 
(e.g. Howard et al.~2014) and the planet search programs at CORALIE (e.g. Marmier et al.~2013) and HARPS (e.g. Moutou et al.~2015). 
An example of a Jupiter-analog planet orbiting a Solar twin is presented in Bedell et al.~(2015).  
While the \textit{Kepler} mission has revolutionized exoplanetary science and provided a first estimate of the frequency of Earth-size planets in Earth-like orbits, 
long-term radial-velocity surveys complement these data with measurements of the frequency of Jupiter-like planets in Jupiter-like orbits.  
This in turn will reveal how common Solar System-like architectures are.

While the idea that Jupiter analogs are required to shield terrestrial planets from impacts has been conclusively dismantled (e.g. Horner \& Jones 2008, 2012; 
Horner, Jones \& Chambers~2010), 
the presence of Jupiter analogs might be critical for the delivery of water to planets that would otherwise have formed as dry, 
lifeless husks (Horner \& Jones~2010, Raymond~2006). The early dynamical evolution of Jupiter and Saturn might also be responsible for a depletion of
the inner planetesemial disk, and for the subsequent formation of small, low-mass terrestrial planets, instead of large, massive super-Earths (Batygin \& 
Laughlin~2015). The search for Jupiter analogs thus provides a key datum for models of planetary formation and evolution -- attempting to answer the question 
``how common are planetary systems like our own?''

The McDonald Observatory Planet Search (Cochran \& Hatzes~1993)
is a high precision RV survey of hundreds of FGKM stars, begun in 1987 using the 2.7~m Harlan J. Smith telescope.  
Since our migration to our current instrumental configuration in 1998 (``Phase III'', described in Hatzes et al. 2003), we achieve routine long-term Doppler velocity 
precision of $\sim$4-8 m\,s$^{-1}$.  With this precision and an observational time baseline approaching 17 years, we are now sensitive to Jovian analogs.  
In this paper, we present two new long-period planetary companions (HD~95872b and $\psi^{1}$~Dra~Bb).  
We also report two cases (HD~10086 and $\beta$~Virginis) of Keplerian-like signals that mimic a Jupiter-type planet but are probably the result of stellar activity akin to the 
11-year Solar cycle.

\section{Observations} \label{obs}

Our radial velocity measurements were obtained using the 2.7~m Harlan J. Smith and 10~m Keck~I telescopes.  The specific instruments/observations are described below.

\subsection{Harlan J. Smith Telescope Observations} \label{2.7m}

For the 2.7~m Harlan J. Smith Telescope (HJST), we utilize the cross-dispersed Echelle Tull Coude spectrograph (Tull et al.~1994). Our configuration uses a 1.2 arcsec slit, 
an Echelle grating with 52.67 groove mm$^{-1}$, and a 2048$\times$2048 Tektronix CCD with 24 $\micron$ pixels, yielding a resolving power ($R=\lambda$/$\Delta\lambda$) of 
$R=60,000$. The wavelength coverage extends from 3,750~$\AA$ to 10,200~$\AA$, and is complete from the blue end to 5,691~$\AA$, after which there are 
increasingly large inter-order gaps.

\subsection{Keck Telescope Observations} \label{keck}

For HD~95872, we also obtained 10 precise RV measurements using Keck I and its HIRES spectrograph (Vogt et al.~1994), during three observing runs allocated to the NASA CoRoT key 
science project, during times when the CoRoT fields were unobservable.

The spectra for HD~95872 were taken with HIRES with a resolving power of $R = 50,000$, using an instrumental setup similar to the 
California Planet Search (e.g. Howard et al.\,2010). Also for HIRES we used an iodine cell to monitor real time instrumental variations 
relevant to measuring precise radial velocities. 

\subsection{Data Reduction} \label{reduce}

The raw CCD data were reduced using a pipeline implemented in the Image Reduction and Analysis Facility (IRAF) using 
standard routines within the \texttt{echelle} package. The process includes overscan trimming, bad pixel processing, 
bias frame subtraction, scattered light removal, flat field division, order extraction, and wavelength solution application using a Th-Ar calibration lamp spectrum. 
Most cosmic rays are successfully removed via IRAF’s interpolation routines; however, particularly troublesome hits are removed by hand.

\section{Analysis} \label{analysis}

\subsection{Radial Velocity Measurements} \label{rvs}

Our radial velocity measurements were obtained using our standard iodine cell RV reduction pipeline \texttt{Austral} (Endl, K\"urster, \& Els~2000). 
Our approach follows the standard iodine cell data analysis methodology: the stellar RV is calculated by comparing all spectra of the target star, 
taken with the iodine cell, with a high signal-to-noise (S/N) stellar template spectrum free of iodine lines. 
During regular RV observations the temperature-controlled iodine cell is inserted 
in the light path and 
superimposes a dense reference spectrum onto the stellar spectrum. The iodine lines thus provide a simultaneous wavelength calibration and 
allow the reconstruction 
of the shape of the instrumental profile at the time of observation. The iodine cell at the Tull spectrograph 
has been in regular operation for more than two decades. 

\subsection{Stellar Activity Indicators} \label{activities}

As a check against photospheric activity masquerading as planet-like Keplerian motion, we measure the Ca H and K Mount Wilson $S_{HK}$ index 
(Soderblom, Duncan \& Johnson~1991, Baliunas et al.~1995, Paulson et al.~2002) 
simultaneously with each RV data point.  In addition, we have calculated the line bisector velocity spans (BVSs; e.g. Hatzes, Cochran \& Johns-Krull~1997) 
of lines outside the region 
of iodine cell absorption.  These time-series measurements are then checked for any possible correlation(s) with the RV measurements.  

\subsection{Stellar Characterization} \label{stars}

We determined stellar atmospheric parameters for all four stars using a traditional absorption line curve-of-growth approach, following a procedure similar 
to that outlined in Brugamyer et al. (2011).  The method utilizes an updated list of suitable Fe and Ti lines, the local thermodynamic equilibrium (LTE) line analysis 
and spectral synthesis code \texttt{MOOG}\footnote{available at http://www.as.utexas.edu/$\sim$chris/moog.html}, and a grid of 1-D, plane-parallel \texttt{ATLAS9} 
(Kurucz 1993) model atmospheres.  We first manually measured the equivalent widths of 132 Fe~I and 41 Ti~I lines, along with 18 Fe~II and 8 Ti~II lines, in our template 
spectra (without the reference iodine cell in the light path).  With these measurements in hand, the stellar effective temperature is constrained by assuming and enforcing 
excitation equilibrium -- by varying the model atmosphere temperature until any trends in derived abundances with temperature are removed.  
Surface gravity is constrained by assuming and enforcing ionization equilibrium -- by varying the model atmosphere gravity until the derived abundances of neutrals and ions agree.  
Microturbulent velocity is constrained by forcing the derived abundances for stronger lines to match those for weaker lines.  
For these processes, we used a weighted average of Fe (2x) and Ti (1x) when computing the relevant slopes/offsets (as there are
approximately twice as many Fe than Ti lines).
This process is repeated iteratively until all conditions are satisfied simultaneously with a self-consistent set of stellar parameters.

The results of our stellar characterization are summarized in Table \ref{startable}.  
Spectral types, photometric data, and parallaxes are taken from the ASCC-2.5 catalog (Version 3; Kharchenko \& Roeser 2009).  
We also include mass and age estimates from Yonsei-Yale model isochrones (Yi et~al.~2001, Kim et~al.~2002).

Using the stellar parameters $\teff$, $\logg$, $\feh$, and their errors, we determined the masses and ages of our stars using the 
procedure outlined in Ramirez et al.~(2014) (their Section~4.5). Briefly, the location of each star on stellar parameter space was compared to that of 
stellar interior and evolutionary model predictions. The Yonsei-Yale isochrone grid was used in our implementation. 
Each isochrone point was given a probability of representing an observation based on its distance from the measured stellar parameters and weighted by the observational errors. 
Then, mass and age probability distribution functions were computed by adding the probabilities of individual isochrone points binned in mass and age, respectively. 
The peaks of these distributions were adopted as the most probable mass and age, while the 1\,$\sigma$-like widths were used to estimate the errors.

Contrary to a more common practice, we did not use parallaxes in our mass and age determinations. 
This is because one of our stars, HD\,95872, does not have a reliable measurement of trigonometric parallax; this star was not included in the 
{\it Hipparcos} catalog. To maintain consistency in our analysis, we employed the spectroscopic $\logg$ values as luminosity indicators instead of absolute magnitudes 
computed using measured parallaxes. If we had used the {\it Hipparcos} parallaxes for the three stars which have those values available, their masses would be only 
about $0.01\pm0.01\,M_\odot$ smaller.

\begin{deluxetable}{lcccccccccccr}
\tabletypesize{\footnotesize}
\tablecaption{Stellar Properties \label{startable}}
\tablehead{
\colhead{Star} & \colhead{Spectral} & \colhead{V} & \colhead{B--V} & \colhead{M$_{V}$} & \colhead{Parallax} & \colhead{Dist.} & \colhead{T$_{eff}$} & \colhead{log g} & \colhead{[Fe/H]} & \colhead{Mass} & \colhead{Age}                              \\
& \colhead{Type} &  &  &  & \colhead{(mas)} & \colhead{(pc)} & \colhead{(K)} &  &  &  \colhead{(M$_\sun$)}  & \colhead{(Gyr)}	}
\startdata
HD 95872 & K0V & 9.895 & 0.827 & 10.50 & 132.30 & 7.56 & $5312\pm100$ & $4.43\pm0.15$ & $0.41\pm0.09$ & $0.95\pm0.04$ & $10.0\pm3.7$  \\
\pdb & G0V & 5.699 & 0.562 & 3.97 & 45.13 & 22.16 & $6212\pm75$ & $4.20\pm0.12$ & $0.01\pm0.06$ & $1.19\pm0.07$ & $3.3\pm1.0$  \\
HD 10086 & G5IV & 6.610 & 0.688 & 4.97 & 46.99 & 21.28 & $5722\pm65$ & $4.43\pm0.10$ & $0.10\pm0.04$ & $1.01\pm0.03$ & $5.5\pm2.3$  \\
$\beta$ Vir & F8V & 3.589 & 0.568 & 3.40 & 91.65 & 10.91 & $6145\pm75$ & $3.98\pm0.12$ & $0.15\pm0.05$ & $1.34\pm0.10$ & $3.2\pm0.7$ \\
\enddata
\end{deluxetable}

\subsection{Planetary Orbit Modeling} \label{orbits}

We performed our planetary orbit modeling using the \texttt{Systemic Console}\footnote{available at http://www.stefanom.org/systemic/} package (Meschiari et al. 2009), a software 
application for the analysis and fitting of Doppler RV data sets.


\section{The planet around HD 95872} \label{hd95872}

The star HD~95872 was originally selected for RV monitoring from a sample of 22 thin disk stars observed on the 2.7~m HJST in 1998 for a 
project to characterize the metal rich end of chemical evolution of the Galactic disk.  
The sample of 22 stars were selected by M. Grenon (Observatorie de Geneve) for Sandra Castro (ESO) and Matthew Shetrone on the basis of their 
extreme kinematic (perigalactica $\sim$ 3 kpc) and photometric properties.

\subsection{Keplerian solution}
Table \ref{tab:hd95872_data} presents the complete set of our RV measurements for HD~95872 from the 
HJST/Tull survey, as well as 10 additional measurements obtained with Keck/HIRES.
The RV coverage spans approximately 11 years of monitoring over 44 measurements.
The median internal uncertainty for our observations is $\approx$ 6 \ms, and the peak-to-peak velocity is $\approx$ 137 \ms.
The velocity scatter around the average RV is $\approx$ 32.1 \ms. 

\begin{table}[ht]
\centering
\begin{tabular}{lccc}
  \hline
 & BJD & dRV & Uncertainty \\ 
 & & (m\,s$^{-1}$) & (m\,s$^{-1}$) \\
  \hline
1 & 2453073.8686 & 66.2 & 5.3 \\ 
  2 & 2453463.7752 & 98.5 & 5.9 \\ 
  3 & 2453843.7736 & 109.2 & 5.9 \\ 
  4 & 2454557.7575 & 63.5 & 6.7 \\ 
  5 & 2455286.7123 & -22.2 & 4.1 \\ 
  6\tablenotemark{a} & 2455366.7841 & 15.5 & 1.9 \\ 
  7\tablenotemark{a} & 2455368.7876 & 13.3 & 3.5 \\ 
  8 & ... & ... & ... \\ 
   \hline
\end{tabular}
\caption{Differential radial velocity observations for HD\,95872 (sample)}\label{tab:hd95872_data}
\tablenotetext{a}{Observed with Keck/HIRES; all others with the HJST/Tull.}
\end{table}

\begin{figure}
\centering
\plotone{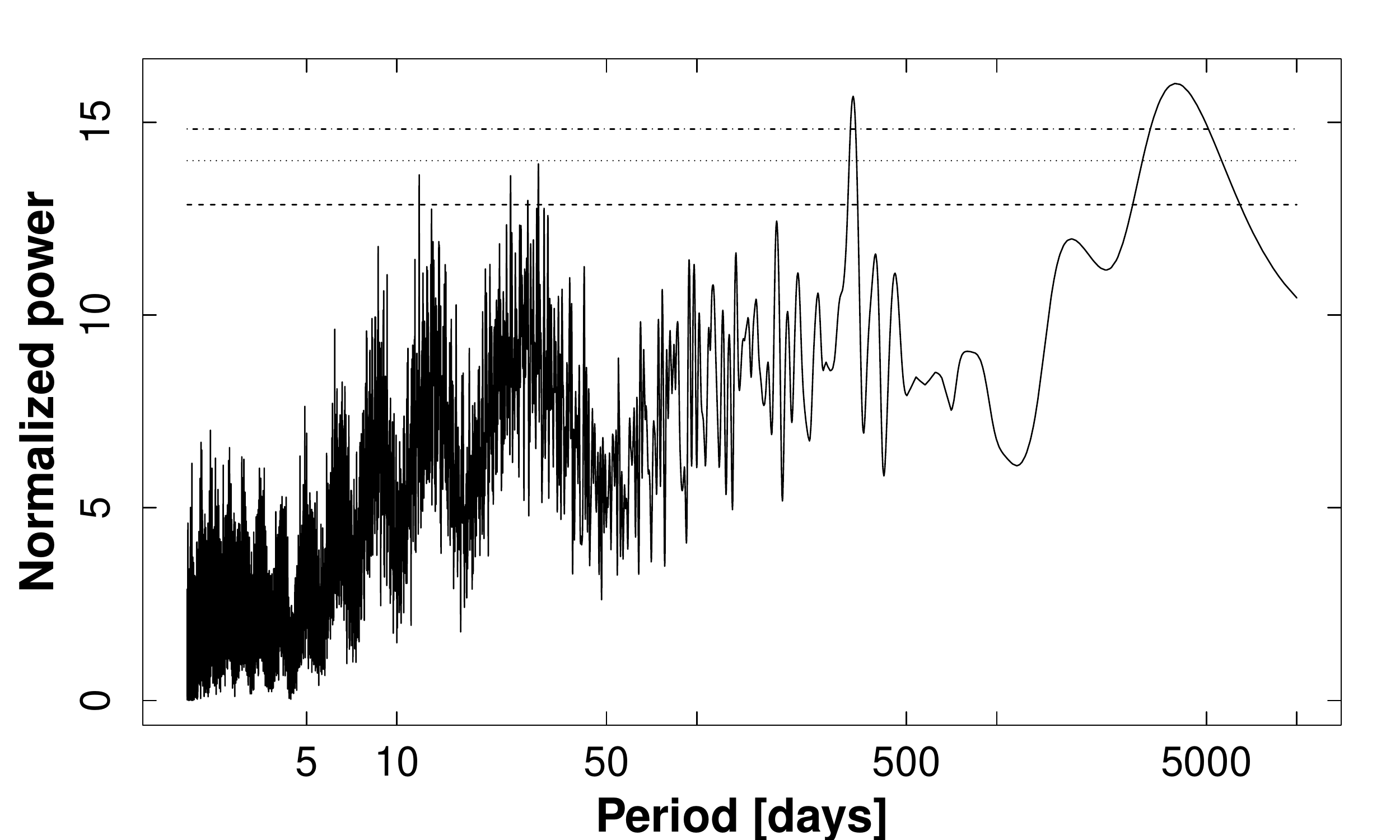}\\
\plotone{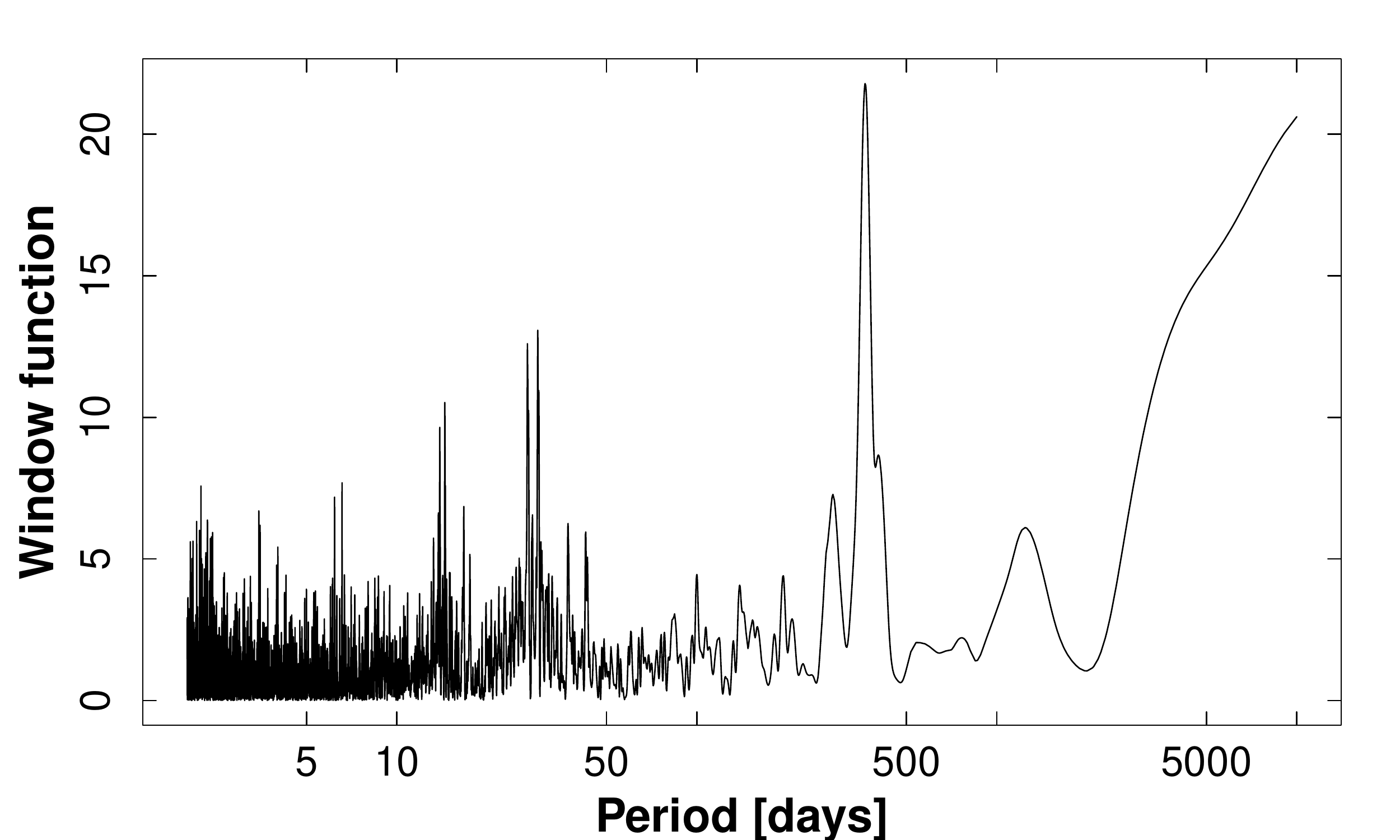}\\
\plotone{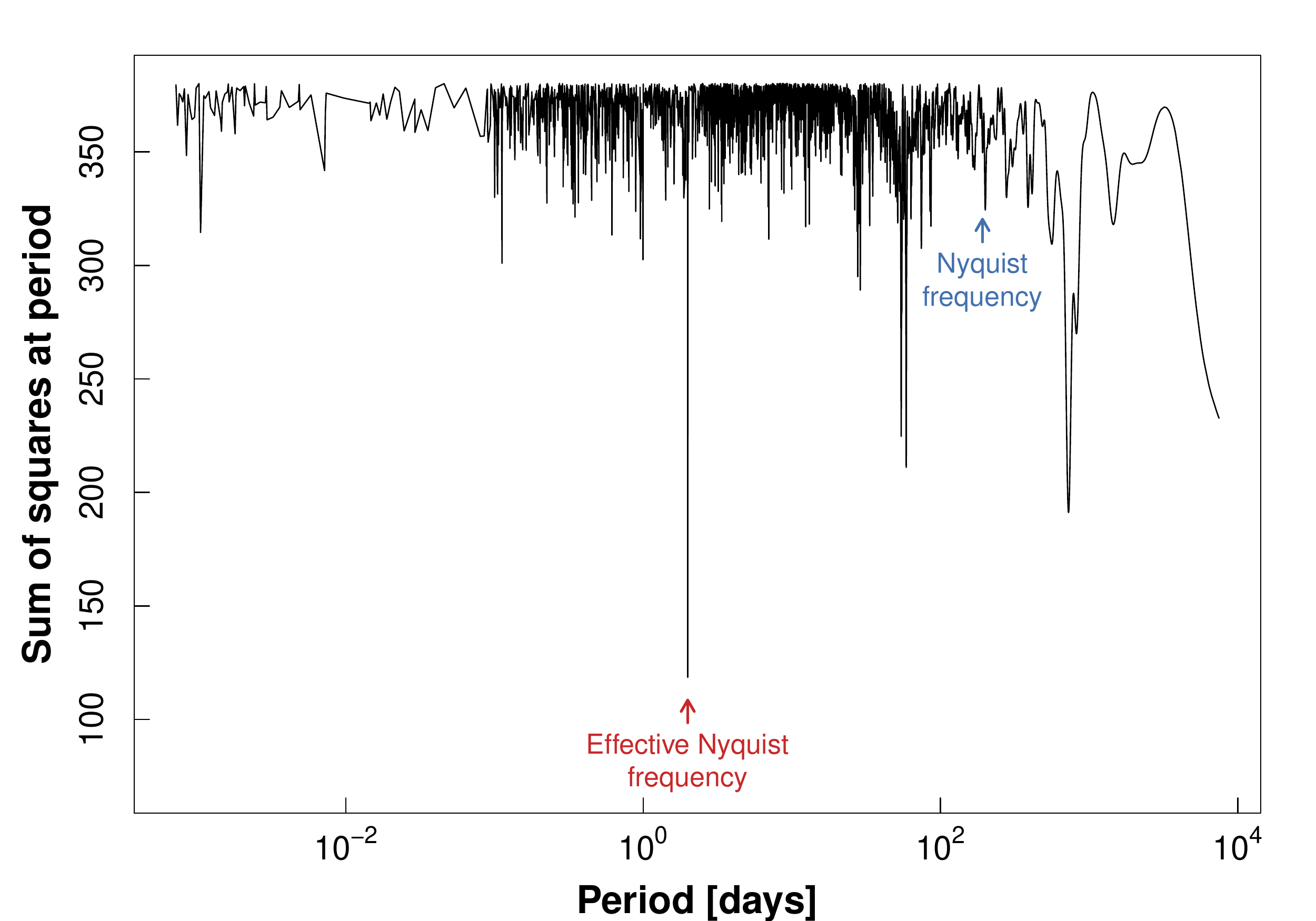}
\caption{\label{fig:hd95872_data} \textit{Top panel:}
Generalized Lomb-Scargle periodogram for the HD~95872 RV data.
False-alarm probability levels are shown at the 10\%, 1\% and 0.1\% level.
\textit{Middle panel:} Periodogram of the window function.
\textit{Bottom panel:} Determination of the ``effective'' Nyquist frequency for the data set.
Both the effective Nyquist frequency, and the corresponding Nyquist frequency for a regularly spaced data set are marked.}
\end{figure}

The second panel shows the error-weighted, normalized Lomb-Scargle periodogram (Zechmeister \& K\"urster 2009).
The three horizontal lines in the plot represent different levels of false alarm probability (FAP; 10\%, 1\% and 0.1\%, respectively).
The FAPs were computed by scrambling the data set 100,000 times, in order to determine the probability that the power at each
frequency could be exceeded by chance (e.g. K\"urster et al.~1997, Marcy et al.~2005). Computing the FAPs for this sparse data set required scanning
only frequencies that were effectively sampled by the set of observation times.
We determined an ``effective'' Nyquist frequency for the data set using the calculation formula of Koen~(2006).
For irregularly spaced data sets, the effective Nyquist frequency is much higher than the
corresponding Nyquist frequency of a regularly spaced data set of the same size.
The algorithm of Koen~(2006) finds a clear minimum at $P \approx 2$ days (bottom panel of Figure \ref{fig:hd95872_data}),
corresponding to the effective Nyquist frequency for the data. Accordingly, we exclude periods shorter than 2 days from our calculations.

\begin{figure}
\centering
\plotone{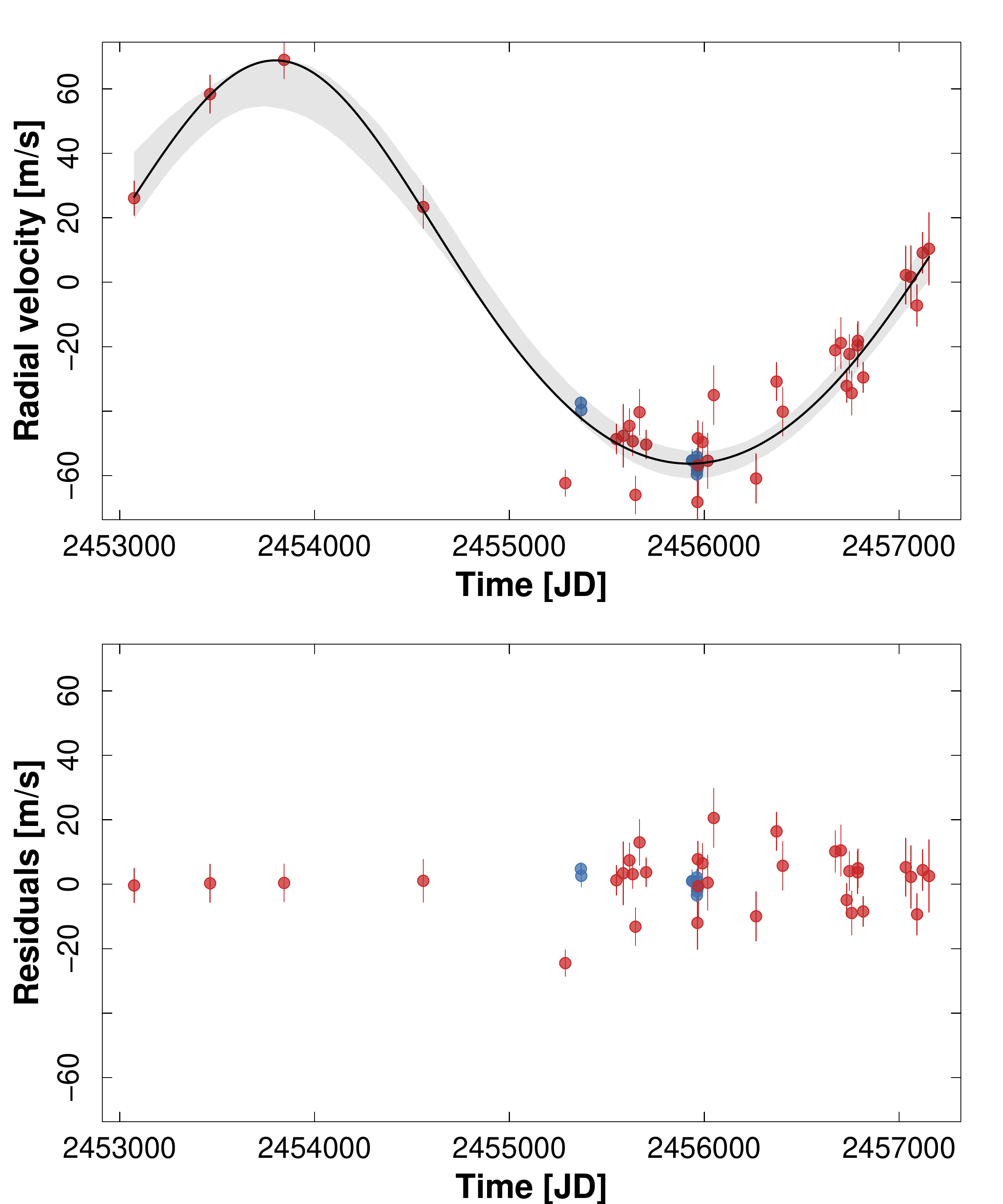}\\
\caption{\label{fig:hd95872_fit} \textit{Top panel:} 
Our RV data from HJST/Tull (red) and Keck/HIRES (blue) and the best-fit Keplerian model. The shaded area marks the 10\%-90\% percentiles of the radial 
velocity curves sampled from the MCMC trials, and indicates the range of the models compatible with the data. \textit{Bottom panel:} Radial velocity residuals.}
\end{figure}


\begin{figure}
\centering
\fbox{\plotone{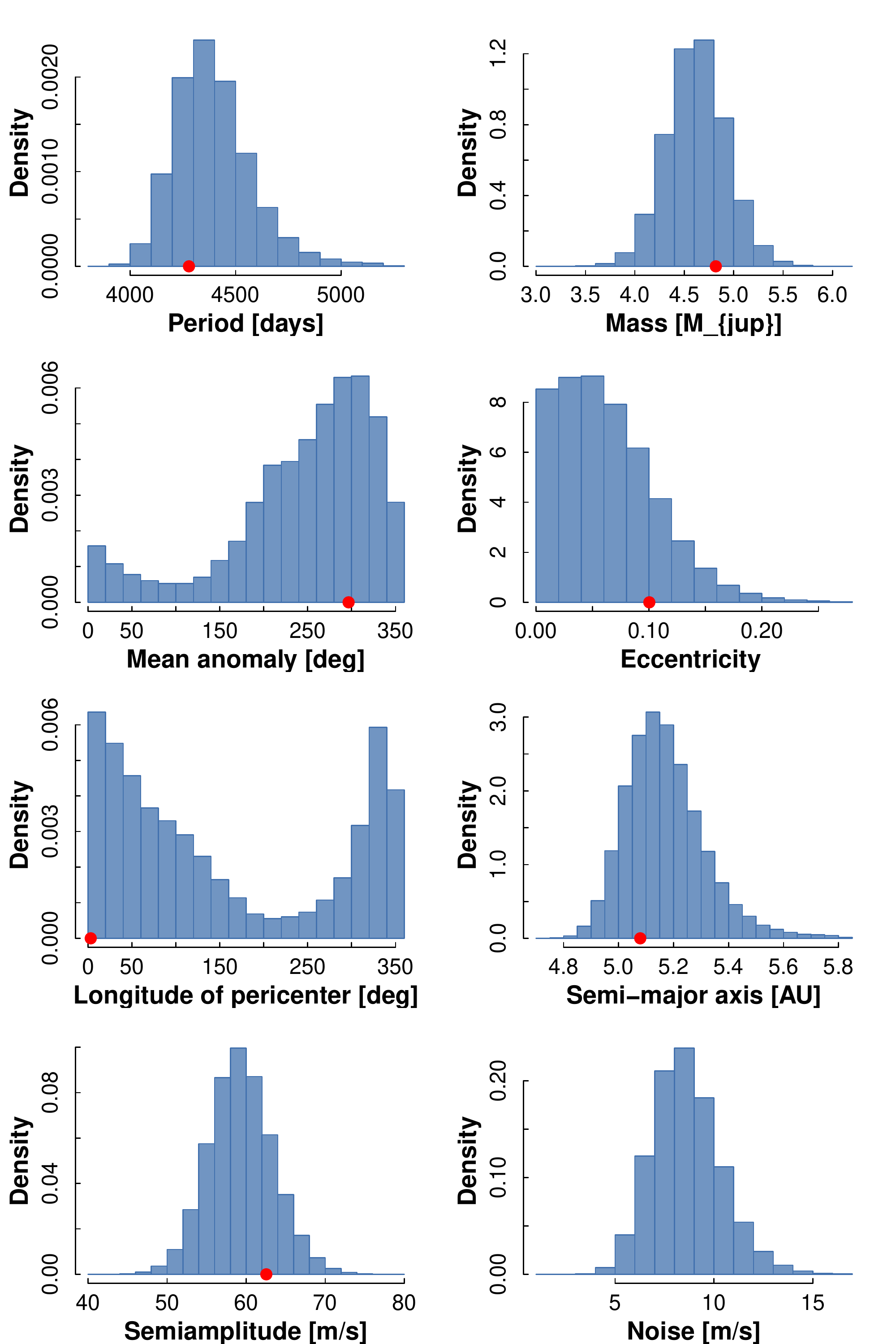}}
\caption{\label{fig:hd95872_dist} Marginal distributions of the orbital elements of HD~95872b, as computed by the Markov-Chain Monte Carlo algorithm. 
The red dot marks the value of the best-fit solution.}
\end{figure}

Visual inspection of the 44 individual RV measurements suggests the presence of a sparsely sampled, long-period signal (see top panel of 
Figure\,\ref{fig:hd95872_fit}). The Lomb-Scargle periodogram (Figure \ref{fig:hd95872_data}) bears this out.
The two strongest signals, at $P = 29.6$ days
(FAP $<\ \ensuremath{4.0 \times 10^{-5}}$) and $P = 331.2$ days (FAP = \ensuremath{1.1 \times 10^{-2}})
have significant power in the window function,
and they are likely related to the periodicities in the observational cadence (the lunar synodic month and the Solar year).
The remaining peak is at $P = 3922.05$ days (FAP = \ensuremath{2 \times 10^{-4}}).
This signal is well fit with a Keplerian orbit of period $P = 4278 \pm{} 169$ days and
semi-amplitude $K = 59 \pm{} 4$ \ms (Figure \ref{fig:hd95872_fit}).
Together with the assumed stellar mass of 0.95 \msun, this implies a minimum mass of $\msini = 4.6 \pm{} 0.3 \mjup$
and a semi-major axis $a = 5.2 \pm{} 0.13$ AU.
The best-fit orbit for the planet shows a small amount of eccentricity ($e = 0.06 \pm{} 0.04$, broadly consistent with circular).
Orbital uncertainties were derived by running a Markov Chain Monte Carlo (MCMC) algorithm 
(Ford~2005,~2006, Meschiari et al.~2009, Gregory~2011) 
on the data set.
Non-informative priors were adopted over all parameters (uniform in logarithm for mass and period).
Marginal distributions of the parameters are shown in Figure \ref{fig:hd95872_dist}; no significant correlations among
parameters were observed.
A summary of the astrocentric orbital elements of HD~95872b is reported in Table \ref{tab:hd95872}.

\begin{figure}
\centering
\plotone{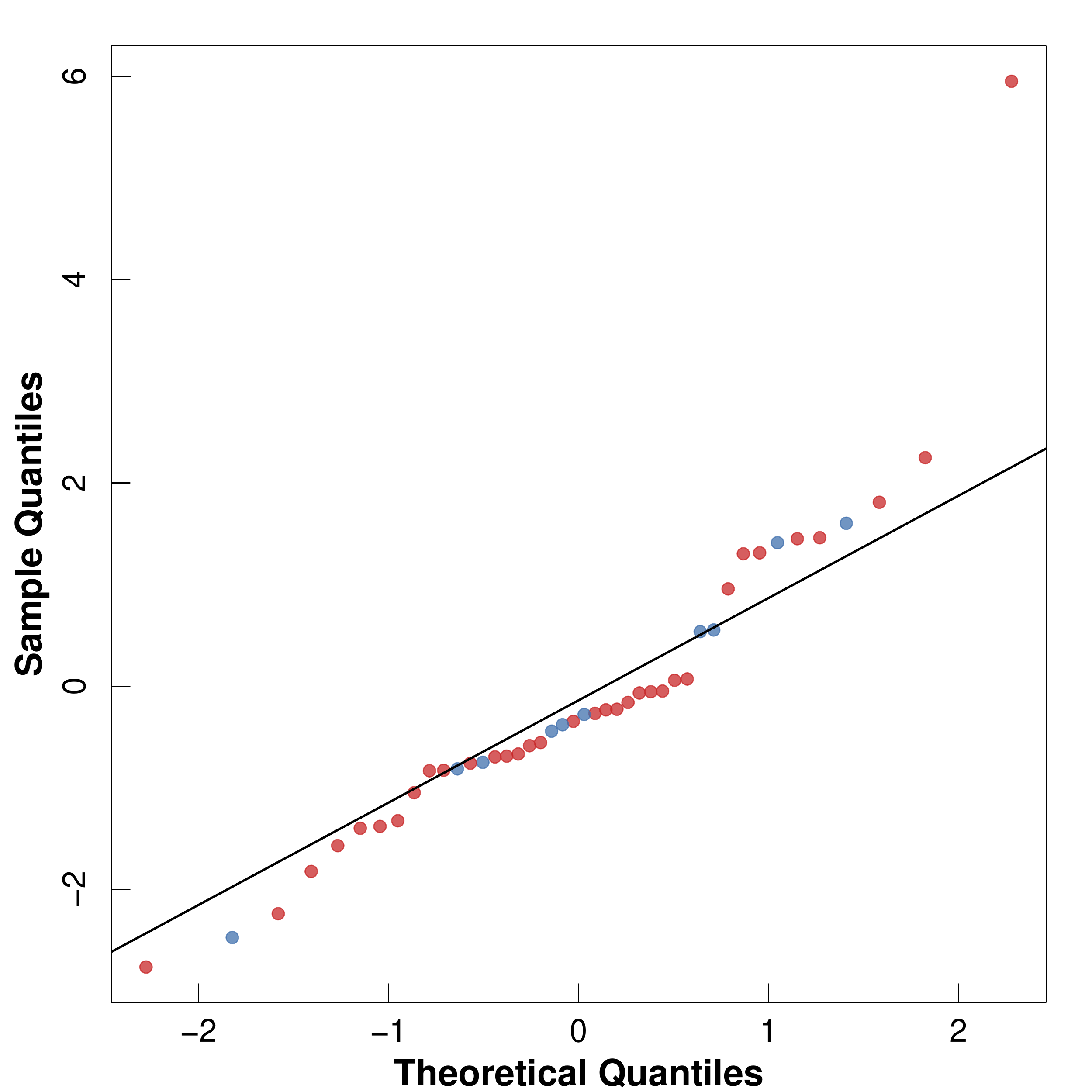}\\
\caption{\label{fig:hd95872_qq} Quantile-quantile plot of the residuals from the 1-planet model for HD~95872. Perfectly normally distributed residuals would fall on the solid line.}
\end{figure}

The one-planet fit reduces the root mean square (RMS) of the data from 46.8 \ms{} to 8.1 \ms{}. The stellar jitter for HD 95872
(that is, the amount of noise added in quadrature to the formal uncertainties required in order to completely fit the residuals)
is $8 \pm{} 2$ \ms, and is derived self-consistently from the MCMC analysis.
We note that the normalized residuals are very nearly normally distributed, aside from a single outlier (Figure \ref{fig:hd95872_qq}).


\begin{figure}
\centering
\plotone{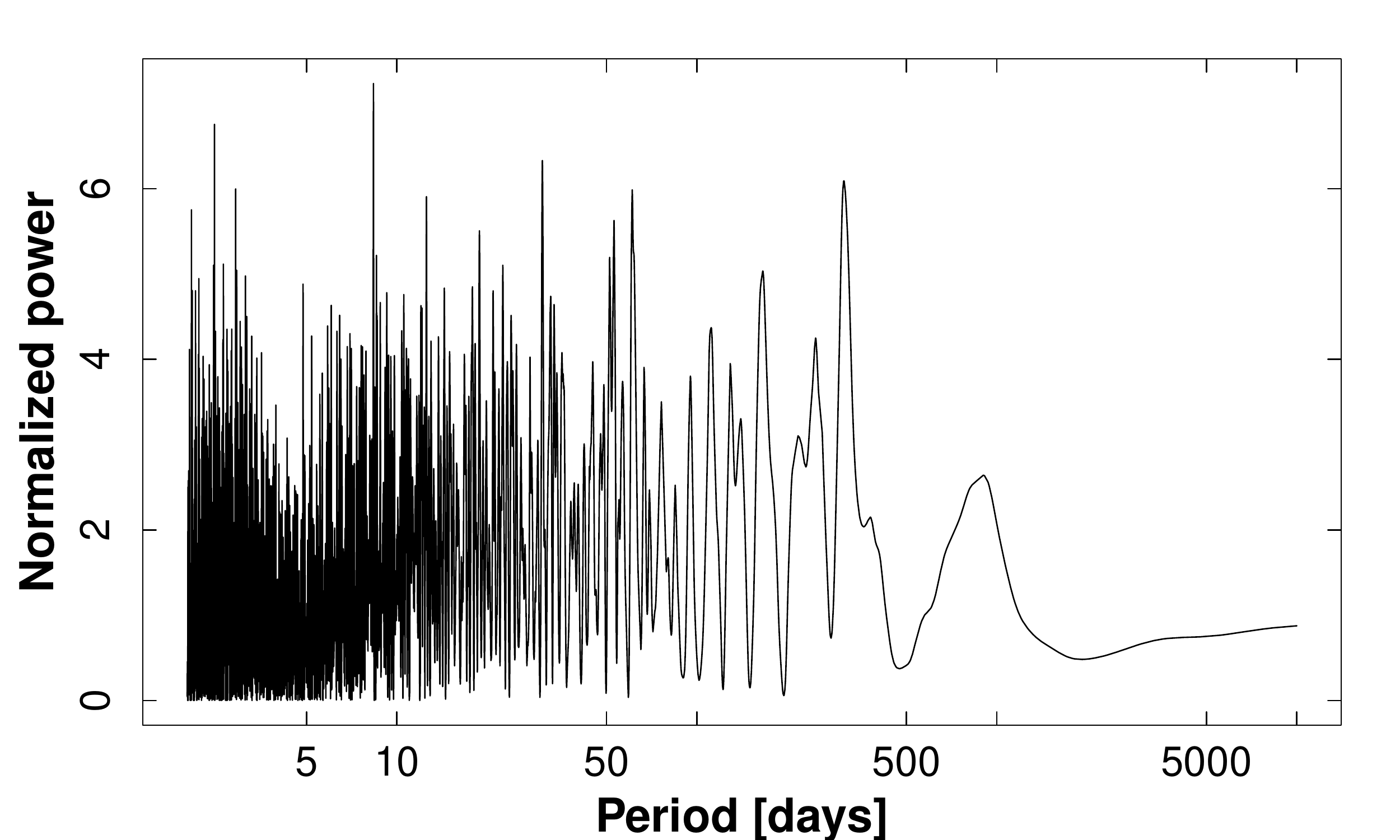}\\
\caption{\label{fig:hd95872_per} Lomb-Scargle periodogram of the residuals. All peaks have a FAP $\>$ 10\%.}
\end{figure}
Figure \ref{fig:hd95872_per} shows the Lomb-Scargle periodogram of the RV residuals from the 1-planet best fit. There is no strong periodicity (FAP $<$ 10\%) in the residuals supporting the presence of additional planets in the system.

\begin{table}[ht]
\centering
\begin{tabular}{rl}
  \hline
 & HD~95872b \\
  \hline
 Period [days] & 4375 [169] \\
   Mass [$M_{jup}$] & 4.6 [0.3] \\
   Mean anomaly [deg] & 283 [65] \\
   Eccentricity  & 0.06 [0.04] \\
   Longitude of pericenter [deg] & 17 [67] \\
   Semiamplitude [m/s] & 59 [4] \\
   Semi-major axis [AU] & 5.2 [0.1] \\
   Periastron passage time [JD] & 2449869 [744] \\
   Noise parameter, KECK data [m/s] & 0.5 [0.6] \\
   Noise parameter, McDonald data [m/s] & 8 [2] \\
   RV offset Keck/McDonald [m/s] & 19 [2] \\
   \hline
   Stellar mass [$M_{sun}$] & 0.95 \\
   RMS [m/s] & 7.90 \\
   Jitter (best fit) [m/s] & 4.80 \\
   Epoch [JD] & 2453073.87 \\
   Data points  & 44 \\
   Span of observations [JD] & 2453073.87 (Apr 2004)\\
   & 2457153.62 (May 2015)\\
   \hline
\end{tabular}
\caption{Orbital elements for HD~95872b. For parameters derived from the MCMC analysis, we report
their median values and their mean absolute deviation (in brackets).}
\label{tab:hd95872}
\end{table}

\subsection{Stellar Activity Check} \label{hd95872ch}

\begin{figure}
\centering
\plotone{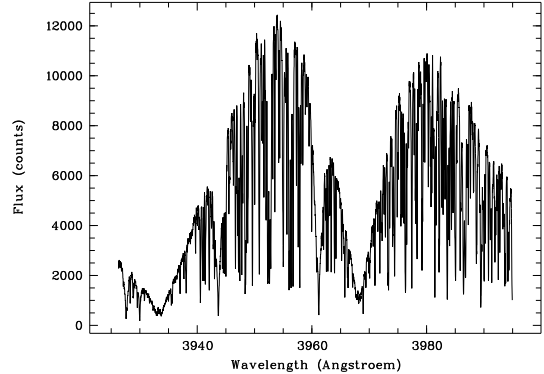}\\
\caption{\label{fig:hd95872_act} The Ca II H \& K lines of HD~95872 in our best Keck/HIRES spectrum.
The very low level of chromospheric emission in the line kernel shows that this star is quiet and inactive
with $R^{'}_{\rm HK}=-5.46\pm0.044$.}
\end{figure}

The large amplitude of $\sim \pm 60$\,m\,s$^{-1}$ of the detected RV signal makes it unlikely that a long-term magnetic cycle is responsible for it. 
The relative faintness of this star ($V=9.9$) leads to very low S/N values in the blue spectral orders that contain the Ca
II H \& K lines at 390 nm. Therefore, we cannot determine a reliable time series of S$_{HK}$ index measurements for this star using the Tull spectra. 
However, 9 HIRES spectra have sufficient S/N to obtain the S$_{HK}$ index value. We calculate the $R^{'}_{\rm HK}$ value following
Paulson et al.~(2002). We find $R^{'}_{\rm HK}=-5.46\pm0.044$ for HD~95872. This means that HD~95872 is an inactive star and that the planetary hypothesis for the 
detected RV variation is the preferred one. Figure\,\ref{fig:hd95872_act} shows the Ca II H \& K lines from the HIRES spectrum with the highest S/N. 
There is nearly no chromospheric emission detectable in the line cores, in agreement with the very low value of $R^{'}_{\rm HK}$.  

Stellar activity can also manifest itself as variation of the average line shape. We therefore 
measured the velocity span of the line bisector (BVS) in the iodine-free regions of our Tull spectra. We find a mean BVS
value of $-0.05$\,km\,s$^{-1}$ with an rms-scatter of $0.077$\,km\,s$^{-1}$. The average 1$\sigma$ error on the
BVS results is $0.052$\,km\,s$^{-1}$. We do not find any gross variability in the line bisectors that would
cast doubt on the planetary origin of the signal. 
The average uncertainty of the BVS measurements is comparable to the detected RV signal which limits the usefullness of this analysis. 
The large uncertainty of $\sim 50 \ms$ is -- again -- due to the low S/N of spectra of this relative faint target star.     

\section{The \pd~System} \label{psi1dra}

The $\psi^{1}$~Draconis system is a visual binary composed of an F5~V primary ($\psi^{1}$ Dra A, 31~Dra A, HR~6636, HD~162003, HIP~86614) 
and an F8~V secondary star ($\psi^{1}$ Dra B, 31~Dra B, HR~6637, HD~162004, HIP~86620) separated by about 30.1 arcsec.  
At a distance of 22.2~pc, this corresponds to a sky-projected separation of approximately 667~AU.  
Previously, Toyota et al.~(2009) reported evidence of an unseen companion orbiting the A component of the system, with a minimum mass of 50 M$_{J}$.  
We have monitored both stars for long-term RV variability and also find evidence for a stellar-mass companion around the A component.
Moreover, we discovered two planetary/sub-stellar companions orbiting the B component. 
Thus, the $\psi^{1}$~Draconis system is at least a hierarchical triple system, with the primary 
having a low-mass, K- or M-dwarf, companion and the secondary having two candidate planetary/sub-stellar companions.

\subsection{$\psi^{1}$~Draconis~A} \label{psi1draa}

\begin{figure}
\centering
\plotone{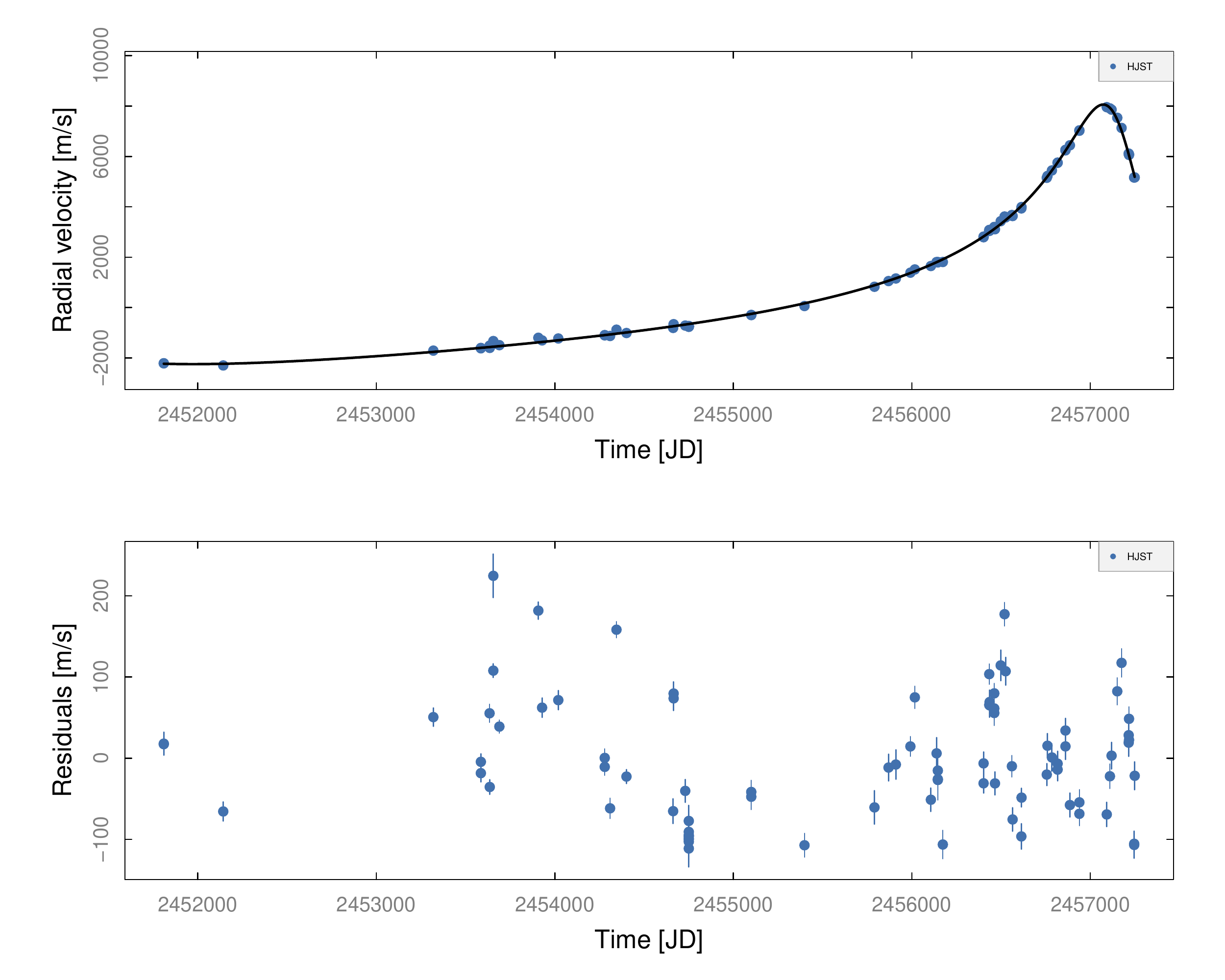}\\
\caption{\label{fig:psi1draa} Top panel: our RV data for $\psi^{1}$ Dra~A showing an eccentric binary orbit with a period of $P\approx6650$~d (nearly 20 years).
Bottom panel: RV residuals from the binary orbit. We find a large excess scatter of $\approx70$\,\ms indicating significant intrinsic stellar variability.}
\end{figure}

Table \ref{tab:psi1draa_data} presents the complete set of our RV observations for the primary star $\psi^{1}$~Dra A. 
The RV coverage spans nearly 15 years of monitoring over 77 RV measurements. 
The median internal uncertainty for our RV data is $\approx$ 15 \ms, and the peak-to-peak velocity change is $>$ 10,000 \ms, typical for a
stellar companion. The most recent RV measurements revealed that the star has passed the maximum of its RV orbit and is now approaching 
a periastron passage (see Figure~\ref{fig:psi1draa}). 

\begin{figure}
\centering
\plotone{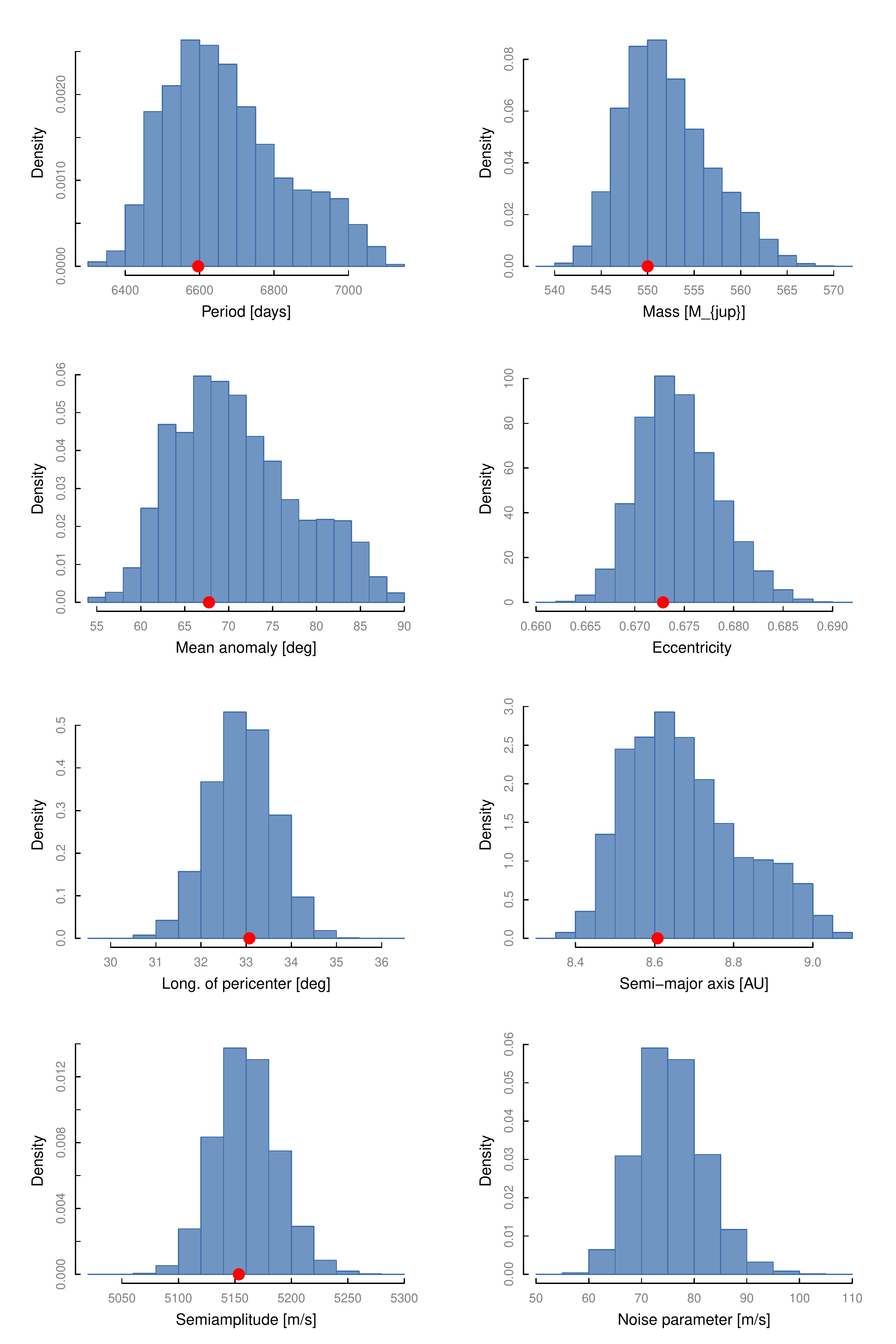}\\
\caption{\label{fig:psi1draa_pars} 
Marginal distributions of the orbital elements for the single-lined spectroscopic binary orbit of $\psi^{1}$~ Draconis~A, as computed by the Markov-Chain Monte Carlo algorithm.
The red dot marks the value of the best-fit solution.}
\end{figure}

We performed a similar orbit fitting analysis as in the case of HD~95872. The marginal distributions of the orbital elements are shown in Figure ~\ref{fig:psi1draa_pars}.
The binary orbit due to the stellar companion to $\psi^{1}$~Dra~A has a period of $P\approx6600$~d, an eccentricity of $e\approx0.67$ and a semi-amplitude of
$K\approx5160$\,\ms. These values are consistent with a low mass stellar companion ($\psi^{1}$~Dra~C) to the primary at an orbital separation of $a\approx9$~AU. 
Table~\ref{tab:psi1draa_tab} summarizes the orbital elements that we determined from the RV data. 

\begin{table}[ht]
\centering
\begin{tabular}{lccc}
  \hline
 & BJD & dRV & Uncertainty \\
 & & (m\,s$^{-1}$) & (m\,s$^{-1}$) \\
  \hline
  1 & 2451809.6596 &  1946.7 &   12.8 \\
  2 & 2451809.6740 &  1947.4 &  14.3\\
  3 & 2452142.6805 &  1862.1 &  11.9\\
  4 & 2453319.6392 &  2450.7 &  11.4\\
  5 & ... & ... & ... \\
   \hline
\end{tabular}
\caption{Differential radial velocity measurements for $\psi^{1}$ Draconis~A (sample)}
\label{tab:psi1draa_data}
\end{table}

\begin{table}[ht]
\centering
\begin{tabular}{rl}
\hline
Parameter & Value [uncertainty]\\
  \hline
   Period [days] & 6649 [160] \\
   Mass [$M_{jup}$] & 551 [5] \\
   Mean anomaly [deg] & 70 [7] \\
   Eccentricity  & 0.674 [0.004] \\
   Long. of pericenter [deg] & 32.9 [0.7] \\
   Semiamplitude [m/s] & 5159 [27] \\
   Semi-major axis [AU] & 8.7 [0.1] \\
   Periastron passage time [JD] & 2450515 [162] \\
             &          \\
   Noise parameter [m/s] & 75 [6] \\
   \hline
   Stellar mass [$M_{sun}$] & 1.430 \\
   Chi-square  & 85.673 \\
   Log likelihood  & 486.353 \\
   RMS [m/s] & 74.231 \\
   Jitter (best fit) [m/s] & 72.617 \\
   Epoch [JD] & 2451809.660 \\
   Data points  & 85 \\
   Span of observation [JD] & 2451809.6596 (Sep. 2000) \\
   & 2457248.6070 (Aug. 2015) \\
   \hline
\end{tabular}
\caption{Orbital elements for the single-lined spectroscopic binary orbit of $\psi^{1}$~Draconis~A \& C.}
\label{tab:psi1draa_tab}
\end{table}

\begin{figure}
\centering
\plotone{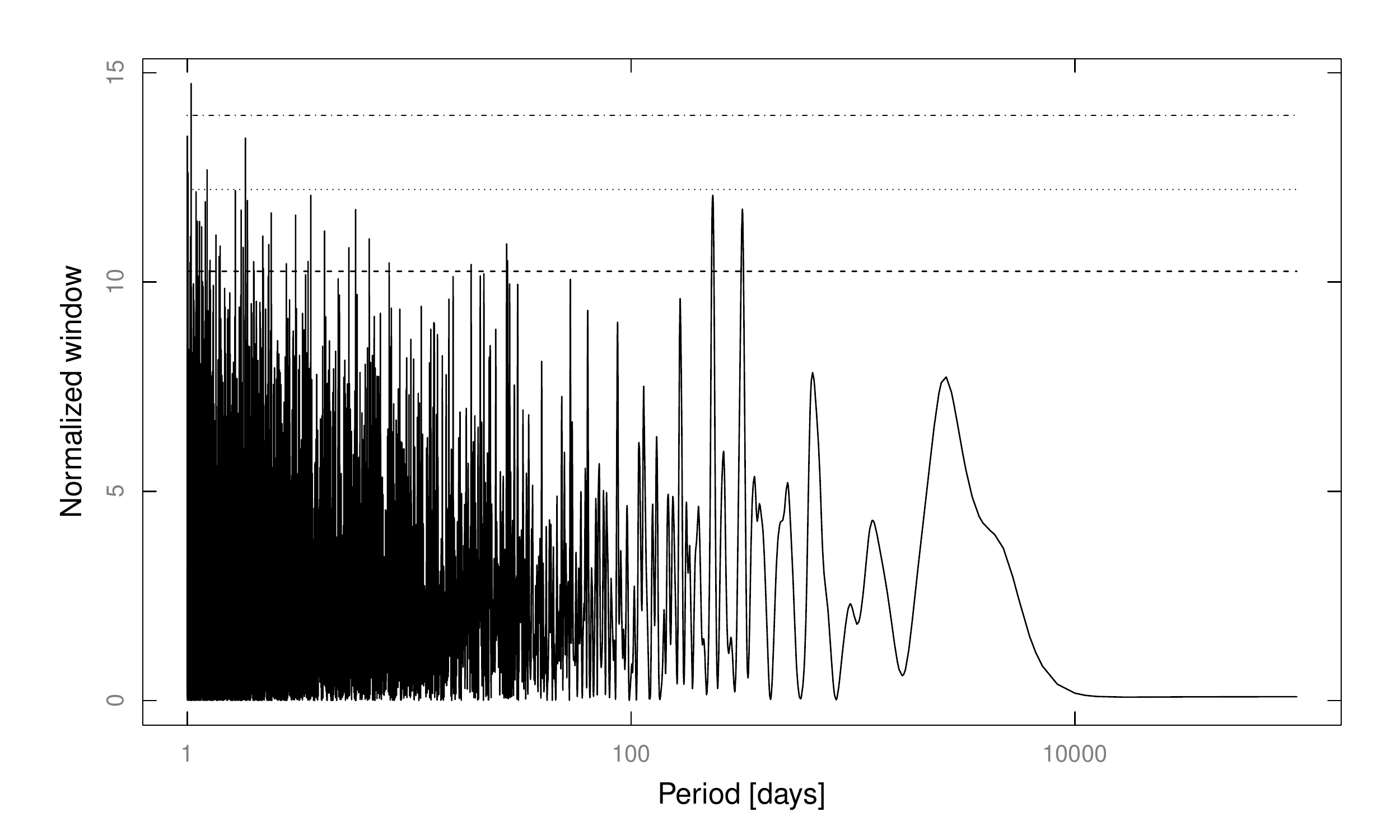}\\
\caption{\label{fig:psi1draa_per}
Generalized Lomb-Scargle periodogram of the best-fit RV residuals from the binary orbital solution. The horizontal dashed lines show
FAP levels of 10\%, 1\% and 0.1\% respectively.}
\end{figure}

\begin{figure}
\centering
\plotone{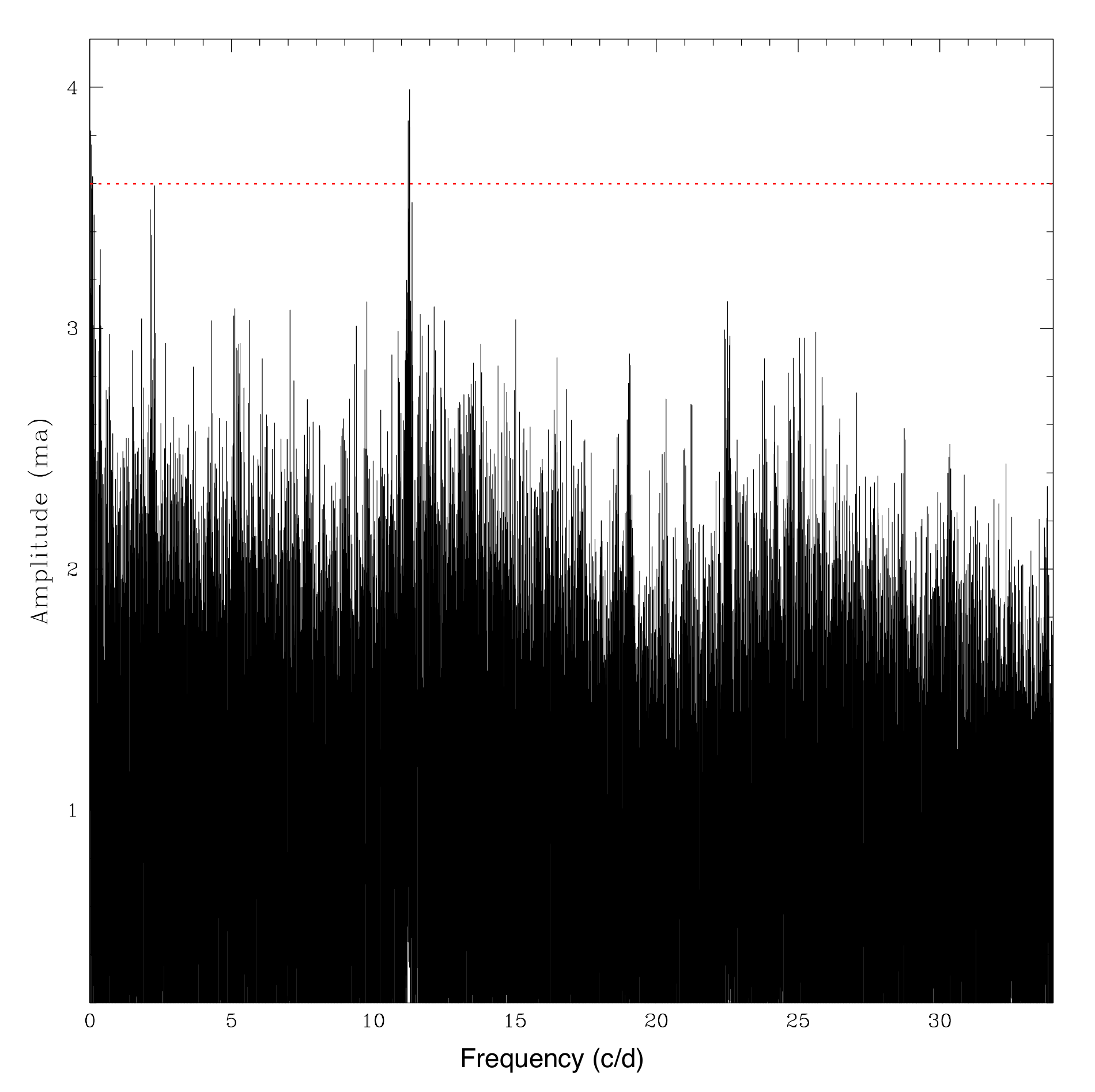}\\
\caption{\label{fig:hipp}
Fourier transform of the {\it Hipparcos} photometry of $\psi^{1}$~ Dra~A revealing a strong peak at a frequency of 11.29 cycles/day ($P = 2.1$ hours).
The horizontal dashed line shows the 4$\sigma$-level (FAP$\approx0.001$) of significance.     
The detected period of 2.1~hours is typical for non-radial stellar oscillations of a classic $\delta$~Scuti variable.}
\end{figure}

One striking feature of these orbital solutions are the large values of residual RV scatter around the fit considering our typical
RV uncertainties of $\approx 15$\,\ms. The models require an astrophysical
noise term of $\approx75$\,\ms to achieve a good fit. The Lomb-Scargle periodogram of the best-fit RV residuals (Figure~\ref{fig:psi1draa_per}) does 
not show any convincing periodic signals that could indicate additional companions in the system. However, with an F5V spectral type classification, $\psi^{1}$~ Dra~A is one 
of earliest spectral types in our target list. In the HR-diagram this star is located close to the red edge of the instability strip. 
We therefore examined the {\it Hipparcos} photometry (ESA~1997) of $\psi^{1}$~Dra~A
to search for stellar pulsations. The Fourier-transform of the photometry is displayed in Figure\,\ref{fig:hipp}. 
We find a peak at a period of 2.1~hours (= 11.29 cycles/day) with a modulation amplitude of $> 4\sigma$.  
This period value falls within the range of a few hours of typical p-mode oscillations of $\delta$~Scuti-type pulsators (e.g. Balona, Daszy\'nska-Daszkiewicz \& Pamyatnykh~2015).
We therefore suspect that these stellar oscillations are responsible for the large observed excess scatter.     
  
In section\,\ref{imaging} we will discuss in more detail the detection of $\psi^{1}$~Dra~C, the stellar companion, by direct imaging. Owing to the small angular separation between 
$\psi^{1}$~ Dra~A and C we expect that some of the residual scatter is also caused by contamination from light from the faint companion star that also entered through the
spectrograph slit. In a companion paper (Gullikson et al.~2015) we successfully retrieve the Doppler signal of the low-level secondary spectrum and thus determine a double-lined
spectroscopic orbital solution for $\psi^{1}$~ Dra~A/C.    

\subsection{$\psi^{1}$~Draconis~B} \label{psi1drab}

\begin{table}[ht]
\centering
\begin{tabular}{lccccc}
  \hline
 & BJD & dRV  & Uncertainty & $S_{HK}$ & Uncertainty\\ 
 & & (m\,s$^{-1}$) & (m\,s$^{-1}$)& & \\
  \hline
1 & 2451066.7344 & -48.8 & 6.1 & 0.155 & 0.0198\\ 
  2 & 2451121.6124 & -48.5 & 3.8 & 0.161 & 0.0213\\ 
  3 & 2451271.9939 & -39.1 & 7.3 & 0.163 & 0.0197\\ 
  4 & 2451329.8559 & -33.4 & 5.5 & 0.162 & 0.0206\\ 
  5 & 2451360.8829 & -39.1 & 4.2 & 0.162 & 0.0214\\ 
  6 & 2451417.7778 & -24.5 & 5.0 & 0.173 & 0.0214\\ 
  7 & 2451451.6921 & -24.9 & 6.1 & 0.167 & 0.0217\\ 
  8 & ... & ... & ...& ...& ...\\
   \hline
\end{tabular}
\caption{Differential radial velocity and Ca H\&K observations for 
$\psi^{1}$ Dra B (sample)} 
\label{tab:psidrab_data}
\end{table}

Table \ref{tab:psidrab_data} presents the complete set of our RV observations for $\psi^{1}$~Dra B. The RV coverage spans approximately 16 years of monitoring over 135
measurements. The median internal uncertainty for our observations is $\approx$ 5.6 \ms, and the peak-to-peak velocity is $\approx$ 62 \ms.
The velocity scatter around the average RV is $\approx$ 14.3 \ms.

\subsubsection{Companion Orbit Models} \label{psi1drabb}

\begin{figure}
\centering
\plotone{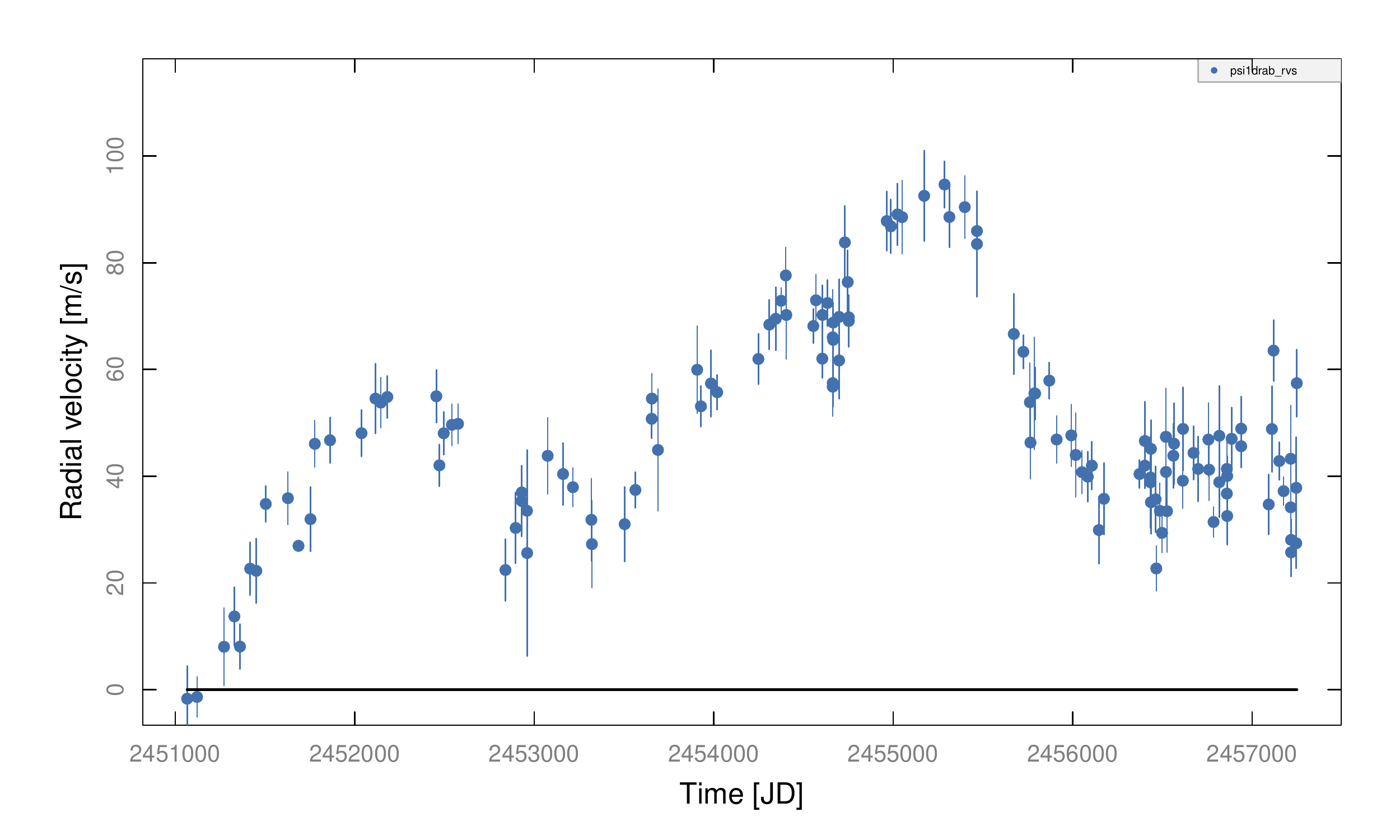}\\
\plotone{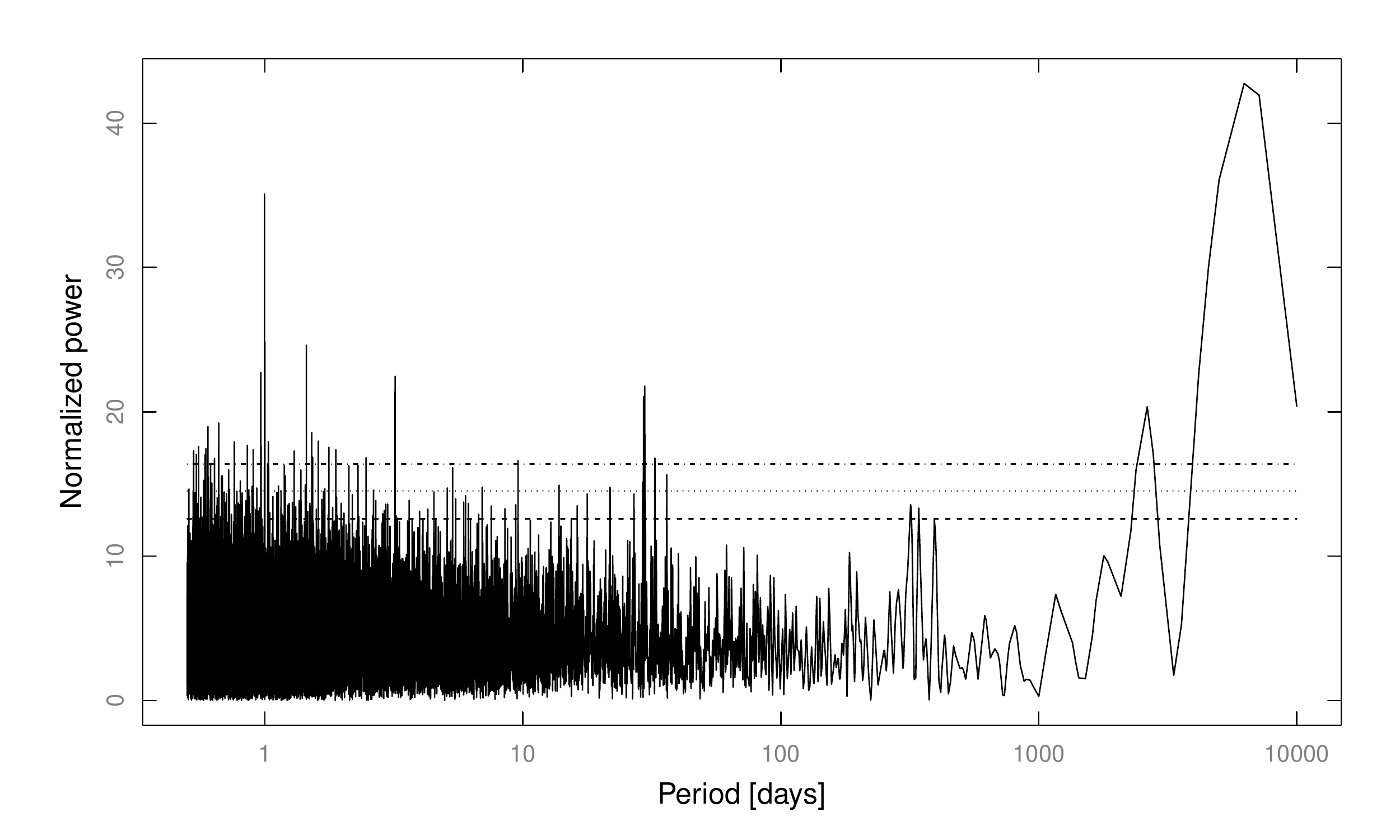}\\
\plotone{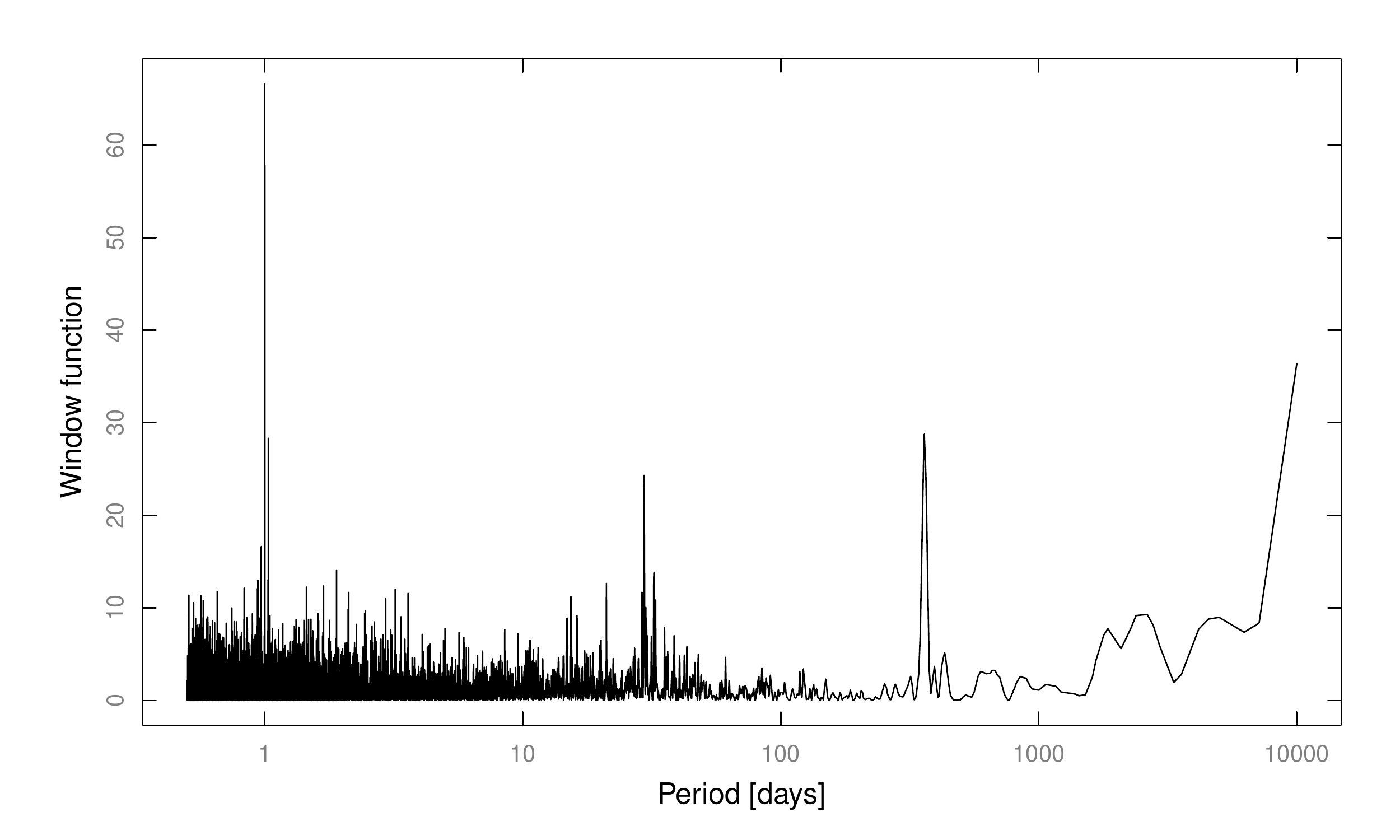}\\
\caption{\label{fig:psidra_data} Radial velocity and Lomb-Scargle periodograms for $\psi^{1}$~Dra B. \textit{Top panel:} Our differential RV data. \textit{Middle panel:}
Error-weighted Lomb-Scargle periodogram for $\psi^{1}$~Dra B. False-alarm probability levels are shown at the 10\%, 1\% and 0.1\% level.
\textit{Bottom panel:} Periodogram of the window function. }
\end{figure}

The differential RV data for $\psi^{1}$~Dra B are plotted in Figure \ref{fig:psidra_data}. The Lomb-Scargle periodogram (Figure \ref{fig:psidra_data}) for the
RV data shows two strong peaks at $P_1 \approx 2381$ days and  $P_2 > 6000$ days (longer than the time span of our observations).
We model the second signal with two parameters representing a linear and a quadratic term (evaluated at the epoch of the fit).

\begin{figure}
\centering
\plotone{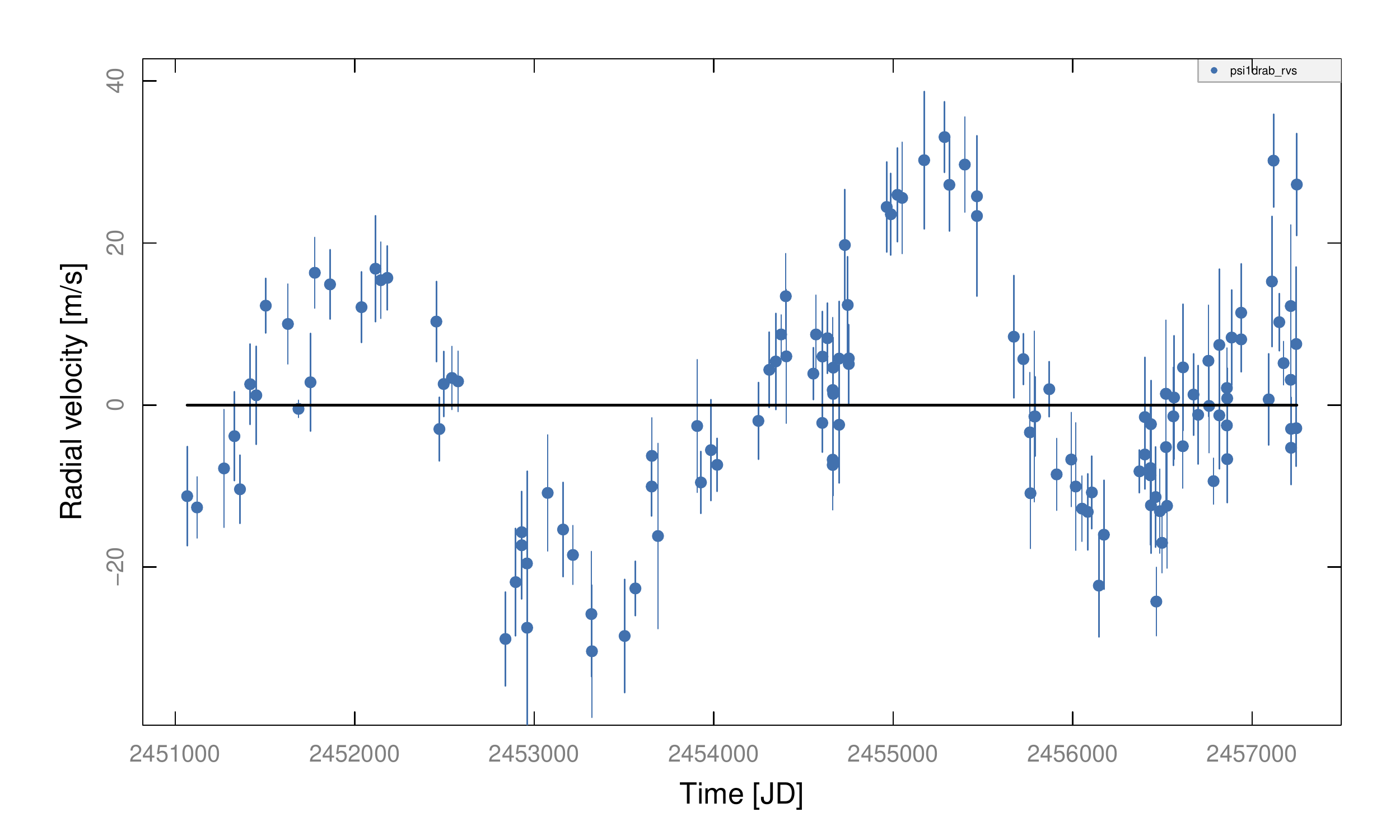}\\
\plotone{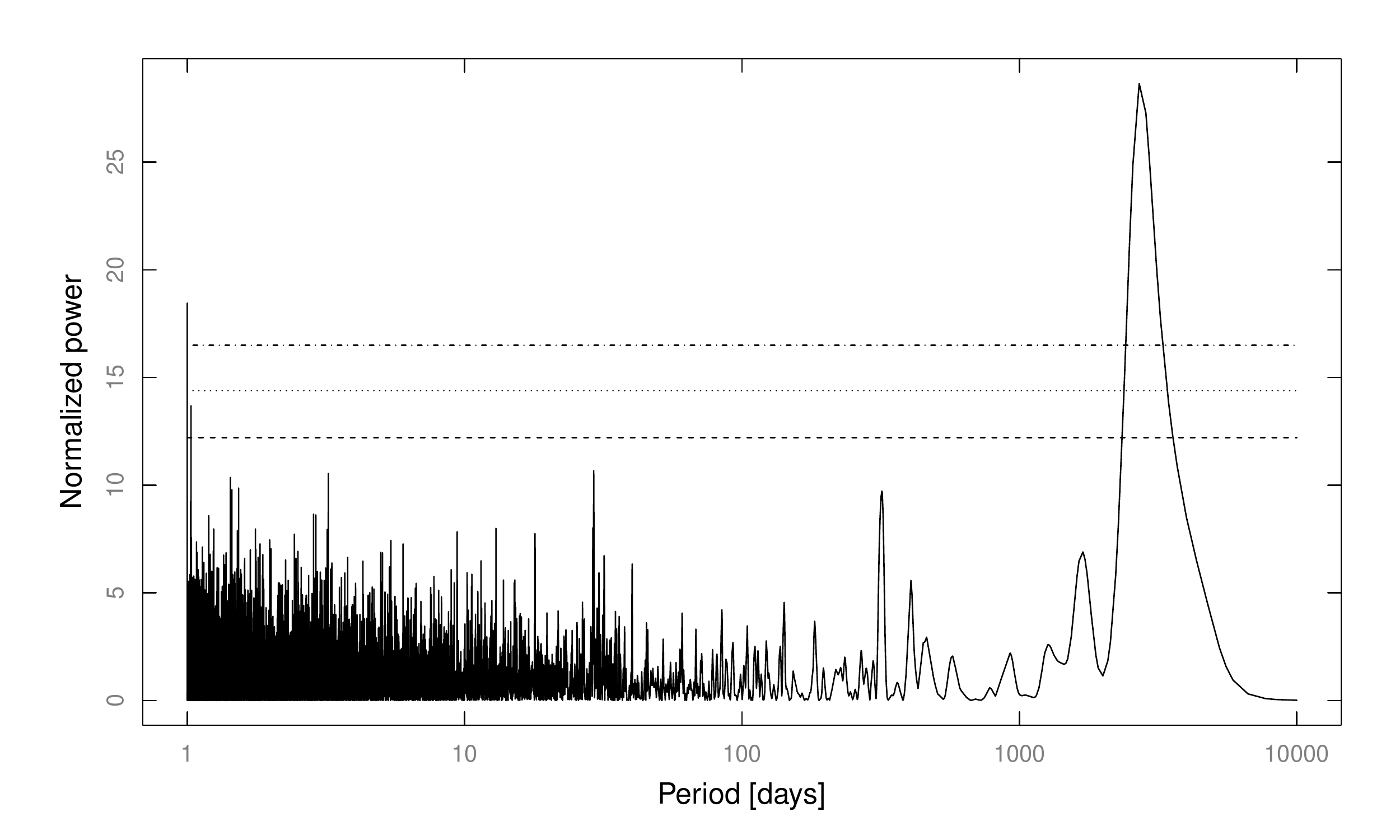}\\
\caption{\label{fig:psidra_lq_data} Radial velocity and Lomb-Scargle periodograms for $\psi^{1}$~Dra B, with the linear and quadratic trends removed.
\textit{Top panel:} Relative
RV data. \textit{Bottom panel:} Error-weighted Lomb-Scargle periodogram for $\psi^{1}$~Dra B.
False-alarm probability levels are shown at the 10\%, 1\% and 0.1\% level.}
\end{figure}

\begin{figure}
\centering
\plotone{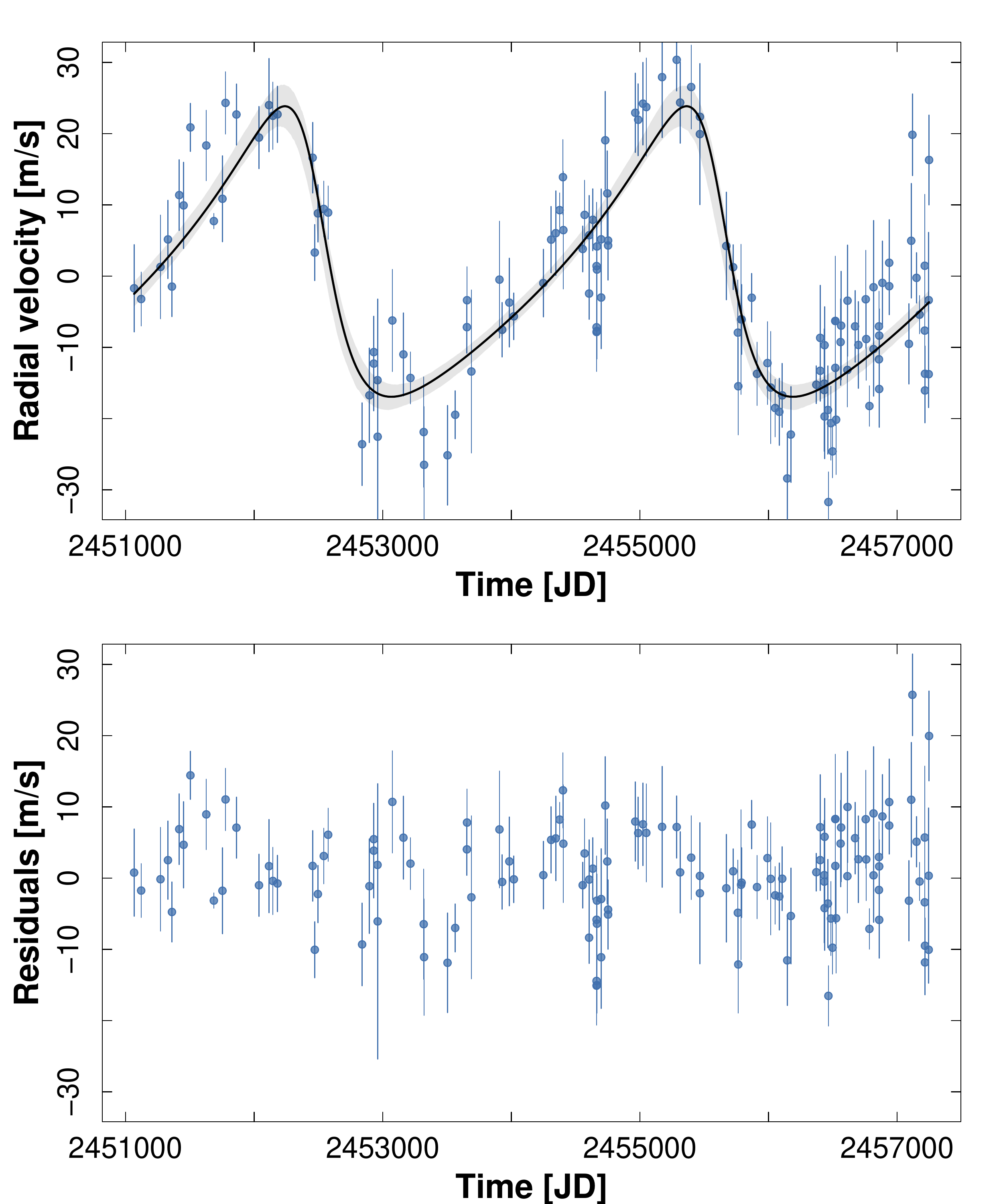}\\
\plotone{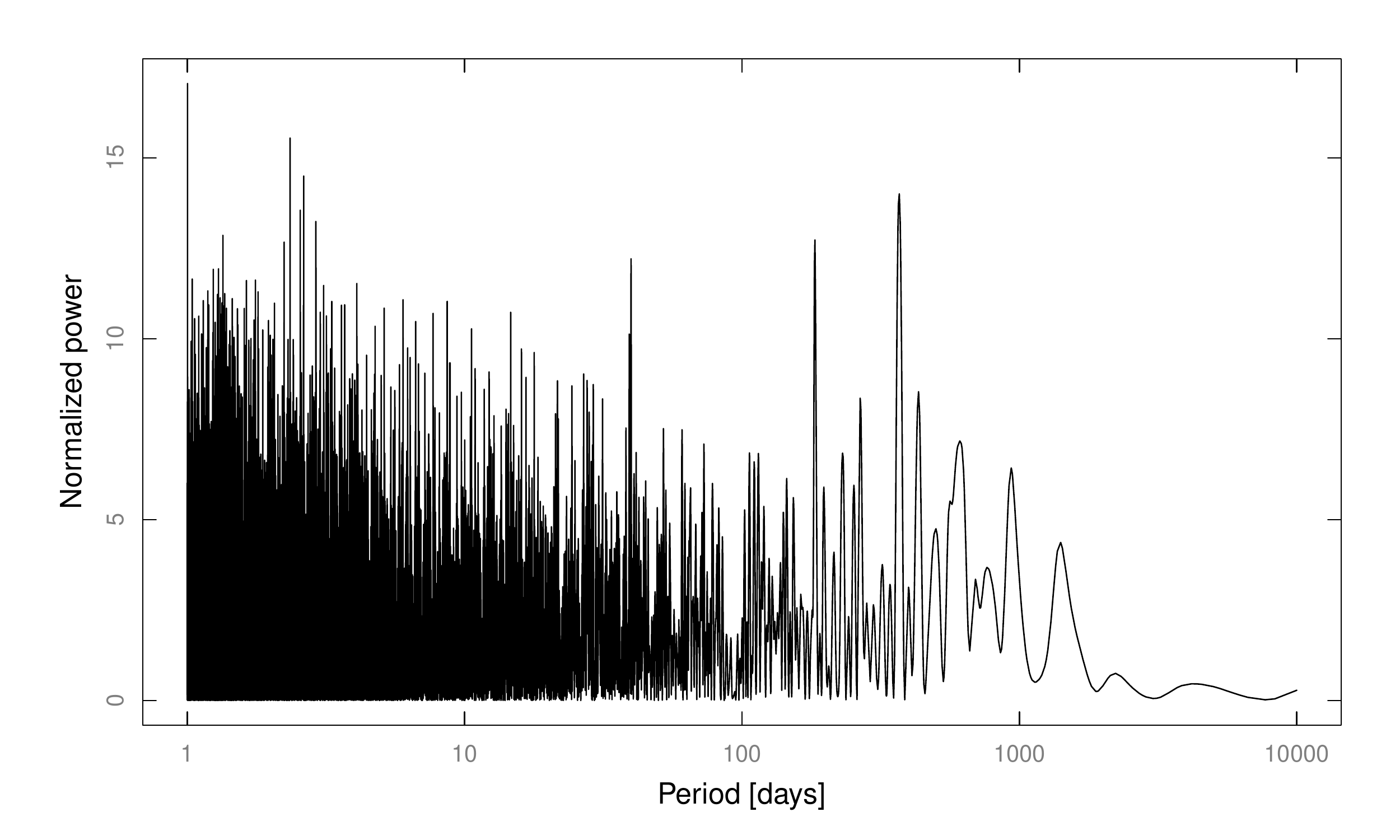}\\
\caption{\label{fig:psidra_1pl_fit} Best 1-planet fit of the RV data set for $\psi^{1}$~Dra B.
\textit{Top:} Radial velocity observations (linear and quadratic term subtracted)
and 1-planet best fit. The shaded area marks the 10\%-90\% percentiles of the radial velocity curves sampled from the MCMC trials,
and indicates the range of the models compatible with the data.
\textit{Middle:} Residuals from the 1-planet best fit. \textit{Bottom:} Periodogram of the residuals from the 1-planet best fit.}
\end{figure}

\begin{figure}
\centering
\plotone{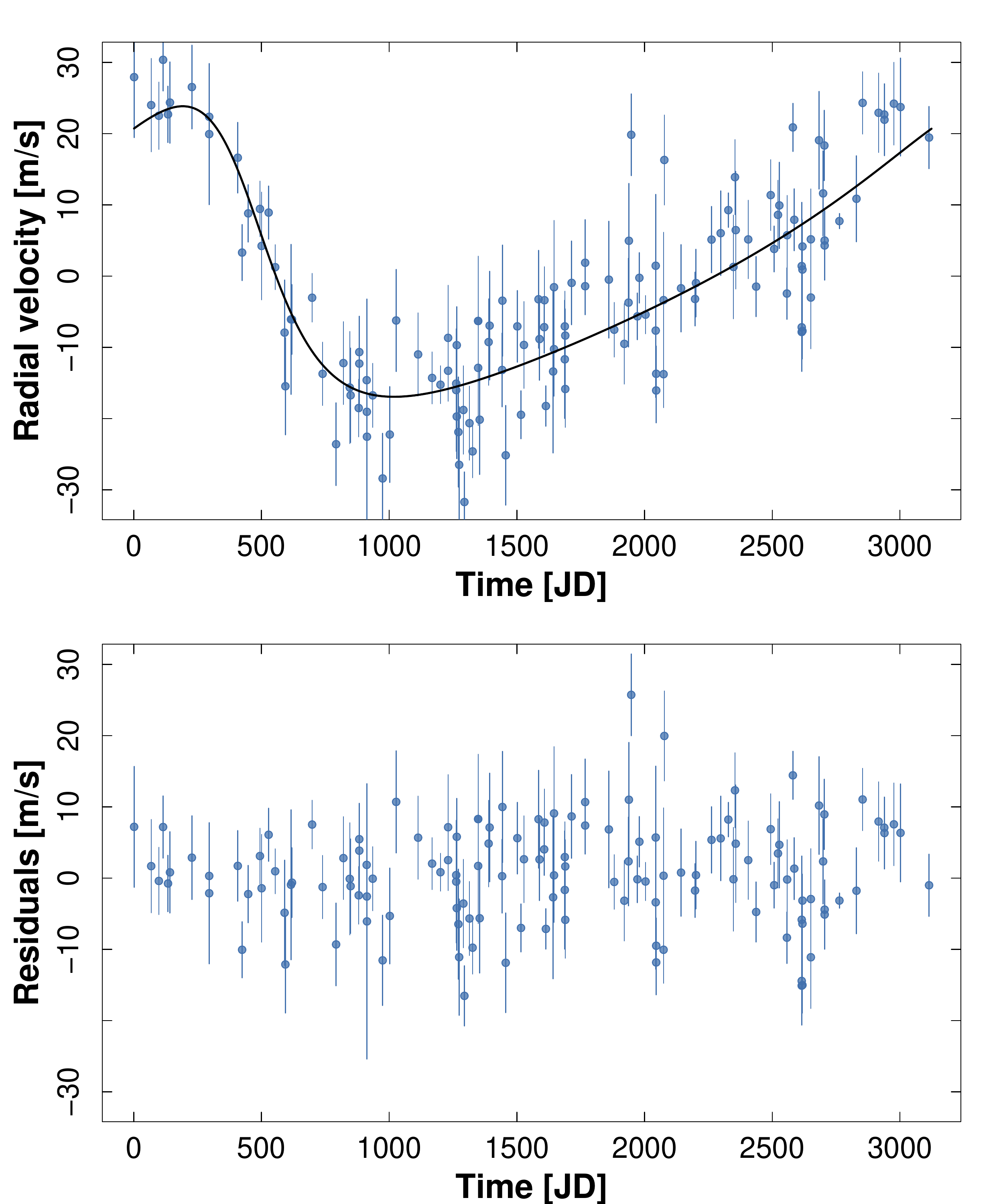}\\
\caption{\label{fig:psidra_1pl_fit_phased} Phased best 1-planet fit of the RV data set for $\psi^{1}$~Dra B. \textit{Top:} Radial velocity observations (linear and
quadratic term subtracted) and 1-planet best fit. \textit{Bottom:} Residuals from the 1-planet best fit.}
\end{figure}

Once the linear and quadratic trend terms are removed (Figure \ref{fig:psidra_lq_data}), a strong periodicity arises at $P \approx 3030$ days. The
bootstrapped FAP probability is very low ($FAP < 2\times 10^{-5}$). We fit this periodicity with a model that simultaneously minimizes the linear and
quadratic trend terms and the five orbital elements describing an eccentric orbit (period, mass, mean anomaly, eccentricity and longitude of periastron). The
best-fit model is shown in Figure \ref{fig:psidra_1pl_fit}. The data are well modeled by a Keplerian orbit of period $P = 3117 \pm{} 42$ days and
semi-amplitude $K = 20.6 \pm{} 1.4$ \ms{}.
Together with the assumed stellar mass of 1.19 \msun, this implies a minimum mass of $\msini = 1.53 \pm{} 0.09 \mjup$ and a semi-major axis $a = 4.43 \pm{} 0.04$ AU.
No compelling peaks are evident in the periodogram of the residuals (bottom panel in Figure~\ref{fig:psidra_1pl_fit}. Figure~\ref{fig:psidra_1pl_fit_phased} displays the
RV data phased to the orbital period of the planet.

The data strongly favor a substantial eccentricity for $\psi^{1}$~Dra~Bb ($e = 0.40 \pm{} 0.05$). The cross-validation algorithm (Andrae et al. 2010) corroborates the clear 
preference for an eccentric model ($\log\mathcal{L}_{circular} \approx 0.02$ vs. $\log\mathcal{L}_{eccentric} \approx -21.3$; lower is better).

\begin{figure}
\centering
\fbox{\plotone{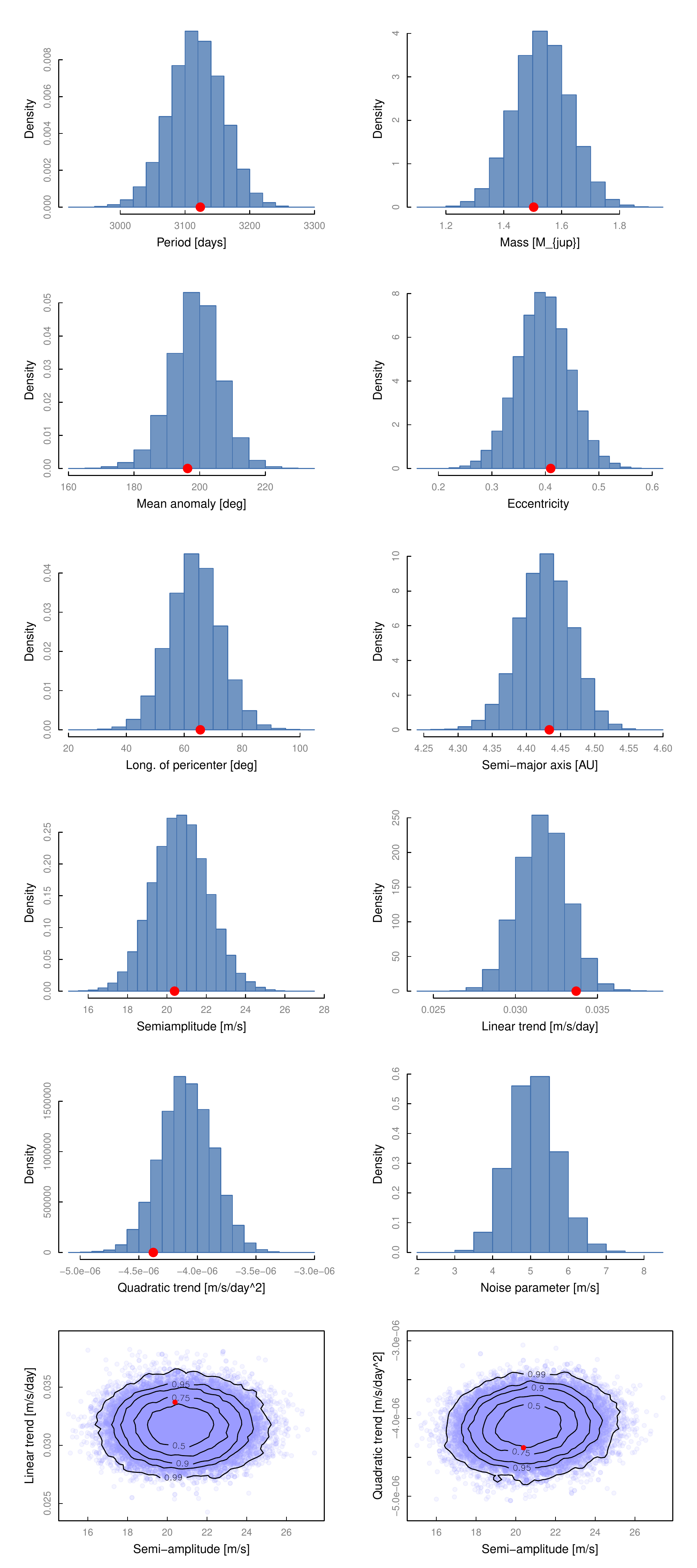}}
\caption{\label{fig:psidra_1pl_dist} Marginal distributions of the orbital elements for the 1-planet model, as computed by the Markov-Chain Monte Carlo algorithm.
The red dot marks the value of the best-fit solution.
The bottom row shows a contour plot of the planet semi-amplitude $K$ versus the linear and quadratic trend parameters.}
\end{figure}

\begin{figure}
\centering
\plotone{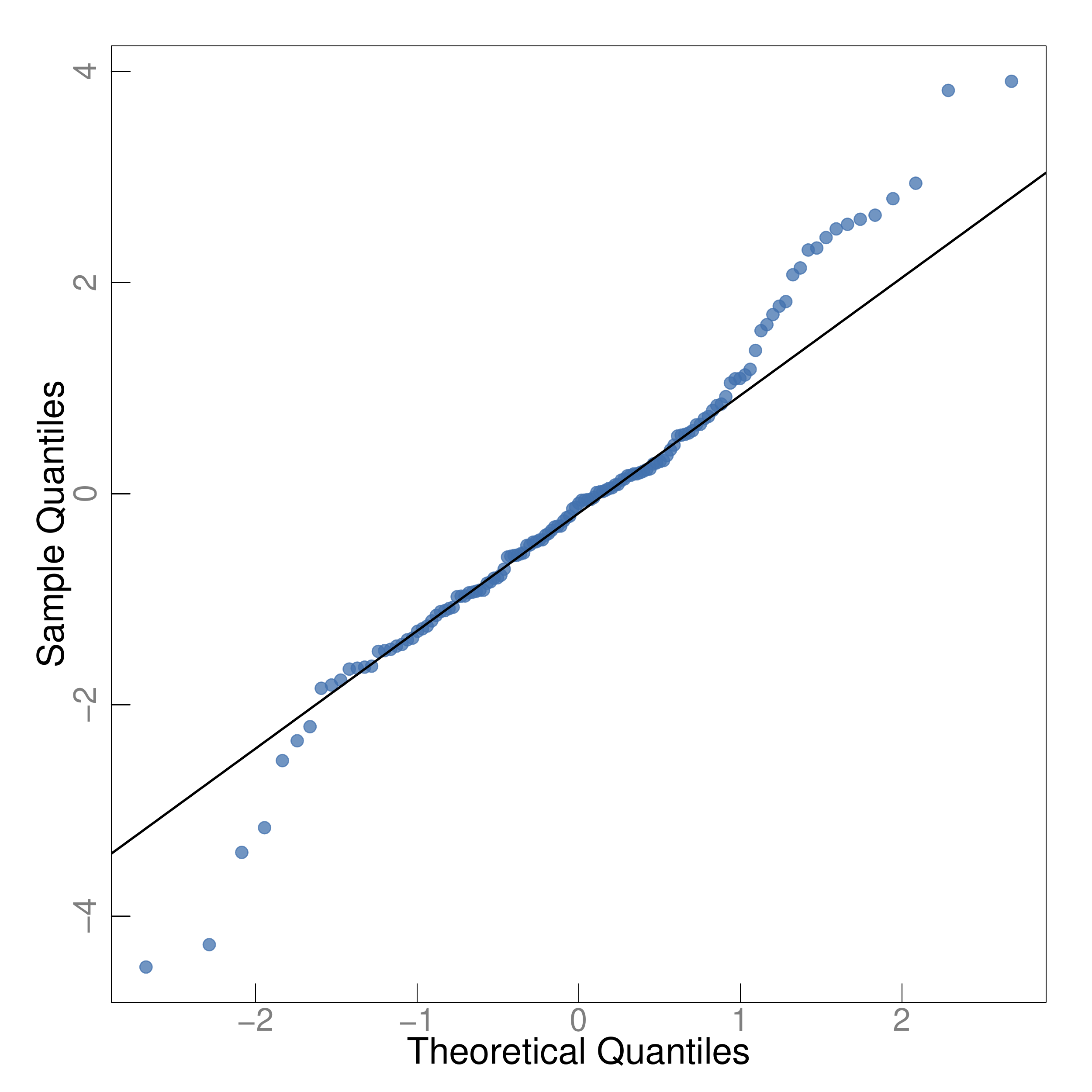}\\
\caption{\label{fig:psidra_1pl_qq} Quantile-quantile plot of the residuals from the 1-planet model.
Perfectly normally distributed residuals would fall on the solid line.}
\end{figure}

The distribution of the orbital elements is shown in Figure~\ref{fig:psidra_1pl_dist}.
There is no strong correlation between any of the parameters of the fit, including between the trend parameters and the semi-amplitude of the planet (bottom row).
The derived stellar jitter is $4.5 \pm{} 0.7$ \ms. The distribution of the residuals shows no evidence for unmodeled periodicities in the data.
Indeed, we note that the normalized residuals are again very nearly normally distributed (Figure \ref{fig:psidra_1pl_qq}).

\begin{table}[ht]
\centering
\begin{tabular}{rl}
  \hline
 & $\psi^{1}$~Dra Bb \\
  \hline
 Period [days] & 3117 [42] \\
   Mass [$M_{jup}$] & 1.53 [0.10] \\
   Mean anomaly [deg] & 199 [7] \\
   Eccentricity  & 0.40 [0.05] \\
   Long. of pericenter [deg] & 64 [9] \\
   Semiamplitude [m/s] & 21 [1] \\
   Semi-major axis [AU] & 4.43 [0.04] \\
   Periastron passage time [JD] & 2449344 [76] \\
  Noise parameter [m/s] & 4.5 [0.7] \\
  Quadratic trend [$\mathrm{m/s^2}$] & -0.0000041 [0.0000002] \\
  Linear trend [m/s] & 0.032  [0.002] \\
\hline
  Stellar mass [$\mass_\mathrm{sun}$] & 1.19 \\
   RMS [m/s] & 7.048 \\
   Jitter (best fit) [m/s] & 3.250 \\
   Epoch [JD] & 2451066.734 \\
   Data points  & 135 \\
   Span of observations [JD] & 2451066.7344 (Oct. 1998)\\
   & 2457248.6109 (Aug. 2015)\\
  \hline
\end{tabular}
\caption{Astrocentric orbital elements for $\psi^{1}$~Dra Bb. For parameters derived from the MCMC analysis, we report their median values and their mean absolute
deviation (in brackets).}
\end{table}

\subsubsection{Origin of the trend}
In this Section, we investigate the nature of the long-term trend observed in the data. In particular, we ascertain whether $\psi^{1}$~Dra A ($a \approx$ 600 AU,
$\mass_A \approx 1.38 \msun$, $P \approx 9.4\times 10^4$ years; Toyota et al.~2009) is the source of the long-term trend.

\begin{figure}
\centering
\plotone{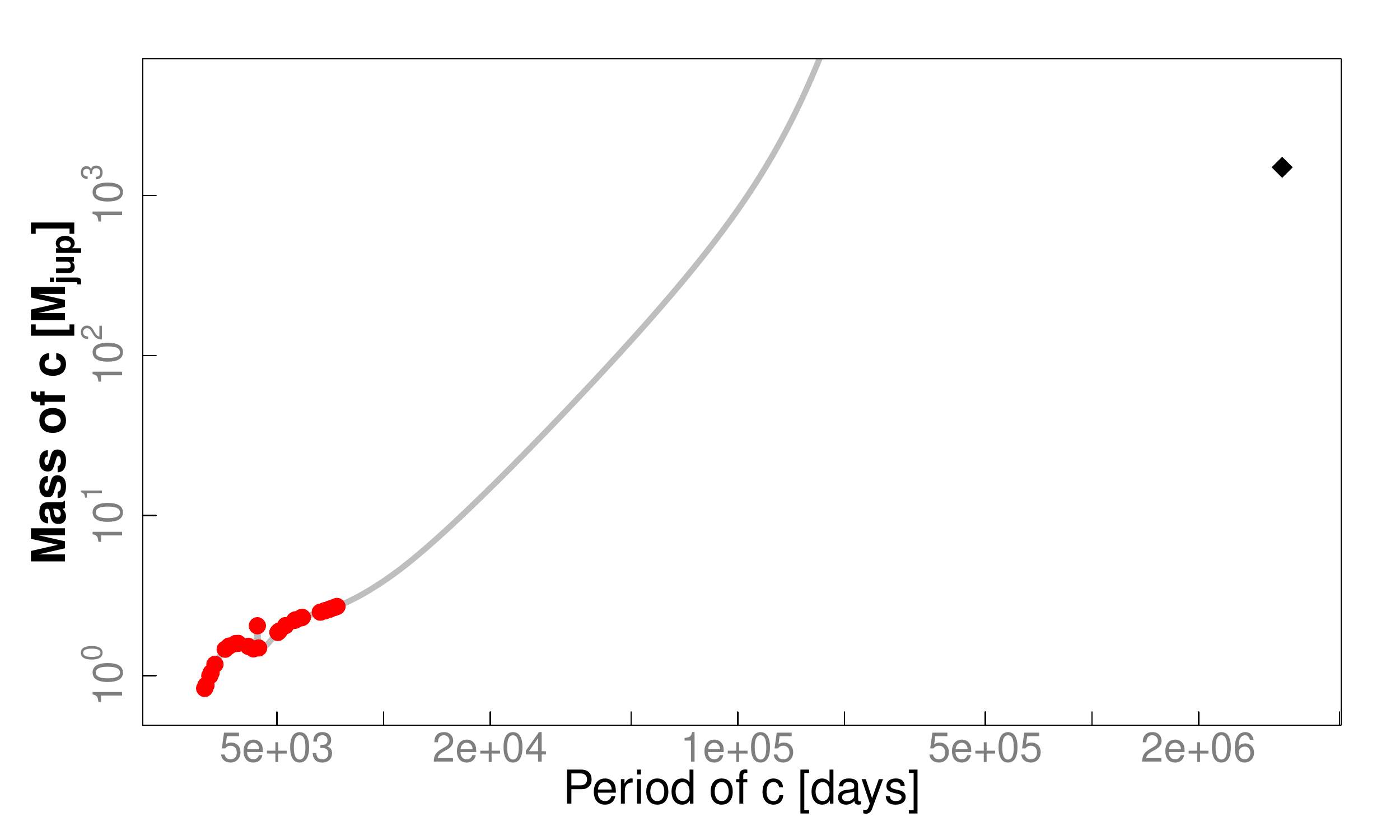}\\
\plotone{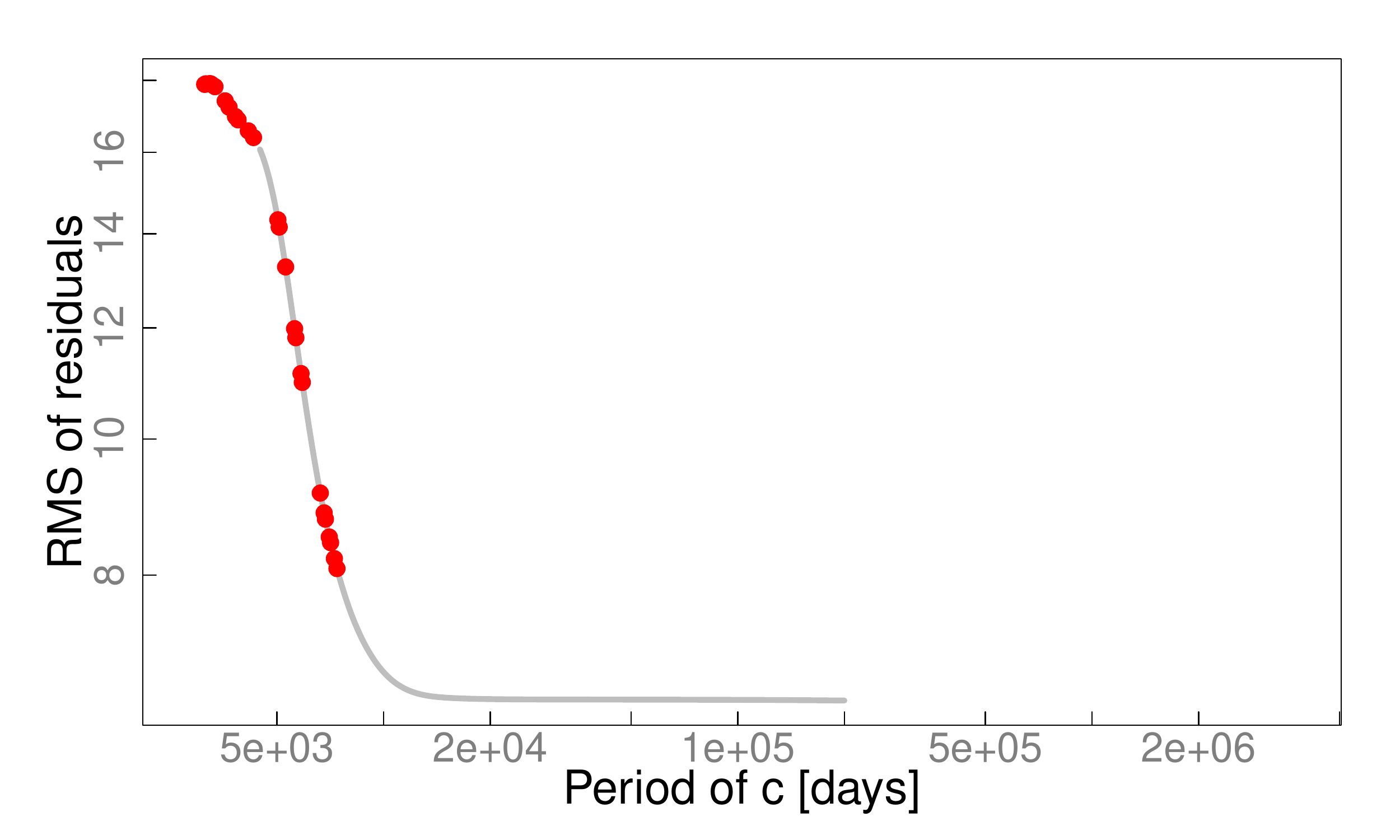}\\
\caption{\label{fig:psidra_trend_circ} \textit{Top:} Correlation between the period and the mass of an outer body in a circular orbit that best fits the trend in the RV
data. The red points correspond to systems that were unstable over a $10^6$ years period. The black diamond marks the semi-major axis and mass of
component$\psi^{1}$~Dra A from Toyota et al. (2009).
\textit{Bottom:} RMS of the residuals for the best-fit at each orbital period of the outer companion.
At periods larger than $\approx 10^4$ days, the marginal distributions of the period and mass of the outer companion are flat.}
\end{figure}

To model the long-term trend, we first assume that the gravitational pull is provided by an external perturber ($\psi^{1}$~Dra~Bc) in a circular orbit. We fit the data
by fixing the eccentricity of the perturber to zero and sampling periods between 4,000 days and 15,000 years.
The top panel of Figure \ref{fig:psidra_trend_circ} shows the best-fit for the mass of the perturber at each period sampled.
The goodness of the fit (as measured by the RMS of the residuals) is shown in the bottom panel.
Beyond approximately $10^4$ days, the RMS is flat and the period and mass of the perturber are degenerate.
We note that component A cannot be the source of the long-term trend, given the minimum mass required for A at the observed binary separation.

\begin{figure}
\centering
\plotone{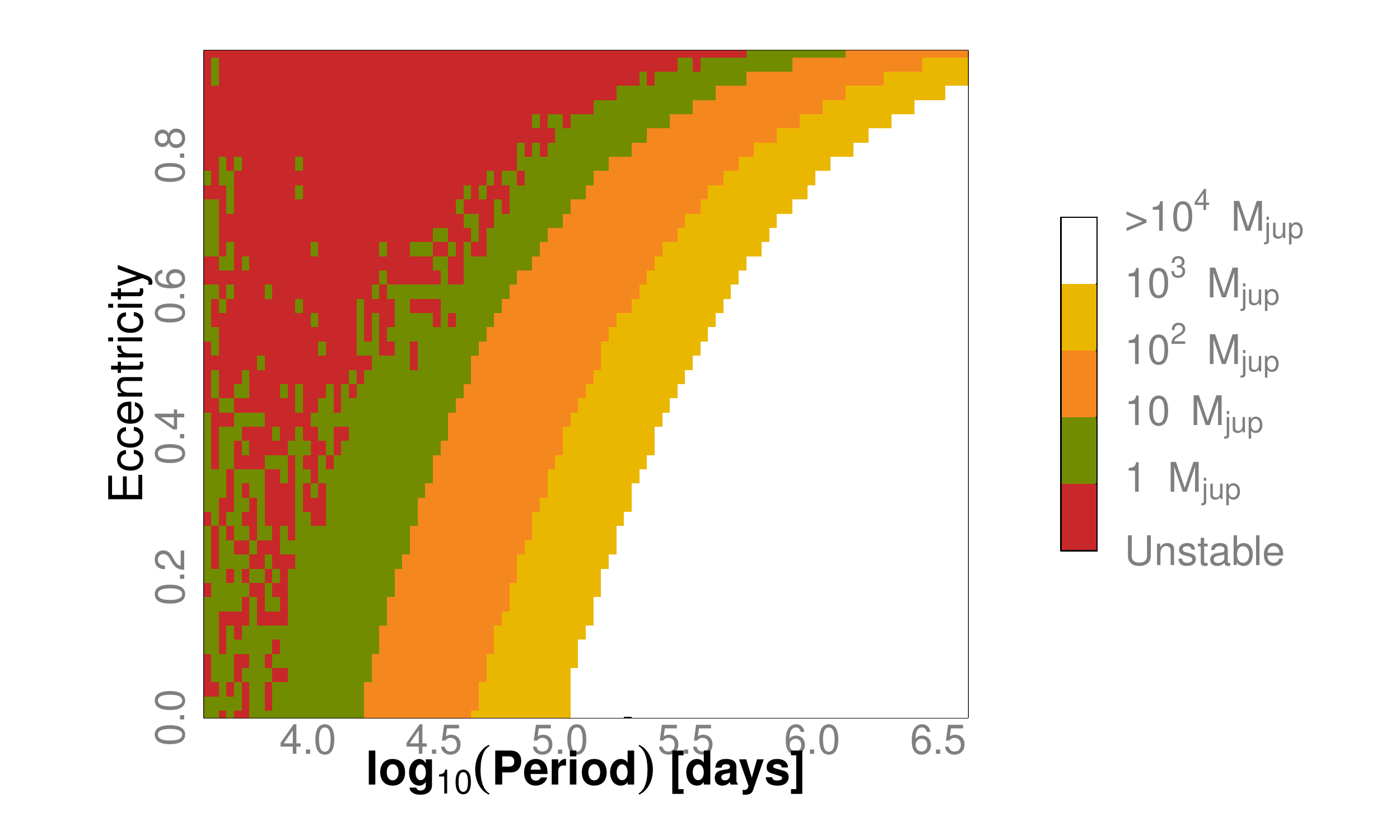}\\
\plotone{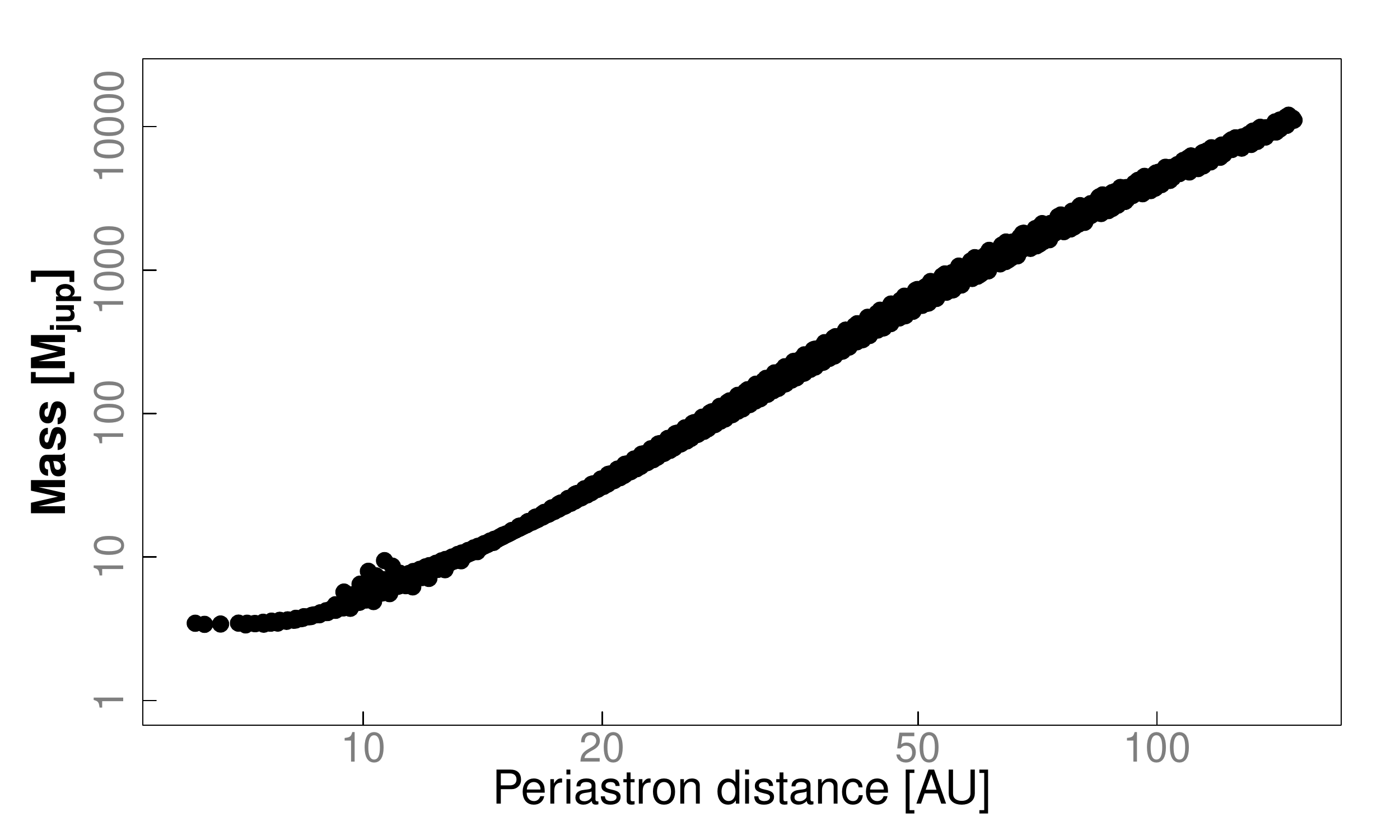}\\
\caption{\label{fig:psidra_trend_ecc} \textit{Top:} Contours of best-fit masses for the outer perturber, computed over a grid of fixed periods and eccentricities.
Systems unstable within $10^5$ years are marked in red.
\textit{Bottom:} Relationship between the periastron distance and the mass of the outer perturber.}
\end{figure}

If we relax the assumption of a circular orbit for the external perturber, then the predicted mass of the perturber at each orbital period will be smaller
at higher eccentricities (Figure \ref{fig:psidra_trend_ecc}). This is because at higher eccentricities and fixed periods, the curvature of the RV signal
will be provided at the pericenter swing of the perturber. Therefore, the mass of the outer companion is determined by the pericenter distance
($q = a(1-e)$; Figure \ref{fig:psidra_trend_ecc}, bottom panel), as expected. Again, component A is not close or massive enough to produce the observed curvature.

\subsubsection{Stellar Activity Check} \label{psi1drabch}

We examined the Ca II S$_{HK}$ values determined from the spectra of $\psi^{1}$~Dra B. The mean $S_{HK}$ index for this star is $0.167\pm0.0008$, which is
a typical value for a magnetically quiet star (e.g. the inactive star $\tau$~Ceti has $S_{HK}=0.167\pm0.0013$ measured from our spectra).
We find a linear correlation coefficient of $0.116$ of the $S_{HK}$ with the RV values. This translates to a probability of $p\approx18\%$ that the null-hypothesis of no
correlation is correct. Very small values of $p$ would indicate a correlation between the two quantities, but they are not strongly correlated.

\begin{figure}
\centering
\plotone{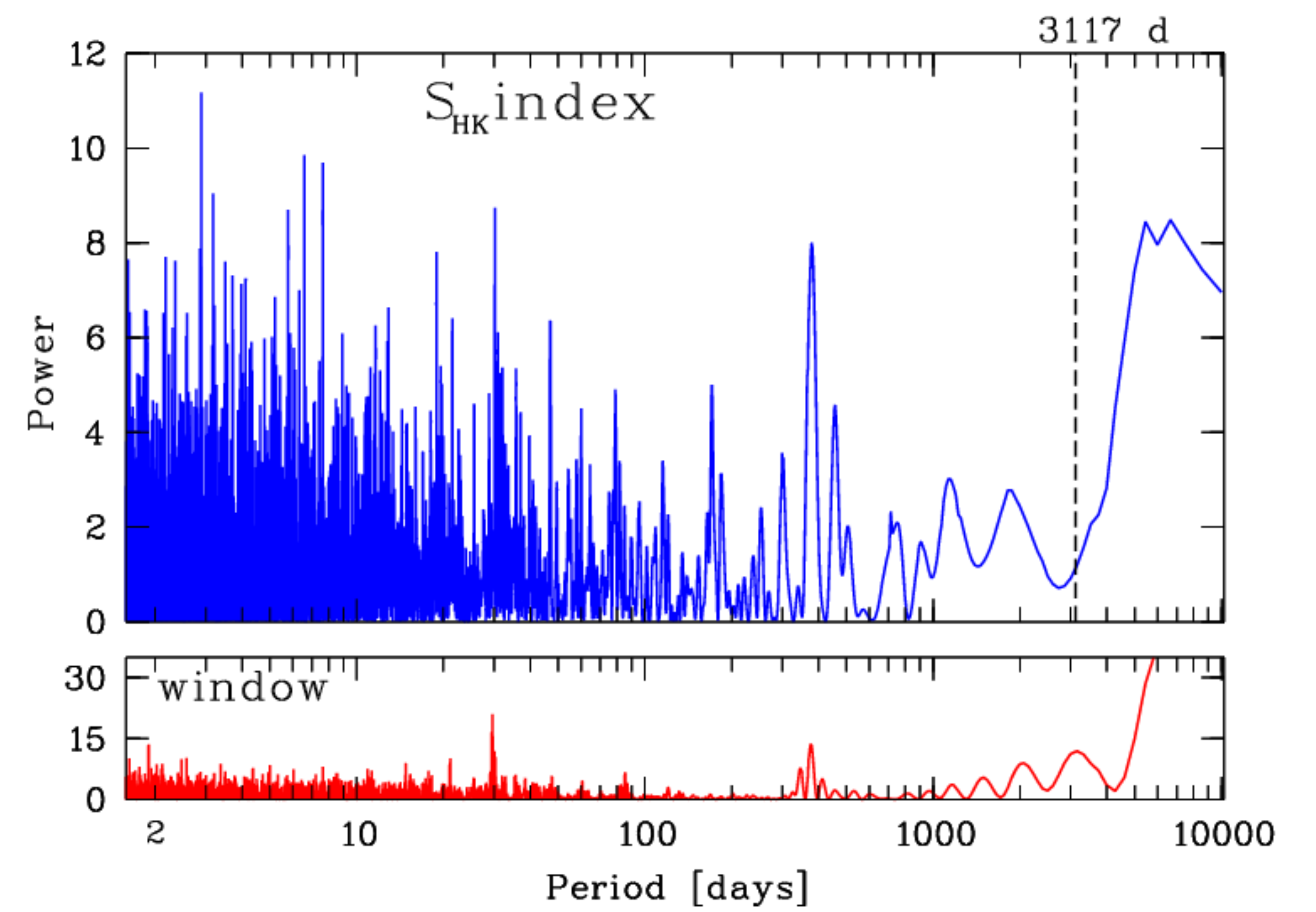}
\caption{\label{fig:psidrab_S} Lomb-Scargle periodogram of the Ca II $S_{HK}$ index values of $\psi^{1}$~Dra B (top) and the
window function of our observations. The vertical dashed line indicates the orbital period of planet b. No power is detected
at the planet's period. There is moderate power at periods exceeding our time baseline ($> 6000$~days). This could be due to
a low-level long-term trend in the magnetic activity of this star.}
\end{figure}

Figure\,\ref{fig:psidrab_S} shows the result of a period search in the $S_{HK}$ data. No strong peaks and thus significant periodicities are detected. We
also note that there is virtually no power at the orbital period of planet b ($P=3117\pm42$\,days, indicated as vertical dashed line in the figure).
This indicates that the RV variations of $\psi^{1}$~Dra B are not caused by stellar activity. There is, however, some moderate power at periods exceeding our
time baseline. This can be due to a low-level trend in magnetic activity of the star, possibly caused by a very long activity cycle.

\subsubsection{Dynamical Stability Analysis} \label{psi1drabstab}

A number of recent studies have highlighted the value of examining the
dynamical behavior of candidate planetary systems as a critical part of
the planet discovery process (e.g. Horner et al. 2012a,b; Robertson et
al. 2012a,b; Wittenmyer et al. 2012a, 2014a).  We therefore chose to
carry out a detailed dynamical study of the stability of the proposed
\pdb\ system, as a function of the orbit of the relatively unconstrained outer body.

\begin{figure}
\epsscale{1}
\plotone{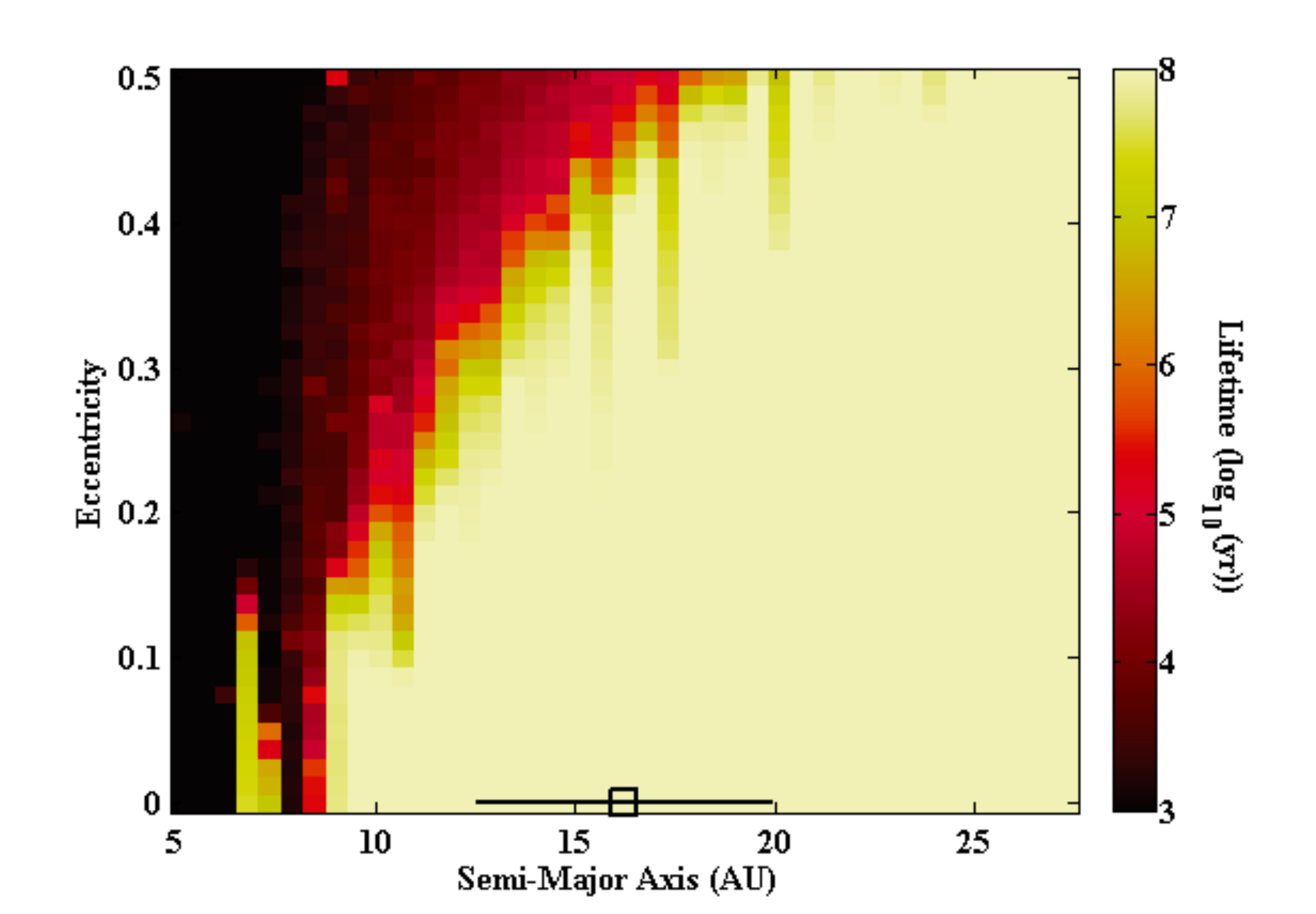}
\centering
\caption{Dynamical stability map in semi-major axis and eccentricity space for the outer
companion in the \pdb\ system. Dark areas represent unstable regions (see text for details).}
\label{fig_psistability}
\end{figure}

As in our earlier work, we carried out a total of 126075 individual
simulations of the \pdb\ planetary system, following the evolution of the
two candidate planets for a period of 100 Myr using the {\it Hybrid}
integration package within the n-body dynamics program \texttt{MERCURY}
(Chambers, 1999).  For these simulations, we have ignored the binary
companion $\psi^1$\,Dra\,A -- with a projected orbital separation of
$\sim$600\,AU, it is expected to have a negligible effect on the
dynamics of the two planets considered here.  In the case that one of
the planets collided with the other, or was either flung into the
central body or ejected from the system, the time at which that event
occurred within the simulation was recorded, and the simulation was then
terminated. This allowed us to create a map of the dynamical stability
of the \pdb\ system as a function of the initial semi-major axis and
eccentricity of the outermost planet, as can be seen in
Figure~\ref{fig_psistability}.

In each of our 126,075 simulations, we used the same initial conditions
for the orbit of the innermost planet, as given in Table~\ref{tab:psidrab_data}.  For \pdb\,c, we systematically varied
the semi-major axis, eccentricity, argument of periastron ($\omega$) and
mean anomaly ($M$) to create a grid of 41x41x15x5 possible orbital
solutions for that planet.  In the case of the planet's semi-major axis,
$\omega$, and mean anomaly, we sampled the full $\pm\,3\sigma$ range
around the nominal best fit values for each parameter. The parameters 
we used for planet c were as follows: $a=16.2\pm$3.7\,AU;
$\omega=299\pm$10 degrees, and mean anomaly $M=299\pm$10 degrees. 
For the eccentricity, we sampled 41 equally spaced values ranging between 0.0
and 0.5.  This allowed us to investigate in some depth the influence
that the eccentricity of the planet's orbit will have on the system's
stability.

The results of our simulations can be seen in Figure~\ref{fig_psistability}.  
At each of the a-e locations in that figure, the lifetime given is the mean
of 75 individual runs, sampling the full $\omega-M$
parameter space.  Most readily apparent in Figure~\ref{fig_psistability} is that the nominal best-fit orbit is located in a broad region of orbital
stability.  Indeed, all solutions within $\pm\,1\sigma$ of the best-fit
semi-major axis are dynamically stable, unless the initial orbital
eccentricity is in excess of 0.2.  This is not surprising: the
relatively sharp delineation between stable and unstable orbits that can
be seen curving upwards from an origin at ($a\sim\,9, e\sim\,0$) is a
line of almost constant periastron distance, and separates those orbits
on which the planets cannot experience close encounters from those on
which they can (and do).  Following Chambers et al.~(1996), we can determine
the mutual Hill radius of the two companions at various semi-major axes
(using their equation 1).  Doing this, we note that when \pdb\,c is
located at $a=9$\,AU, the mutual Hill radius of the two companions is
$\sim$1.02 AU, meaning that their orbits would be separated by less than
5 mutual Hill radii.  More critically, however, this situation would
allow the two companions to approach one another within two mutual Hill
radii should a close encounter happen whilst \pdb\,b (with its moderately
large orbital eccentricity of 0.42) were close to apastron.

A few other noteworthy features can be readily observed in
Figure~\ref{fig_psistability}.  Interior to the broad area of stability lies a
narrow island of stability at $a\sim$7 AU.  Orbits in this region can be
protected from destabilization by the influence of the mutual 2:1
mean-motion resonance between the two companions.  Given an initial
semi-major axis for \pdb\,b of 4.31 AU, a perfect 2:1 commensurability
between the orbits of the two planets would occur at $a_c\sim$ 6.84 AU,
so long as the initial architecture of the system is appropriate, and
the eccentricity of the orbit of \pdb\,c is not too large.  Such islands
of resonant stability are not uncommon, and are thought to ensure the
stability of several known exoplanetary systems (e.g. Wittenmyer et al.~2012b, Wittenmyer et al.~2014b).

Finally, a number of ``bites'' can be seen taken out of the broad region
of stability -- vertical strips of lower-than-average stability dotted at
regular intervals through the whole range of semi-major axes examined
(with the most prominent visible at $a\sim$11 AU).  These represent
locations where resonant interactions between the two companions act to
destabilize, rather than stabilize, their orbits.  These features serve
as a reminder that even when two planets are well separated in their
orbits around a given star, their orbits should still be checked for
dynamical verisimilitude.

\subsection{Direct Imaging} \label{imaging}

We observed both the A and B components of the $\psi^{1}$~ Draconis system separately with the Differential Speckle Survey Instrument (DSSI) at the 
Gemini North telescope on 19 July 2014 UT. DSSI is a two-channel speckle camera described in Horch et al.~(2009), which yields diffraction-limited 
information in two pass bands simultaneously. A 1000-frame sequence was recorded by each channel on each component. 
All frames were 60 ms exposures, and had format of 256 $\times$ 256 pixels. 
The seeing for both observations was 0.65 arc seconds. The image scale is approximately 0.011 arcseconds per pixel for both cameras.

We reduced and analyzed the results as follows. We form the average autocorrelation and average triple correlation of the set of speckle frames, 
and from these we estimate both the magnitude and phase of the Fourier transform of the source. The former must be deconvolved by a point source observation in general; 
in the case of the data here, we constructed a point source matching the elevation and azimuth of the source by taking an observation of a point source at very high elevation 
and correcting it for the atmospheric dispersion expected for the elevation and azimuth of the science target. After the deconvolution, the magnitude and phase are assembled 
in the Fourier plane, low-pass filtered to suppress high-frequency noise, and inverse-transformed to arrive at a reconstructed (i.e. diffraction-limited) image of the target. 
More information about the reduction method with the current EMCCD cameras used with DSSI can be found in Horch et al.~(2011).


\begin{figure}[tb]
\plotone{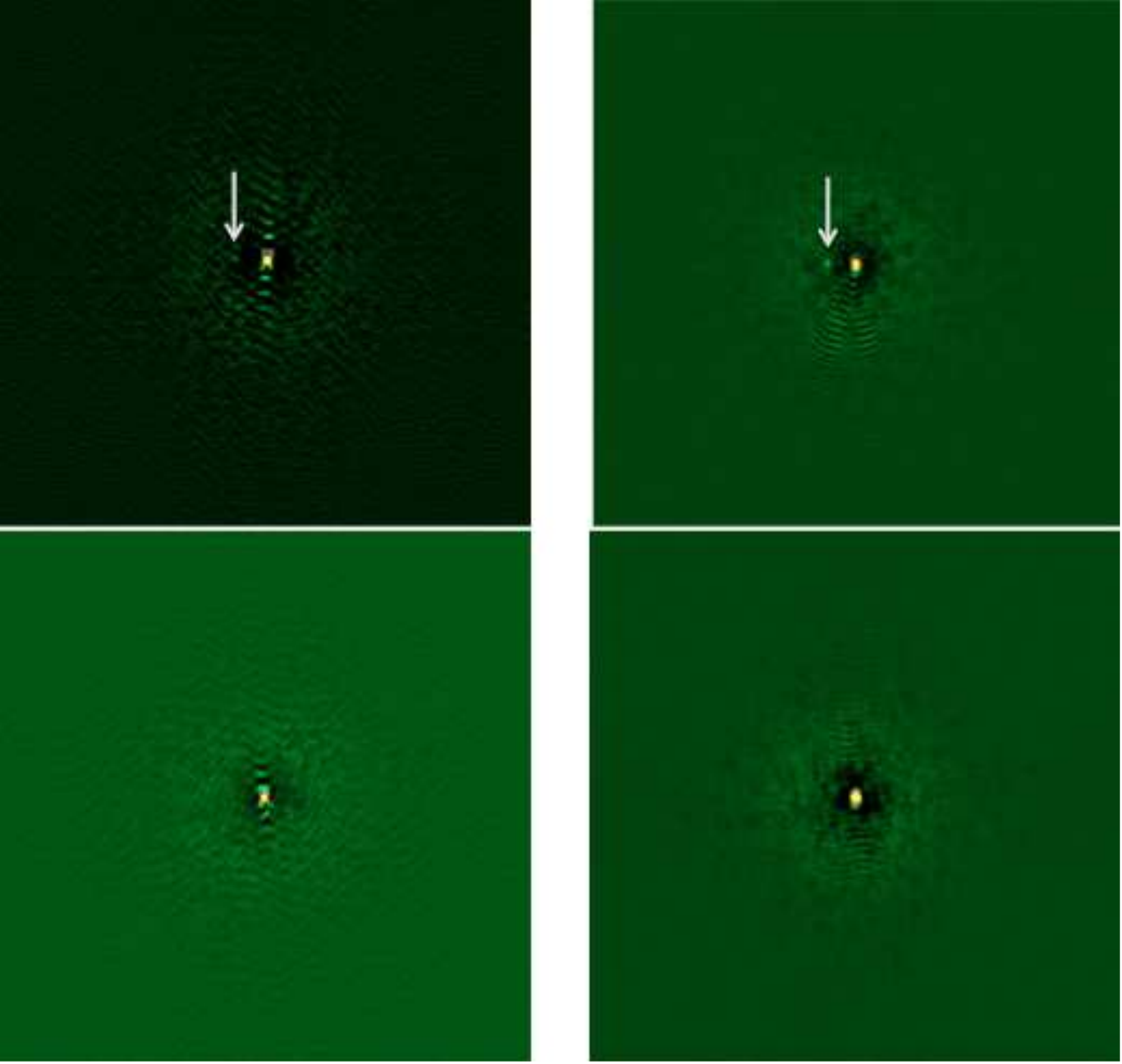}
\caption{
\label{fig:psi_ima}
Reconstructed images of $\psi^{1}$ Draconis A and B taken with the Differential Speckle Survey Instrument on the Gemini North telescope, 19 July 2014 UT.
Each frame is $2.8 \times 2.8$ arc seconds is size.
Top left: $\psi^{1}$ Draconis A at 692 nm. Top right: $\psi^{1}$ Draconis A at 880 nm.
Bottom left: $\psi^{1}$ Draconis B at 692 nm. Bottom right: $\psi^{1}$ Draconis B at 880 nm.
A faint very red companion is visible to the left of the primary star in the A images; the separation is 0.16 arcsec.
}
\end{figure}

With the reconstructed image in hand, we attempt to find companions by first examining the image. 
The images are shown in Figure\,\ref{fig:psi_ima}.
This yielded a strong stellar candidate at approximate separation of 0.16 arcseconds from the primary star for $\psi^{1}$~Dra A, but no candidates for $\psi^{1}$~Dra B.  

We then also computed a detection limit curve for the image; that is, a curve showing the largest magnitude difference that could be detected as a 
function of separation from the central star in the image. To construct the curve, we choose a set of concentric annuli centered on the central star, and 
determine the statistics of the local maxima (peaks) occurring inside the annulus. If a peak in the annulus has a value of more than five times the sigma of all of the 
peaks above the average value of the peaks, we consider it to be a definitive detection of a stellar companion. 
Details of this process for Gemini data can be found in e.g. Horch et al. (2012). 

\begin{figure}
\plotone{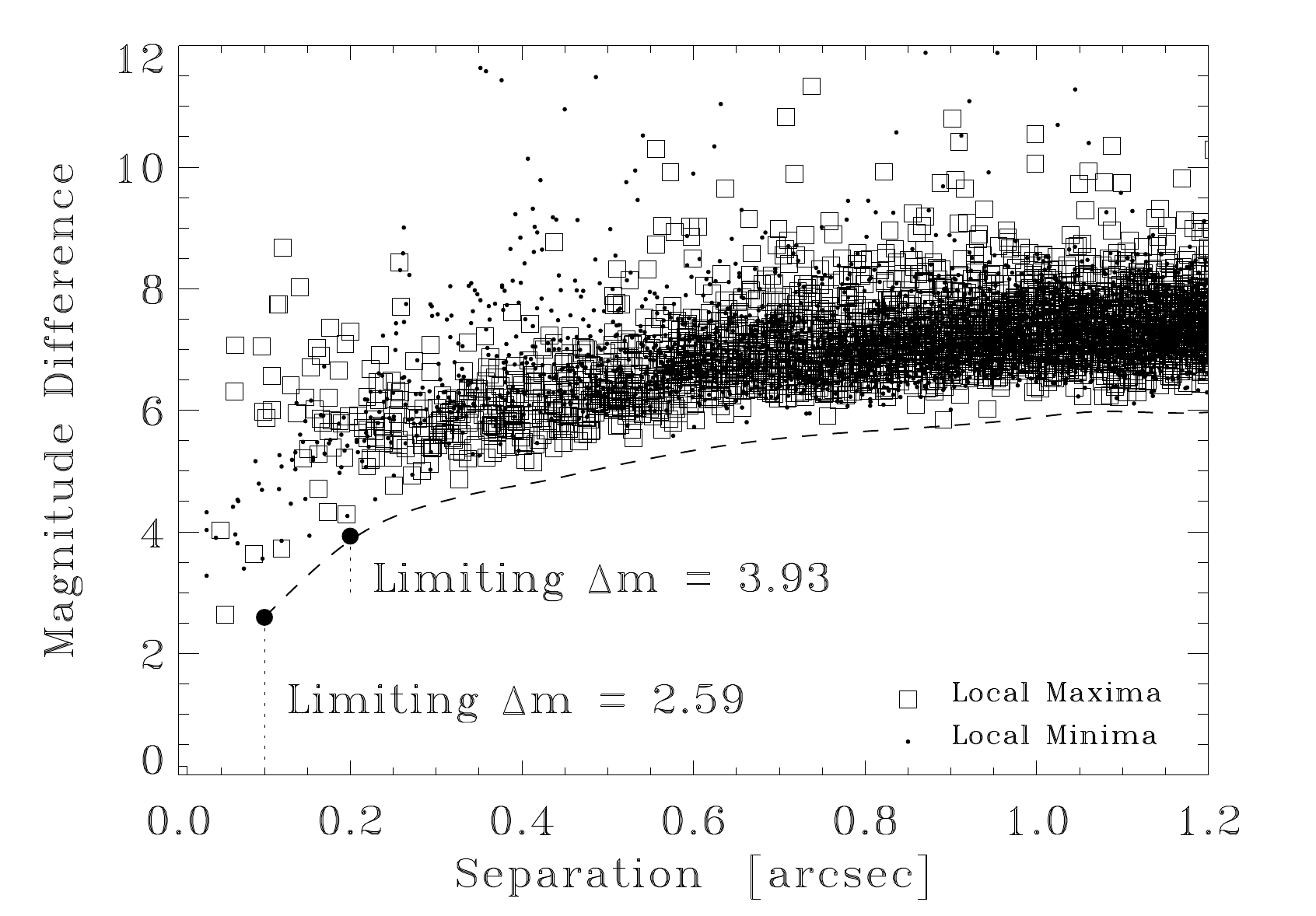}
\plotone{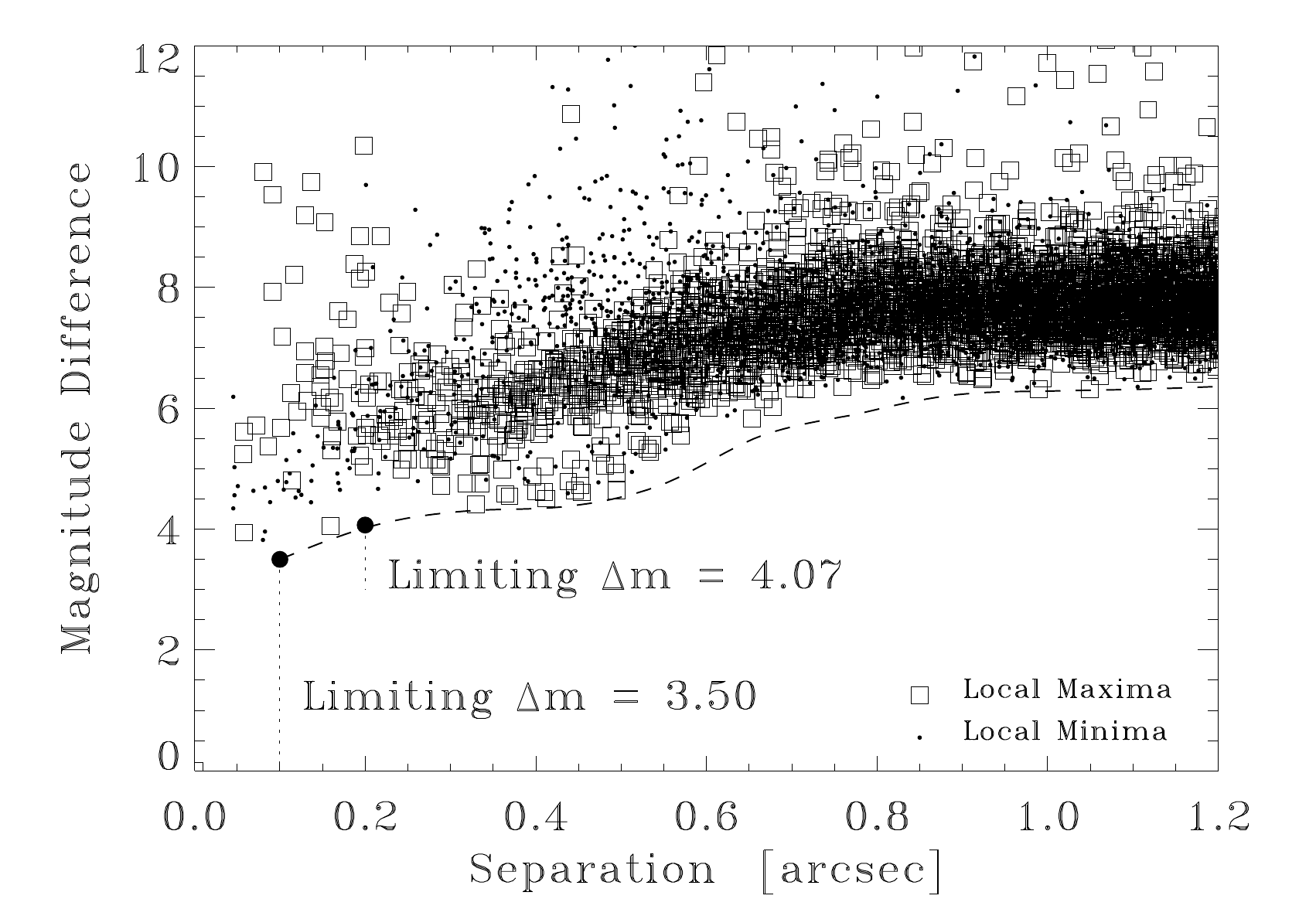}
\plotone{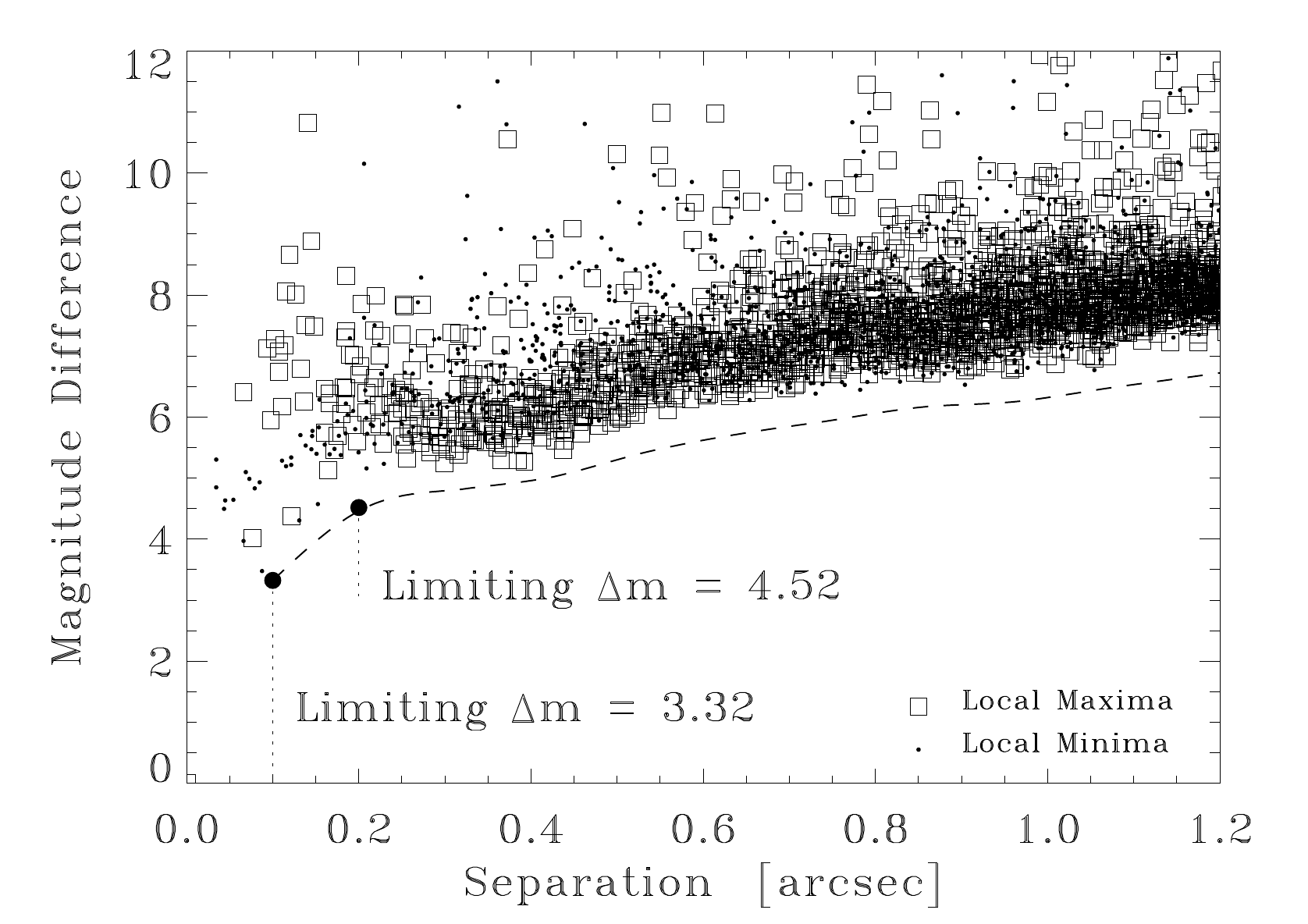}
\plotone{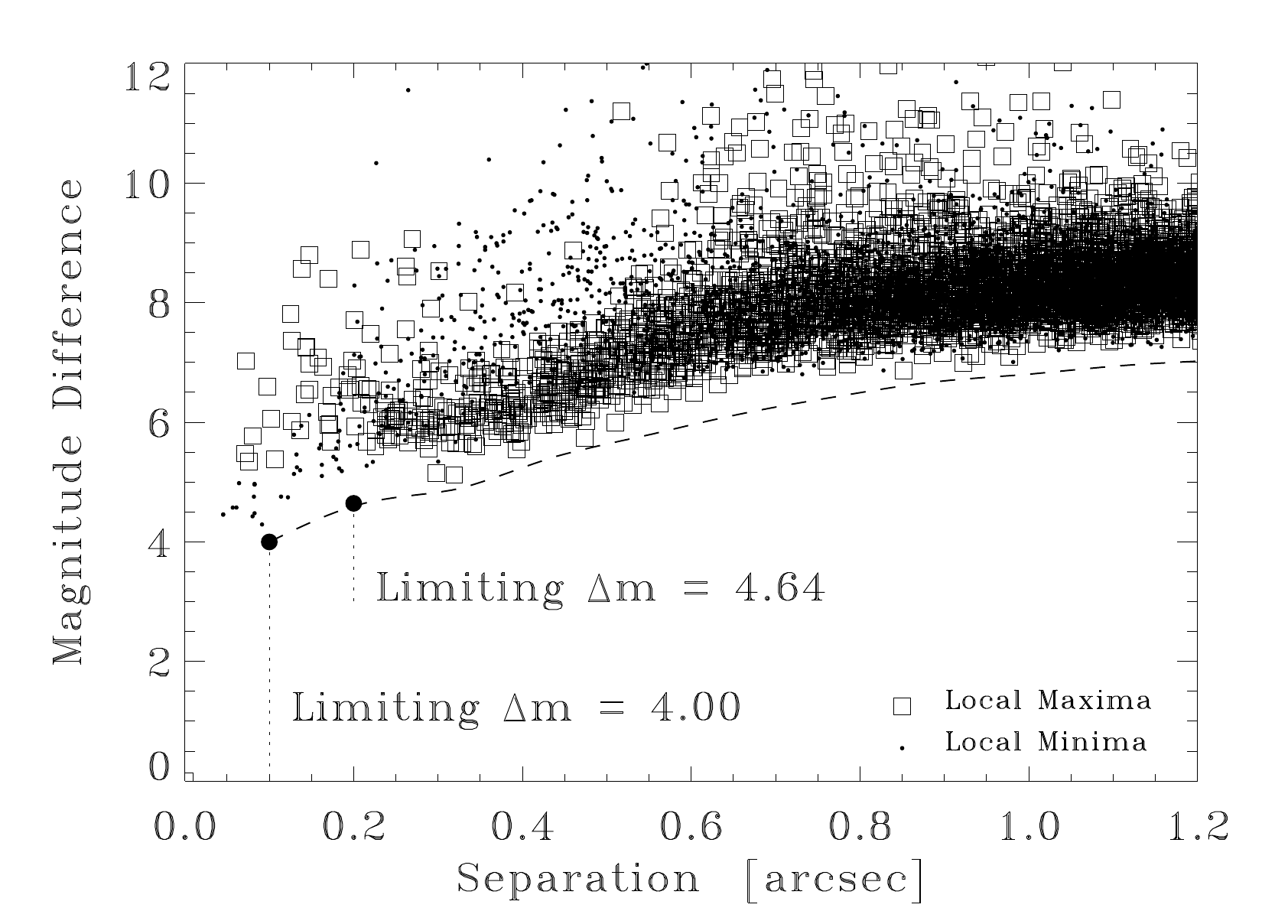}
\caption{
Detection limit curves as a function of separation for (a) $\psi^{1}$~Dra A at 692 nm, (b) $\psi^{1}$~Dra A at 880 nm, (c) $\psi^{1}$~Dra B at 692 nm, and (d) 
$\psi^{1}$~Dra B at 880 nm. These 
data are derived from the speckle image reconstructions as described in the text. In all plots, the squares are the implied magnitude differences from the central star 
for all local maxima in the reconstructed image, and solid points are the same drawn for the absolute values of all local minima (plotted to compare with the 
distribution of maxima). The dashed line indicates the magnitude difference for a peak that would be 5-sigma above the average local maximum value, as judged from the 
statistics of the distribution itself. A detection of a companion at formal significance greater than 5 sigma will lie below the line. \label{fig:psi_lim} } 
\end{figure}

Figure\,\ref{fig:psi_lim} shows the detection limit curves for the four reconstructed images. The line in these figures is the 5-sigma detection limit. 
If a source is a formally above a 5-sigma detection, it would appear as a square data point lying below the curve. 
For $\psi^{1}$~Dra B, there are no plausible sources. For $\psi^{1}$~Dra A, we see a nearly 5-sigma detection of a second star at a separation of about 0.16 arcsec 
from the primary in the 880-nm image. In the 692-nm image, the same peak appears, but it is not as close to the 5-sigma line. 
Looking again at the two images, we see that these data correspond in both cases to the peak at pixel (114, 129). Since both images show the same peak in the same spot, 
we are very confident that this component is stellar in nature. In these images north is essentially up, and east is to the left, although there is a tilt of the 
celestial coordinates relative to pixel axes of about 5 degrees.

We next used our power spectrum fitting routine to determine the separation, position angle, and magnitude difference of the secondary. 
The results are summarized in Table \ref{tab:imaging1}, when deconvolving by the calculated point sources described above.

\begin{table}
\centering
\begin{tabular}{cccc}
\hline
filter & position angle & separation & magnitude difference \\
(nm) & (deg) & (arcsec) & (mag) \\
\hline
692 & 91.8 & 0.155 & 4.13  \\
880 & 91.5 & 0.158 & 3.80  \\
\end{tabular}
\caption{\label{tab:imaging1} Results of imaging for $\psi^{1}$~Dra A, using generic point source deconvolution}
\end{table}

Since $\psi^{1}$~Dra B is not resolved in our images, we also used it as the point source to deconvolve the images of $\psi^{1}$~Dra A, and in doing the power spectrum 
fitting that way, we obtain the results summarized in Table \ref{tab:imaging2}.

\begin{table}
\centering
\begin{tabular}{cccc}
\hline
filter & position angle & separation & magnitude difference \\
(nm) & (deg) & (arcsec) & (mag) \\
\hline
692 & 91.8 & 0.156 & 4.22  \\
880 & 91.3 & 0.158 & 3.71  \\
\end{tabular}
\caption{\label{tab:imaging2} Results of imaging for $\psi^{1}$~Dra A, using $\psi^{1}$~Dra B point source deconvolution. }
\end{table}

The differences between these numbers and the above give an estimate for the internal precision of the measurement technique. 
In looking at the power spectra for each file, we also see clear fringes that match the location shown in the reconstructed image. 
This gives an additional layer of confidence that we have detected a real stellar companion.


\subsection{Comparison of Elemental Abundances}

\subsubsection{Planet signatures in stellar abundances}

An independent stellar parameter and detailed (multi-element) chemical composition analysis for both stars in the $\psi^{1}$\,Draconis system was carried out in order 
to search for chemical abundance anomalies that could be related to planet formation processes, as suggested by a number of recent studies. 
In their highly precise spectroscopic studies of Solar twin stars, Melendez et al.~(2009) and Ramirez et al.~(2009) have found the Sun to be slightly 
deficient in refractory elements, attributing this observation to the formation of rocky bodies in the Solar System. They suggest that these objects captured the 
refractory elements that would have otherwise ended up in the Sun. In related work, Ramirez et al.~(2011) and Tucci Maia et al.~(2014) have found that the two Solar-analog 
components of the 16\,Cygni binary system have slightly different overall metallicities and have attributed this effect to the formation of the gas giant planet that 
orbits 16\,Cygni\,B (Cochran et al.~1997).

The rocky planet formation hypothesis for the refractory element depletion seen in the Sun has been challenged by Gonzalez-Hernandez et al.~(2010, 2013)
while Schuler et al.~(2011) 
have found no chemical abundance differences for the 16\,Cygni stars. Thus, further investigation of other relevant stellar systems 
could shed light on this problem. The $\psi^{1}$\,Draconis system is an interesting target in this context. Although not similar to the Sun, these stars are similar to 
each other, which is favorable to high-precision relative chemical composition analysis. Our $\psi^{1}$\,Dra\,A spectrum is contaminated by the light from the stellar 
companion at the 1\,\% level. 

\subsubsection{Atmospheric parameter determination}

We acquired very high signal-to-noise ratio spectra of the $\psi^{1}$\,Draconis stars with the Tull spectrograph on the 2.7\,m Telescope at 
McDonald Observatory on 21 April 2014. At 6\,000\,\AA, these spectra have $S/N\simeq500$ per pixel and a spectral resolution $R=60\,000$. 
These spectra are not part of the RV planet search data set; they were acquired specifically for the purpose of carrying out a detailed, strict differential 
atmospheric parameter and chemical abundance analysis. As described below, we analyzed $\psi^{1}$\,Dra\,A relative to $\psi^{1}$\,Dra\,B, but we also tested our 
differential calculations using a Solar spectrum as reference. The latter was taken from a previous observing run (18 December 2013) in which reflected sunlight 
from the asteroid Vesta was used to collect a high signal-to-noise ratio ($S/N\simeq350$ at 6\,000\,\AA) Solar spectrum with the same instrument/telescope and 
identical configuration.

Equivalent widths of 73 \fei\ lines and 18 \feii\ lines were measured by fitting Gaussian profiles to the observed spectral lines in the $\psi^{1}$\,Draconis stars' and 
Solar spectra using the \texttt{splot} task in IRAF. The linelist and atomic parameters adopted are those by Ramirez et al.~(2013). The uncertainty of the adopted 
$\log gf$ values and whether those were taken from laboratory measurements or calibrated using reference spectra (i.e., ``astrophysical'' values) are irrelevant 
in the strict differential approach implemented here. As mentioned above, the $\psi^{1}$\,Dra\,A spectrum is contaminated at the 1\,\% level by its stellar companion. 
We noticed this minor contamination in our spectra and attempted to remove it in our equivalent width measurements by using the ``deblend'' feature of \texttt{splot} 
whenever possible or by excluding sections of line wings in the line profile fits. Nevertheless, we expect the equivalent widths measured for $\psi^{1}$\,Dra\,A to be 
less precise than those of $\psi^{1}$\,Dra\,B, not only due to the spectral contamination, but also because of its somewhat faster projected rotational 
velocity.

The equivalent widths of each of the $\psi^{1}$\,Draconis stars and the Sun were employed to calculate iron abundances using the \texttt{abfind} driver of the MOOG 
spectrum synthesis code, adopting Kurucz's \texttt{odfnew} grid of model atmospheres interpolated linearly to the assumed atmospheric parameters of each star. 
Then, on a line-by-line basis, differential iron abundances relative to the Sun were computed for the $\psi^{1}$\,Draconis stars. The stellar parameters of the 
$\psi^{1}$\,Draconis stars were modified iteratively until correlations of the iron abundance with excitation potential and reduced equivalent width 
disappeared and until the mean abundance of iron derived from \fei\ and \feii\ lines separately agreed. 
This procedure is standard in stellar spectroscopy (cf Section~\ref{stars}) and it is sometimes referred to as the excitation/ionization equilibrium method of stellar parameter 
determination. 
To be more specific, hereafter we refer to this technique as the ``iron line only'' method. The particular implementation used here, including the error analysis, is 
described in detail in Ramirez et al.~(2014; Sect.~3.1 and references therein).

\begin{table}
\centering
\caption{Atmospheric Parameters of the $\psi^{1}$\,Draconis Stars}
\begin{tabular}{ccccc} \hline\hline
Star & $\teff$ & $\logg$ & $\feh$\footnote{The error bars in parenthesis correspond to the $1\,\sigma$ line-to-line scatter.} & Ref. \\ \hline
A & $6546\pm56$ & $3.90\pm0.14$ & $-0.10\pm0.04$ ($\pm0.07$) & Sun \\
B & $6213\pm20$ & $4.35\pm0.05$ & $+0.00\pm0.01$ ($\pm0.04$) & Sun \\ \hline
A & $6544\pm42$ & $3.90\pm0.11$ & $-0.10\pm0.03$ ($\pm0.05$) & B \\ \hline
\end{tabular}
\\
\label{t:pars}
\end{table}

The atmospheric parameters of the $\psi^{1}$\,Draconis stars, derived as described above, are given in the first two rows of Table~\ref{t:pars}. The errors listed in 
that table are formal, i.e., they represent the precision with which we are able to find a self-consistent solution for the parameters, but do not take into 
account possible systematic errors. The $\psi^{1}$\,Draconis stars are both significantly warmer than the Sun. Thus, we expect the analysis using the Solar spectrum as 
reference to be affected by systematic errors in a non-negligible way. Since we are interested in the relative elemental abundances of the two $\psi^{1}$\,Draconis 
stars, we could attempt to reduce these formal errors, and also minimize the potential systematics, by analyzing $\psi^{1}$\,Dra\,A using $\psi^{1}$\,Dra\,B as the 
reference star. Adopting the parameters derived for $\psi^{1}$\,Dra\,B using the Sun as reference (row 2 in Table~\ref{t:pars}), we computed the parameters of 
$\psi^{1}$\,Dra\,A given in row 3 of Table~\ref{t:pars}. Note that the formal errors reduced, but the average values of the parameters were not significantly affected. 
This demonstrates that, when using the Sun as reference, systematic errors are introducing line-to-line scatter to the iron abundances of the $\psi^{1}$\,Draconis 
stars.

In the last step we implicitly assumed that the parameters of $\psi^{1}$\,Dra\,B derived using the Sun as reference are reliable. We tested this assumption by computing 
those parameters using independent techniques. For $\teff$, we employed the effective temperature -- color calibrations by Casagrande et al.~(2010). 
For $\logg$, we used the stars' trigonometric parallaxes as given in the {\it Hipparcos} catalog along with the Yonsei-Yale theoretical isochrone grid. 
Details on these techniques and the implementation used here are also provided in Ramirez et al.~(2014; Sects.~5.1 and 5.3 and references therein).

Using the $\feh$ values from Table~\ref{t:pars}, the Casagrande et~al.~(2010)
$\teff$ calibrations for the $(B-V)$, $(b-y)$, and $(B_\mathrm{T}-V_\mathrm{T})$ colors 
provide mean values of $6519\pm20$\,K for $\psi^{1}$\,Dra\,A and $6194\pm21$\,K for $\psi^{1}$\,Dra\,B. Both these values are in agreement within formal error with 
those computed from the iron lines only (i.e., with the parameters given in Table~\ref{t:pars}). 
Moreover, their difference is 325\,K according to the $\teff$-color calibrations and 333\,K according to the iron line analysis. 
This test confirms that the $\teff$ adopted for $\psi^{1}$\,Dra\,B in the strict differential analysis for $\psi^{1}$\,Dra\,A is reliable.

The trigonometric $\logg$ values were computed using the $\teff$ from the color calibrations, thus making them completely independent of the iron line only analysis. 
We calculated $\logg=4.02\pm0.02$ for $\psi^{1}$\,Dra\,A and $\logg=4.32\pm0.02$ for $\psi^{1}$\,Dra\,B. The spectroscopic (iron line only) $\logg$ of 
$\psi^{1}$\,Dra\,A appears slightly low, yet it is still in marginal agreement with the trigonometric value within formal error. 
However, for $\psi^{1}$\,Dra\,B, the agreement is excellent, which also suggests that the $\logg$ adopted for $\psi^{1}$\,Dra\,B in the strict differential analysis of 
$\psi^{1}$\,Dra\,A is reliable.

As part of this calculation, we also computed age probability distributions for these stars using the same Yonsei-Yale isochrone set. We found most-probable 
ages 
of 2.3 and 2.5 Gyr for psi1 Dra A and psi1 Dra B, respectively, both with precision errors of order 0.3 Gyr. The details of this calculation are explained in 
Ramirez et al.~(2014, their Sect. 4.5). Note in particular that the errors do not include systematic uncertainties from the stellar models. The ages quoted above 
are somewhat younger than that given in Table 1 for psi1 Dra B (3.3 Gyr), but note that the latter has a much larger error bar due to the less precise stellar 
parameters employed. In fact, this age is consistent within error with the values listed here. In any case, the ages determined with the parameters measured as 
described in this section, which are based on higher quality spectra, lead to consistent ages for psi1 Dra A and psi1 Dra B, as expected for a binary system.

\subsubsection{Multi-element abundance analysis}

\begin{figure}
\centering
\plotone{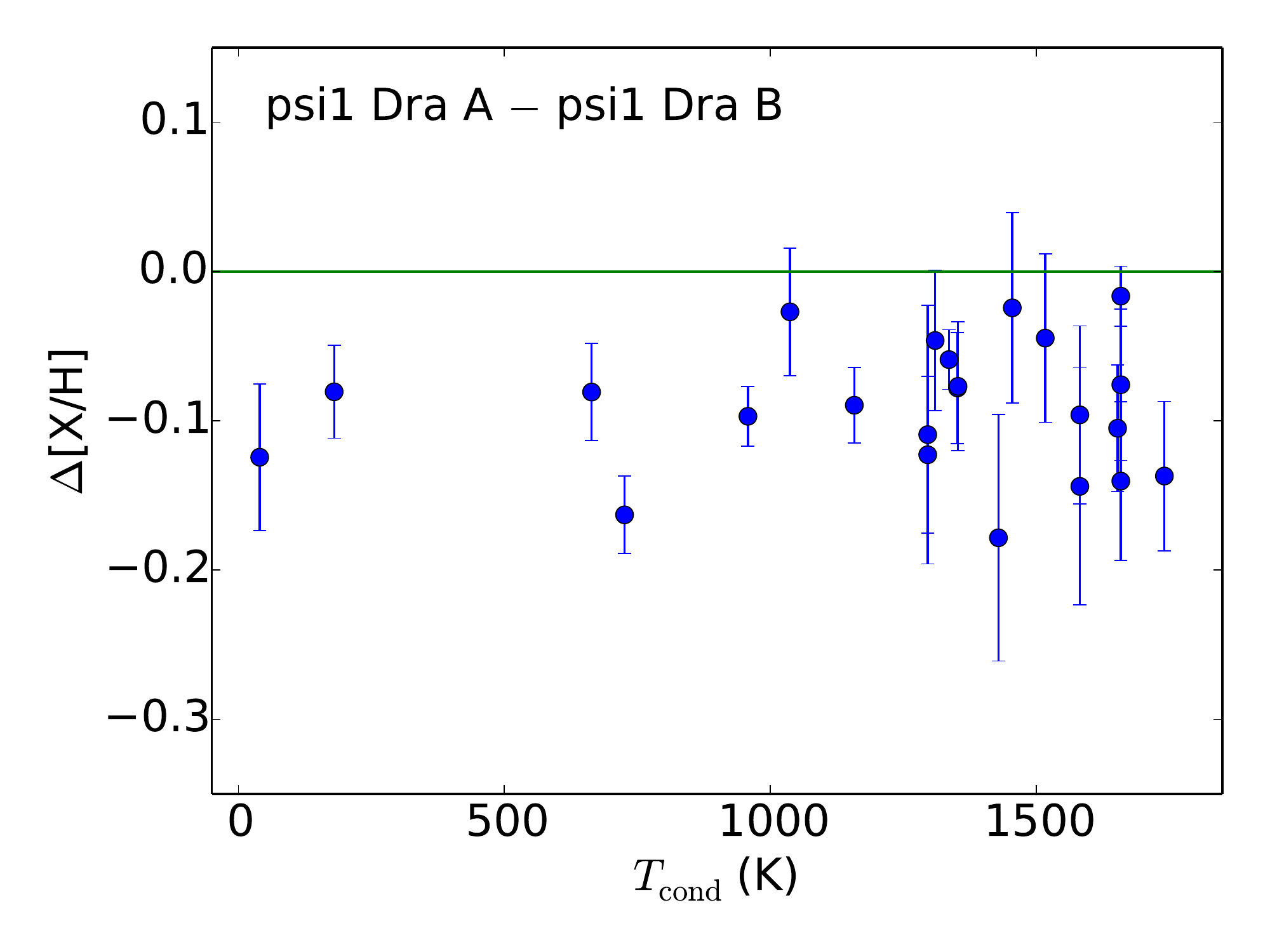}
\caption{Elemental abundance difference between $\psi^{1}$\,Dra\,A and $\psi^{1}$\,Dra\,B as a function of the elements' condensation temperatures.}
\label{f:xh_ab}
\end{figure}

Equivalent widths of spectral lines due to species other than iron were measured to compute differential abundances of 20 chemical elements in the $\psi^{1}$\,Draconis 
stars. The linelist and adopted atomic data, including hyperfine structure parameters when available, are from Ramirez et al.~(2009, 2011). 
Oxygen abundances were inferred using the O\,\textsc{i} triplet lines at 777\,nm, corrected for non-LTE effects using the grid by Ramirez et al.~(2007).

The relative elemental abundances we measured are plotted in Figure~\ref{f:xh_ab} as a function of the elements' 50\,\% condensation temperatures, as 
computed by Lodders et al.~(2003) for a Solar composition gas. Note that this ``A--B'' difference in chemical abundances was obtained when $\psi^{1}$\,Dra\,A was 
directly analyzed with respect to $\psi^{1}$\,Dra\,B in a strict line-by-line differential manner. The error bars are significantly smaller compared to the case in 
which the elemental abundances are first determined with respect to the Sun and then subtracted. This is a consequence of reducing the systematic errors of the 
analysis by avoiding a reference that is very dissimilar to either one of the $\psi^{1}$\,Draconis stars.

Figure~\ref{f:xh_ab} shows that $\psi^{1}$\,Dra\,A is metal-poor relative to $\psi^{1}$\,Dra\,B. On average, the metallicity difference is $-0.09\pm0.04$\,dex. We do 
not detect a statistically significant correlation with the condensation temperature, but this is likely due to the relatively large errors in the abundance differences. 
In the Mel\'endez, Ram\'irez, et~al.\ works the precision of relative abundances is of order 0.01\,dex. In our case those errors are about 0.04\,dex instead. 
Thus, we cannot rule out possible trends based on our data.

One may be tempted to attribute the elemental abundance discrepancy shown in Figure~\ref{f:xh_ab} to uncertain stellar parameters. 
The derived chemical abundances are most affected by the adopted $\teff$ values, and we have shown above that those of $\psi^{1}$\,Dra\,B are reliable. Thus, we can 
explore this potential systematic error by simply calculating the relative abundances for different $\teff$ values for $\psi^{1}$\,Dra\,A and keeping everything else 
constant. Increasing the $\teff$ of $\psi^{1}$\,Dra\,A by 200\,K would make the average elemental abundance difference nearly zero, but only for refractory elements 
($T_\mathrm{cond}\gtrsim1\,000$\,K). The abundances of C and O in this case would differ by about $-0.2$\,dex. 
On the other hand, decreasing the $\teff$ of $\psi^{1}$\,Dra\,A by 200\,K would make the C and O abundances difference nearly zero, but then the refractories would 
differ by about $-0.2$\,dex. In both cases, we note that the element-to-element scatter as well as the line-to-line relative abundance scatter for individual elements 
increase relative to the case when our derived $\teff$ value is adopted instead. 
In other words, the elemental abundance differences are more internally consistent for our derived parameters, suggesting that the hotter or cooler temperatures 
of $\psi^{1}$\,Dra\,A are not realistic (within our modeling assumptions, of course). Thus, it is not possible to reconcile the chemical abundance difference between 
$\psi^{1}$\,Dra\,A and B by assuming that the $\teff$ of the former is either underestimated or overestimated. The spectral contamination of $\psi^{1}$\,Dra\,A can not 
explain the observed abundance difference either. Since only 1\,\% of the flux is from the low-mass stellar companion, the equivalent widths and abundances derived could have 
been underestimated by 1\,\% at most. This corresponds to less than about 0.005\,dex in [X/H]. We are led to conclude that the offset seen in Figure~\ref{f:xh_ab} is real.

\subsubsection{Does $\psi^{1}$~Dra~A have $\delta$~Scuti abundances?}

The chemical composition of $\delta$~Scuti stars may be peculiar. The prototype star of this class shows a severe enhancement, up to about 1.0 dex, in the abundances of 
elements with atomic number above 30 (Yushchenko et al.~2005). A very similar abundance pattern is observed in $\rho$~Puppis, a very bright $\delta$~Scuti star, as shown 
by Gopka et al.~(2007). Note, however, that both $\delta$~Scuti and $\rho$~Puppis are about 500 K warmer than $\psi^{1}$~Dra~A.

A $\delta$~Scuti star with known detailed chemical abundances which is more similar in stellar parameters to $\psi^{1}$~Dra~A is CP Boo (Galeev et al. 2012). This star is about 
200 K cooler than $\psi^{1}$~Dra A, and a bit more metal rich ([Fe/H]=+0.16). The abundances measured by Galeev et al. also reveal enhancements of the heavy metals, but not as 
dramatic as in $\delta$~Scuti and $\rho$~Puppis. On average, the abundances of elements with atomic number above 30 are higher by about 0.3 dex. Such level of enhancement would 
be easily detected in our spectra. Figure~\ref{f:xh_ab} shows that at least the abundances of Zn (Z=30, Tc=726K), Y (Z=39, Tc=1659K), Zr (Z=40, Tc=1741K), and Ba (Z=56, Tc=1455K) 
are not enhanced in $\psi^{1}$~Dra A relative to $\psi^{1}$~Dra B. They are also not enhanced when the abundances are measured relative to the Solar abundances.

To provide further evidence for this finding, we measured the abundances of Nd (Z=60, Tc=1594K) and Eu (Z=63, Tc=1347K) using spectrum synthesis following the procedure 
described in Ramirez et al.~(2011). For both of these species we found an A-B difference of -0.10+/-0.06 dex. In other words, the Nd and Eu abundances of $\psi^{1}$~Dra~A 
relative to $\psi^{1}$~Dra~B fit perfectly the pattern seen in Figure~\ref{f:xh_ab}. In particular, they are also not enhanced in the former. An enhancement of 1.0 dex, as in 
$\delta$~Scuti or $\rho$~Puppis, or 
even the mild enhancement of 0.3 dex seen in CP Boo would have been trivial to detect in our analysis. In fact, in that case some points would be found out of the chart in 
Figure~\ref{f:xh_ab}.

Although the abundance pattern of $\psi^{1}$~Dra A does not look like that of some prototypical $\delta$~Scuti stars, it should be noted that these peculiarities may not 
correlate perfectly with the stars' pulsation characteristics. In fact, Fossati et al.~(2008) argue that $\delta$~Scuti stars have abundance patterns that are indistinguishable 
from a sample of normal A- and F-type stars. Thus, the non-enhancement of heavy metal abundances that we find for $\psi^{1}$~Dra A does not necessarily rule it out as a 
candidate for a star of the $\delta$~Scuti class.

\subsubsection{Possible interpretations}

In the 16\,Cygni system, the secondary hosts a gas giant planet whereas the primary has not yet shown evidence of sub-stellar mass companions. 
Ramirez et al.~(2011) found that 16\,Cyg\,B is slightly metal-poor relative to 16\,Cyg\,A and explained this observation as a signature of planet formation. Briefly, 
they suggested that the missing metals of 16\,Cyg\,B are currently located inside its planet. Considering that hypothesis, possible explanations for our results 
regarding the $\psi^{1}$\,Draconis system include:

\begin{enumerate}

\item The 16\,Cygni planet signature hypothesis is incorrect because in $\psi^{1}$\,Draconis, the secondary, which is a gas giant planet host, is actually more 
metal-rich than the primary, which does not show evidence of hosting planets in our RV data. 
Metals should have been taken away from the planet-host star $\psi^{1}$\,Dra\,B and that star should be metal-poor relative to $\psi^{1}$\,Dra\,A, which is the opposite 
of what we observe. In this case, the cause of the observed abundance differences seen in both 16\,Cygni and in $\psi^{1}$\,Draconis remains unknown.

\item Planet-like material and perhaps even fully-formed planets were once present in orbit around $\psi^{1}$\,Dra\,A, with a total mass greater than that of 
$\psi^{1}$\,Dra\,B's 
planet or planets combined. However, the stellar companion $\psi^{1}$\,Dra\,C made the planetary environment unstable, ejecting all of the planet material away from 
$\psi^{1}$\,Dra\,A. In this scenario, the outer layers of $\psi^{1}$\,Dra\,A would have accreted metal-poor gas during the planet-formation stage. The metals missing 
from $\psi^{1}$\,Dra\,A would have been locked-up in the material that was ejected later. The late ejection of that material is required to explain our non-detection 
of planets around $\psi^{1}$\,Dra\,A. Since $\psi^{1}$\,Dra\,B has a planet (or two), its atmosphere is also depleted in metals relative to the initial metallicity of 
the system, but the metal depletion suffered by $\psi^{1}$\,Dra\,A was greater. The latter would be easily explained by a larger total mass of planet-like material, but 
it could also be in part due to the thinner convective envelope of this warmer star, which did not dilute the chemical signature as much as $\psi^{1}$\,Dra\,B.

\item $\psi^{1}$\,Dra\,A never formed planets due to the influence of its low-mass stellar companion. On the other hand, $\psi^{1}$\,Dra\,B formed much more planet-like 
material than seen today in the planet or planets that we have detected. A fraction of this material, that which is not in the planet(s) detected by us, 
was accreted into the star at a later stage, increasing the metallicity of its atmosphere. 
The amount of planet material accreted that is necessary to explain our observations had to have been larger than the total mass of the planet or planets detected. 
This is because the formation of those planets imply that metals were already taken away from the star and this needs to be first compensated in order to result 
in a stellar atmosphere that is more metal rich than the birth cloud of the system. In this scenario, the metallicity of $\psi^{1}$\,Dra\,A is that of the gas cloud 
from which both stars formed whereas $\psi^{1}$\,Dra\,B's atmosphere became metal-rich at a later stage.

\item Planets did also form around $\psi^{1}$\,Dra\,A, but we did not detect them because the $\delta$~Scuti pulsations and spectral contamination from
$\psi^{1}$\,Dra\,C lead to the observed large RV-jitter that prevents the detection of the RV-signature of any planet around this star. Another way the planets
could avoid detection by the RV method is if the angle between the planetary orbital planes and the sky is small.   

\end{enumerate}

Finally, we note that the potential post-main-sequence nature of $\psi^{1}$\,Dra\,A could have allowed dredge-up to occur in that star. However, this 
mechanism is 
expected to enhance the metallicity of the stellar convection zone and photosphere while the effect that we observe is that of a depletion of metals in 
$\psi^{1}$\,Dra\,A. Thus, dredge-up can also be ruled out as a possibility to explain the abundance offset seen in Figure~\ref{f:xh_ab}. 


\section{Two Cold Jupiter ``False Alarms'' Related to Stellar Activity} \label{false}

\subsection{HD 10086} \label{hd10086}

\begin{figure}
\centering
\plotone{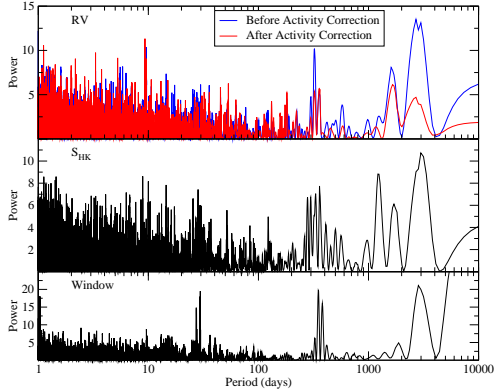}
\caption{Generalized Lomb-Scargle periodograms of our RV data for HD 10086 before (blue) and after (red) correcting for stellar activity, along with the corresponding 
periodogram of $S_{HK}$.}
\label{fig:hd10086_pscomp}
\end{figure}

We have obtained 84 RV measurements of HD 10086 over approximately 16 years, as listed in Table \ref{tab:hd10086}. 
The RVs have an RMS of 13.1 \ms \,with a mean uncertainty of 6.3 \ms, indicating the potential presence of a periodic signal.  
The periodogram of the velocities (Figure \ref{fig:hd10086_pscomp}) shows a broad and significant peak centered at 2800 days. 
This signal may be modeled as a circular Keplerian orbit with period 2800 days and a semi-amplitude $K = 11$ \ms, which would correspond 
to a planet with a minimum mass $M \sin i = 0.74M_{\textrm{Jup}}$ at $a = 3.9$ AU. Incidently, HD~10086 was also included in the 
Lick Observatory RV survey (Fischer et al.~2014). The 40 Lick RVs have an RMS of 18.6 \ms and a mean uncertainty of 3.6 \ms.
This independently confirms the apparent RV variability of this star. 

\begin{table}[ht]
\centering
\begin{tabular}{lccccc}
  \hline
 & BJD & dRV & Uncertainty & $S_{HK}$ & Uncertainty \\ 
 & & (m\,s$^{-1}$) & (m\,s$^{-1}$) & & \\
  \hline
1	&	2451152.7300	&	7.0	&	7.2	&	0.312	&	0.020	\\
2	&	2451213.6509	&	-0.8	&	5.4	&	0.302	&	0.018	\\
3	&	2451240.6034	&	37.4	&	10.9	&	0.299	&	0.017	\\
4	&	2451452.8818	&	38.6	&	6.0	&	0.335	&	0.020	\\
5	&	2451503.7088	&	21.5	&	7.1	&	0.327	&	0.021	\\
6	&	2451530.7712	&	4.9	&	6.2	&	0.290	&	0.019	\\
7	&	2451558.6106	&	20.5	&	6.6	&	0.304	&	0.020	\\
8	&	2451775.9029	&	-2.8	&	5.7	&	0.273	&	0.020	\\
9    &  ...   &  ...   &  ...   &  ...   & ...  \\
   \hline
\end{tabular}
\caption{Differential radial velocity and Ca~H\&K observations for HD~10086 (sample).}
\label{tab:hd10086}
\end{table}

\begin{figure}
\centering
\plotone{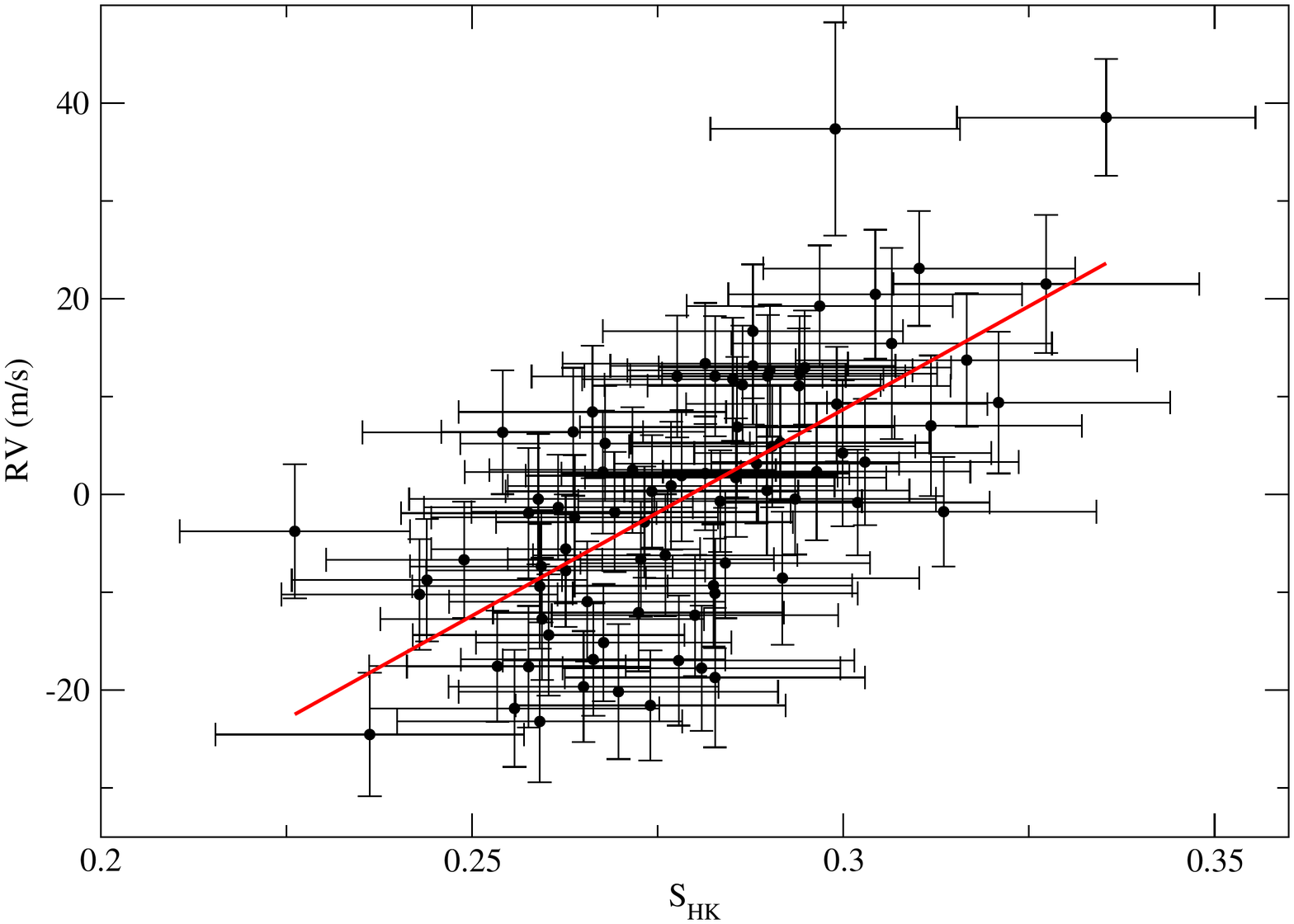}\\
\plotone{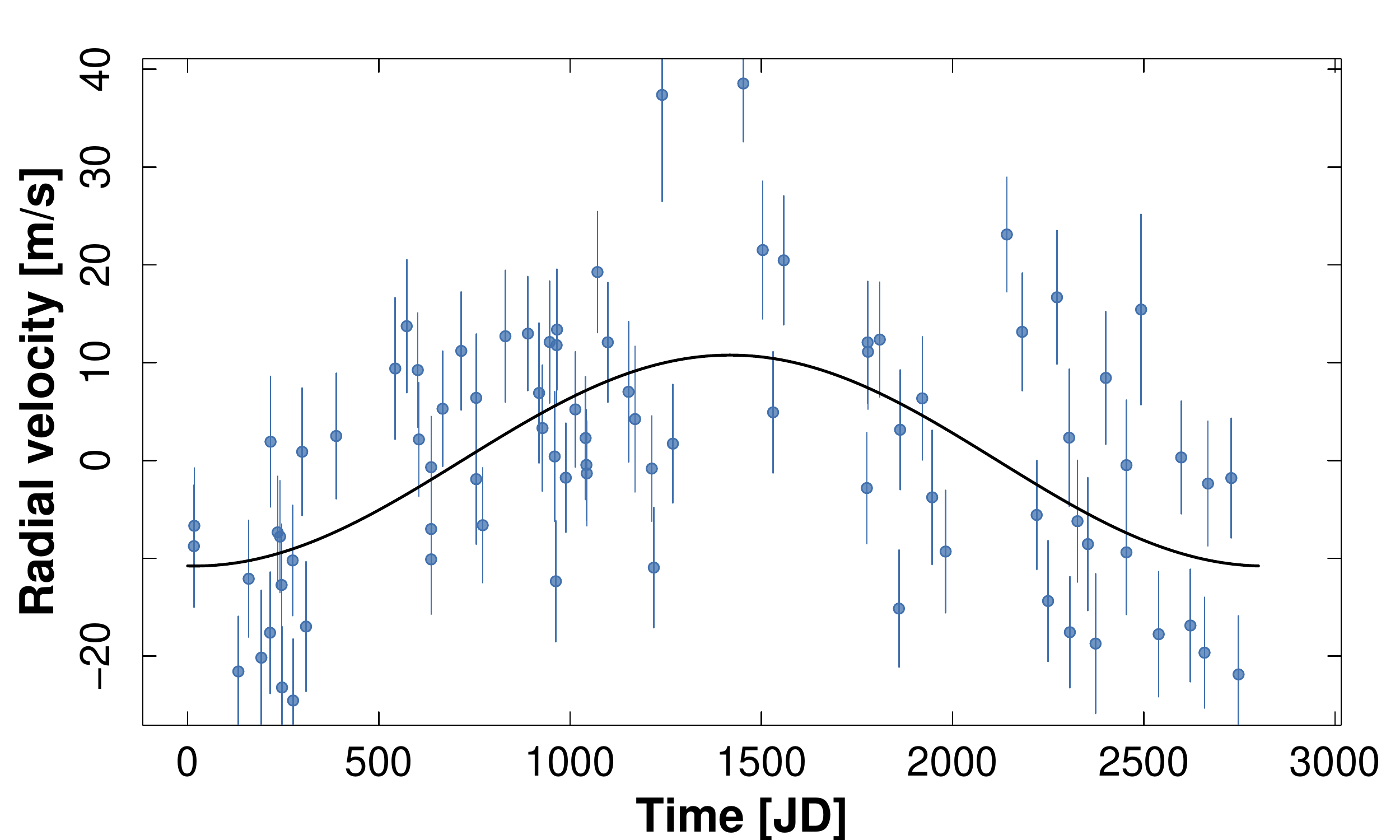}\\
\plotone{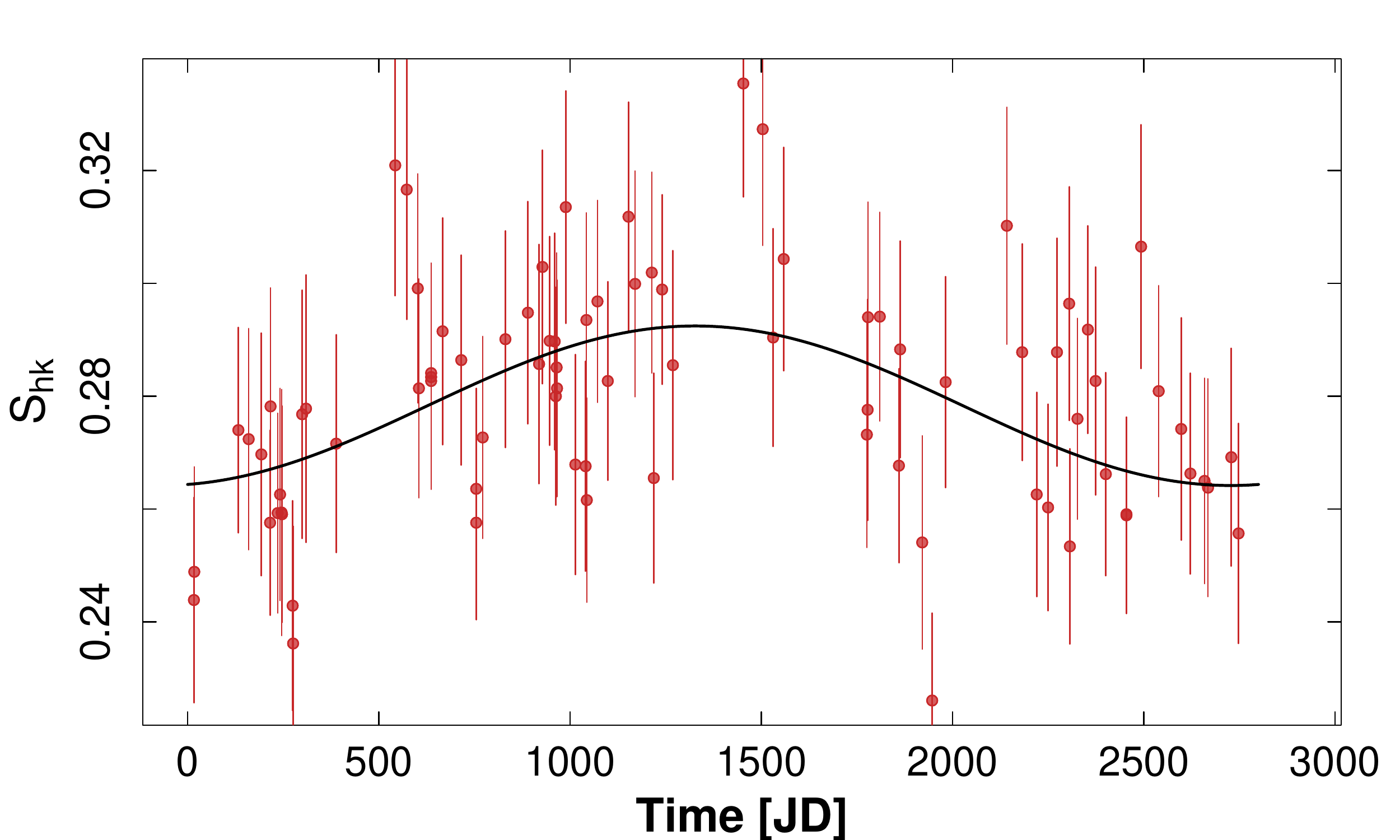}\\
\caption{\emph{Top Panel}: RV for HD 10086 as a function of $S_{HK}$ at the time of each observation.  
The linear least-squares fit to the relation is given as a solid red line.  
\emph{Middle and Bottom Panels}: RV and $S_{HK}$, respectively, folded to the 2800-day period of the stellar activity cycle.  
Sinusoidal models to each data set are shown as solid black curves.}
\label{fig:hd10086_act}
\end{figure}

However, in our analysis of stellar activity indicators for HD 10086, we see a similar 2800-d peak in the periodogram of the $S_{HK}$ Ca H\&K index, suggesting the RV 
modulation may reflect a stellar activity cycle rather than a giant planet. Considering RV as a function of $S_{HK}$ (Figure \ref{fig:hd10086_act}, top panel) confirms this 
hypothesis. The RVs of HD 10086 are very strongly correlated with $S_{HK}$; a Pearson correlation test yields a correlation coefficient $r = 0.66$ which, 
for a sample size $N = 84$ indicates a probability of $P<10^{-12}$ that we would observe such a correlation if RV and activity were uncorrelated.

Given the tight correlation between RV and Ca H\&K emission for this star, we attempted to perform a simple stellar activity correction by 
fitting and removing a linear least squares model for RV versus $S_{HK}$.  
We find a linear fit of $v_r = -120(15) + 420(50) \times S_{HK}$.  
Upon subtracting this model from the RVs, we see from the activity-corrected periodogram that the 2800-day signal is almost completely eliminated, 
providing final confirmation that this signal is caused by Doppler shifts associated with a 7.7-year activity cycle.  
We show both RV and $S_{HK}$ folded to the period of this cycle in Figure \ref{fig:hd10086_act}.  
We see no statistically significant additional signals in RV, and conclude we have not discovered any exoplanets around this star to date.

\subsection{$\beta$ Virginis} \label{betvir}

\begin{figure}
\centering
\plotone{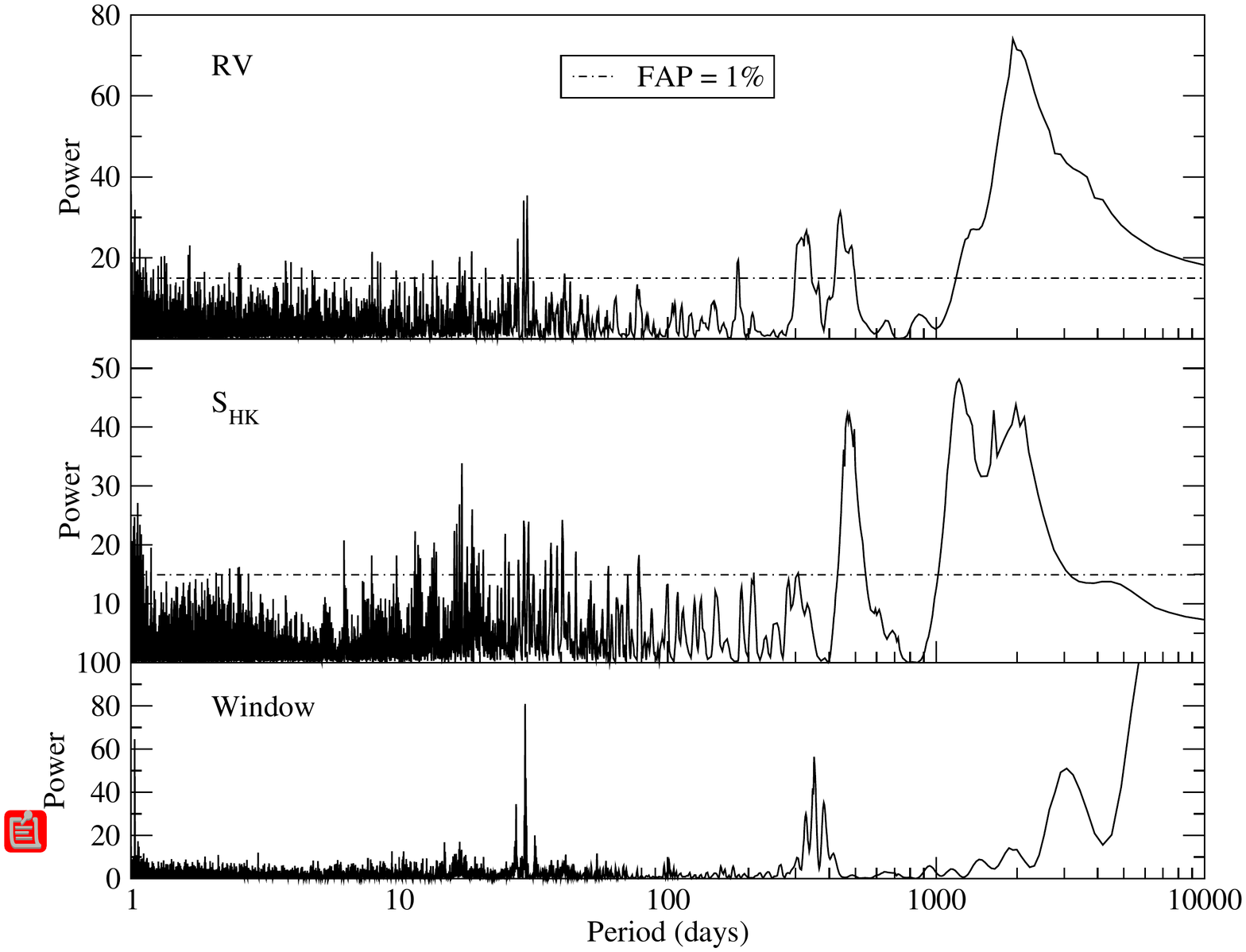}
\caption{Generalized Lomb-Scargle periodograms of RV and $S_{HK}$ for $\beta$ Virginis.  
The threshold for a false alarm probability (FAP) of 1\% is shown as a dash-dotted line.}
\label{fig:betvir_ps}
\end{figure}

We have observed $\beta$ Virginis (hereafter $\beta$ Vir) for approximately 16 years, obtaining a total of 311 RV measurements, as listed in Table \ref{tab:betvir}.  
These velocities have an RMS of 9.0 m\,s$^{-1}$ with a mean uncertainty of just 3.7 m\,s$^{-1}$.  
In Figure \ref{fig:betvir_ps}, we show the Lomb-Scargle periodogram of the RVs, 
which includes a broad, highly significant peak at approximately 2000 days.  
The best Keplerian model to the data produces an eccentric ($e = 0.26$) orbit with $P = 2040$ days and $K = 9$ m\,s$^{-1}$. 

\begin{table}[ht]
\centering
\begin{tabular}{lccccc}
  \hline
 & BJD & dRV & Uncertainty & $S_{HK}$ & Uncertainty \\ 
 & & (m\,s$^{-1}$) & (m\,s$^{-1}$) & & \\
  \hline
1	&	2451009.6241	&	10.9	&	2.4	&	0.158	&	0.014	\\
2	&	2451153.9622	&	0.7	&	6.0	&	0.164	&	0.020	\\
3	&	2451213.0360	&	-4.4	&	2.5	&	0.159	&	0.020	\\
4	&	2451241.8748	&	-11.3	&	3.2	&	0.176	&	0.020	\\
5	&	2451274.7687	&	3.1	&	4.2	&	0.179	&	0.019	\\
6	&	2451326.7453	&	1.6	&	3.1	&	0.168	&	0.017	\\
7	&	2451358.6645	&	8.2	&	2.0	&	0.162	&	0.016	\\
8	&	2451504.0169	&	2.0	&	2.2	&	0.166	&	0.023	\\
9	&	...	&	...	&	...	&	...	&	...	\\
   \hline
\end{tabular}
\caption{Differential radial velocity and Ca~H\&K observations for $\beta$~Virginis (sample).}
\label{tab:betvir}
\end{table}

\begin{figure}
\centering
\plotone{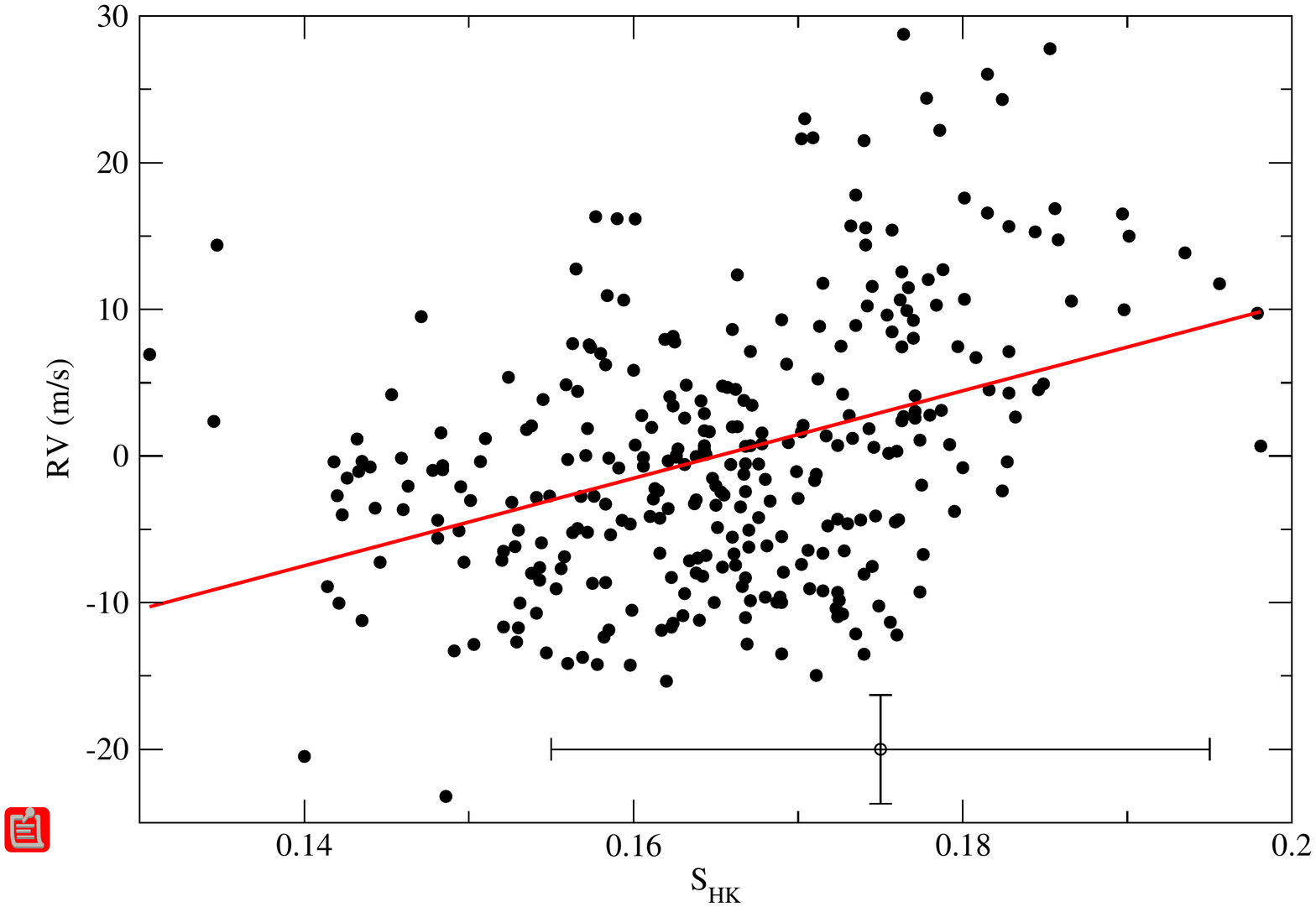}\\
\plotone{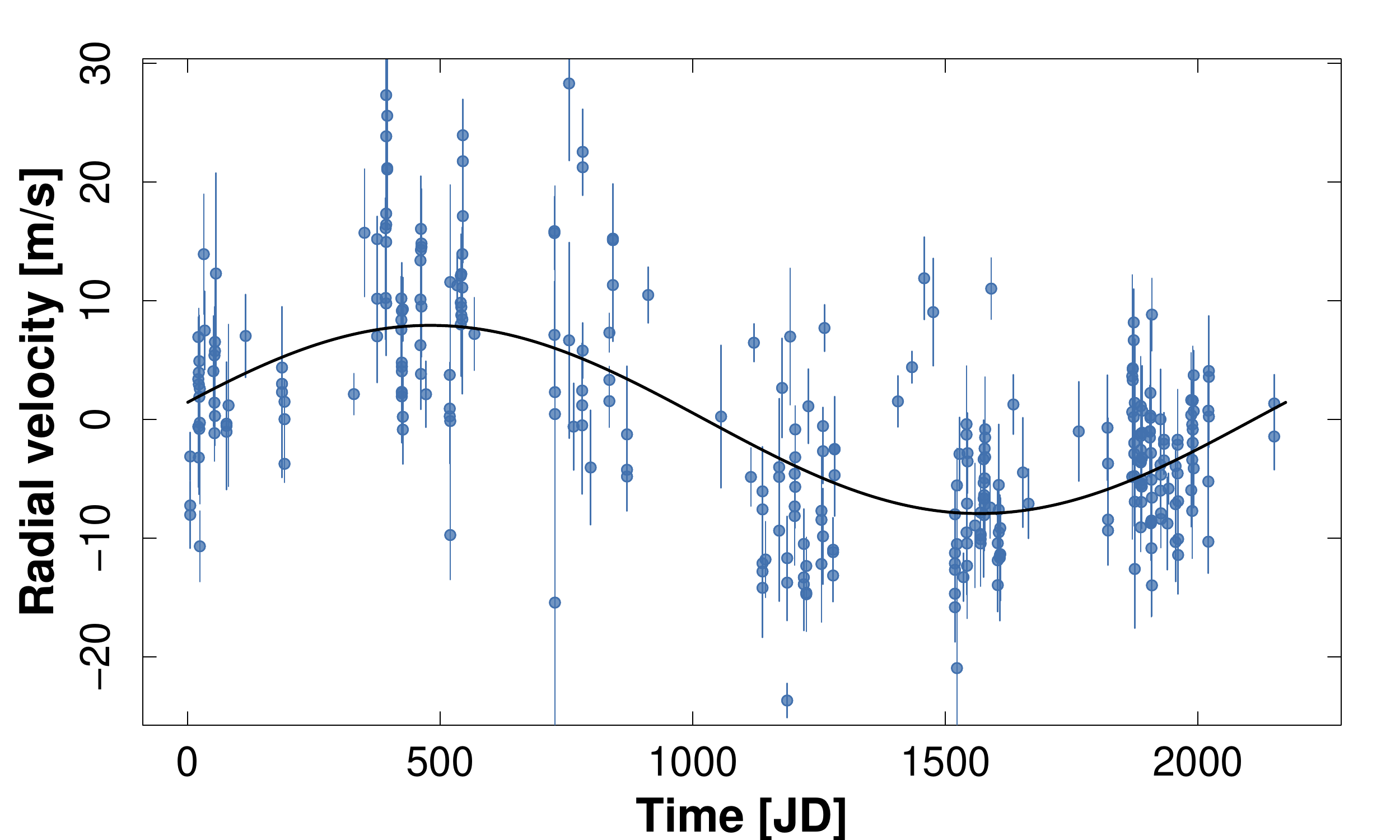}\\
\plotone{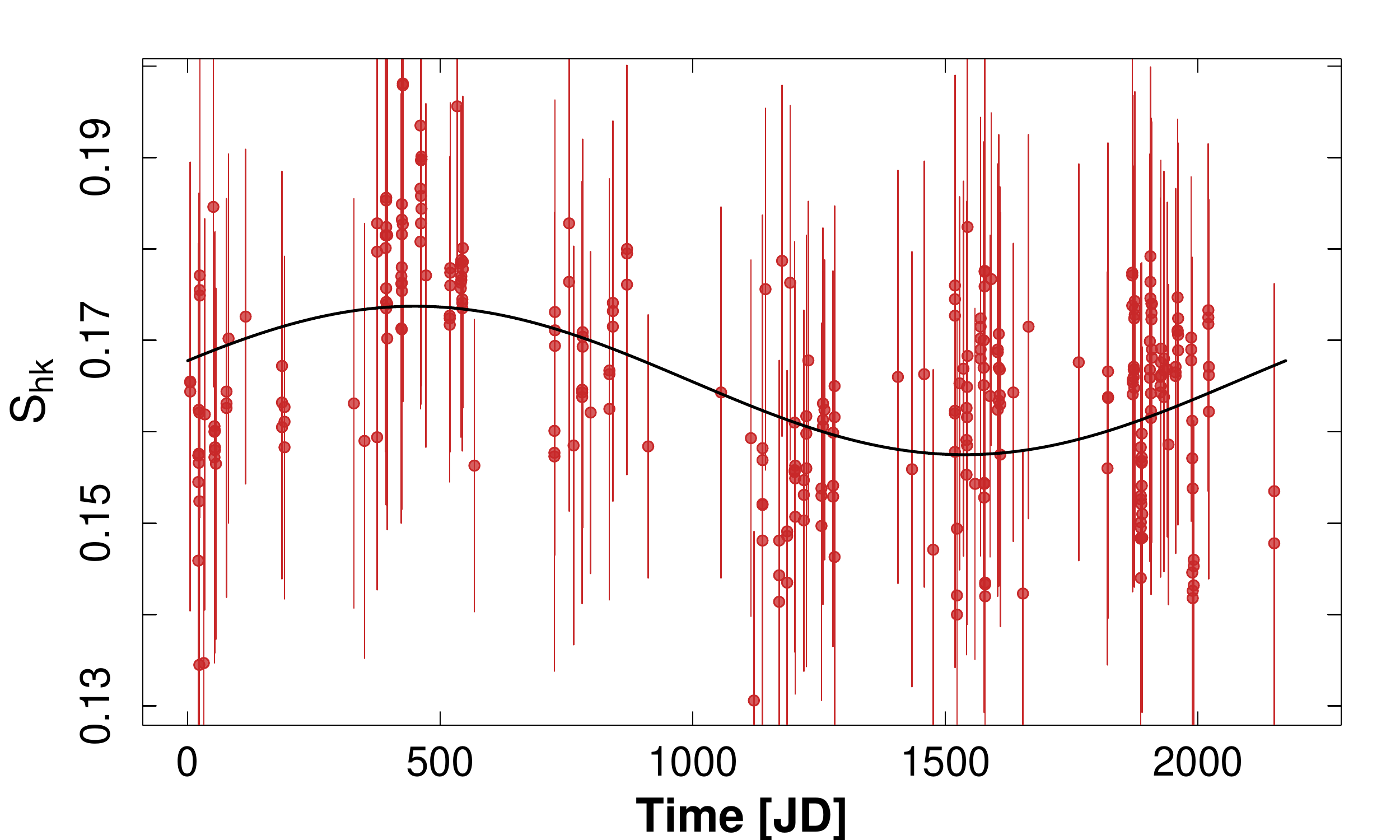}\\
\caption{\emph{Top Panel}: RV as a function of $S_{HK}$ for $\beta$ Virginis.  
The linear least squares fit to the data is shown as a solid red line.  For the sake of visibility we do not show error bars on the 
individual points, but indicate the mean 1$\sigma$ uncertainty on each variable.  \emph{Middle and Bottom Panels}: 
RV and $S_{HK}$, respectively, folded to the 2200-day period of the stellar activity cycle.  Sinusoidal models to each data set are shown as solid black curves.}
\label{fig:betvir_VvS}
\end{figure}

If the RV modulation is indeed produced by an exoplanet, this Keplerian model corresponds to a gas giant orbiting at $a = 3.5$ AU with minimum mass 
of $M \sin i = 0.65M_{\textrm{Jup}}$.  As with HD 10086, though, the Ca H\&K emission of $\beta$ Vir suggests the observed signal is not associated with a planet.  
We include the periodogram of $S_{HK}$ in Figure \ref{fig:betvir_ps}, which also includes a very broad peak between 1000 and 2500 days.  
RV as a function of $S_{HK}$ (Figure \ref{fig:betvir_VvS}) again shows a highly significant correlation; we compute a Pearson correlation coefficient $r = 0.39$ and a 
$P$-value of $2 \times 10^{-12}$.  We therefore suspect the periodicity observed for $\beta$~Vir is also a stellar activity cycle which mimics 
a Doppler shift such as would be expected for a Jupiter analog planet.

A number of features of our data set for $\beta$ Vir prevent the application of a simple stellar activity correction analogous to the one we performed for HD 10086.  
First, the weaker calcium emission (mean $S_{HK} = 0.17$, versus 0.28 for HD 10086) leads to lower signal to noise in the Ca H\&K measurements.  
Furthermore, our RVs show significant short-term scatter (5.1 m\,s$^{-1}$ over the 2013 observing season) and possibly a long-term linear acceleration in addition 
to the activity-induced periodicity.  Finally, the ``eccentricity" of the RV signal created by the activity cycle suggests the activity signal may be non-sinusoidal, 
and the RV-activity relationship may therefore not be linear. These factors make it especially difficult to fit and subtract a simple activity-RV dependence, and we 
therefore do not attempt a stellar activity correction for $\beta$ Vir.  The matching periodicities in RV and Ca H\&K and the correlation between RV and $S_{HK}$ lead 
us to conclude the observed signal is due to a stellar activity cycle, but the evaluation of any additional (possibly planetary) signals in the velocities must be 
postponed pending a more sophisticated stellar activity analysis, which is beyond the scope of this paper.

\section{Discussion \& Conclusion} \label{conclusion}

We present two cases (HD~95782 and $\psi^{1}$~Dra~B) for gas giant planets at large orbital separations, in the  
Jupiter-analog range. Owing to the very long time baseline of over a decade or more, the RV discoveries of such planets are still relatively rare.
Long-term precise RV surveys, like the McDonald Observatory planet search, still represent the current best capability to find these planets.
These planets cannot be found by {\it Kepler}, nor {\it K2}, nor TESS (Ricker et al.~2015), due to the short time span of monitoring, coupled with a very low transit 
probability of planets at 5~AU. And, despite that the best direct imaging instruments like {\it Gemini Planet Imager} (GPI, Macintosh et al.~2014) and 
{\it SPHERE} (Beuzit et al.~2008) reach the small inner working angles for nearby stars, the
low luminosities of mature, old, Jupiters makes them virtually undetectable, even for these instruments. 
Discoveries of giant planet candidates like 51~Eri~b (Macintosh et al.~2015) at very young ages of $\sim20$~Myrs and separations of $a>10$~AU will
eventually allow us to find a complete picture for these type of planets in time and orbital separation space.   

Another advantage of long-term RV surveys is the
fact that we can use the RV data for all stars, where we do not detect a planet, to set tight constraints for the presence of these gas giants. These will
allow to determine the occurrence rate of Jupiter analogs, and even of Solar System-type architectures with two gas giants.

The possibly crucial role of Jupiter, as well as of Saturn, for the formation of the terrestrial planets in our Solar System has been highlighted recently by         
Batygin \& Laughlin~(2015). These authors present a model, within the context of the ``Grand Tack'' model (Walsh et al.~2011), that explains why we do not
have super-Earths in the inner Solar System, like the numerous {\it Kepler} systems. In their model, the migration of Jupiter (which is halted and reversed by 
Saturn's dynamical evolution) depletes the interior planetesimal disk, possibly driving all existing short-period super-Earths into the Sun. The low-mass terrestrial
planets then subsequently formed in this depleted disk in the inner region of our Solar System. This model would therefore predict that planetary systems similar to ours 
can only form 
with at least two gas giants that end up at large separations after the early phase of migration has finished. The search for long-period gas giants therefore gains 
importance also in the search for Earth-like planets. The $\psi^{1}$~Dra~B planetary system could be an excellent candidate for 
a system with a planetary architecture very similar to our own, with two gas giants at large multi-AU separations, which possibly also helped the formation of lower 
mass rocky planets in the inner few AUs.    

The other two stars, HD~10086 and $\beta$~Vir, are stark reminders that stellar activity can mimic also the presence of Jupiter-analogs. Long-term magnetic cycles can
present themselves as slow RV modulations very similar to a Jupiter. Given that our Sun's magnetic field cycle is comparable to Jupiter's orbital period, we need to 
develop techniques that can correct for these effects and make planet detection possible, even in the presence of a stellar activity cycle. Our first approach to
correlate the chromospheric emission in the Ca II H \& K lines with the RV signal, works in a simple case like HD~10086. The need for a more
sophisticated model is obvious in the case of $\beta$~Vir. The relatively large RV amplitudes of the activity signal of several m\,s$^{-1}$ is somewhat unexpected, 
especially for the relatively inactive stars $\beta$~Vir.
However, the SARG binary planet survey also found an activity cycle of the star HD~200466~A, that produces an RV signal with a semi-amplitude of 
$\approx20$\,m\,s$^{-1}$ (Carolo et al.~2014). It is thus clear that in our search for Jupiter-analogs, we need to expect activity-induced RV signals that can mimic 
even massive 
gas giants. Fortunately, all our spectra from the Tull instrument, contain the Ca II H \& K lines and permit us an immediate test for possible activity 
cycles. The very long duration McDonald Observatory planet search at the 2.7\,m HJST/Tull will therefore provide a unique data set for its entire sample of over 200 
stars to find planetary systems similar to our own.

\acknowledgments

This work has been made possible through the National Science Foundation (Astrophysics grant AST-1313075) and various NASA grants over the years.  
We are grateful for their generous support. We also thank the McDonald Observatory Time Allocation committee for its continuing support of this program.

This research has made use of the SIMBAD database, operated at CDS, Strasbourg, France.

We would also like to thank all the observers who have helped gather data over the years for the 2.7\,m radial velocity planet search program at McDonald Observatory, 
including Diane Paulson, Stuart I. Barnes, Candice Gray and Anita Cochran.

Finally, we would like to thank the anonymous referee: her/his comments helped to improve this paper.

{\it Facilities:} McDonald Observatory (Tull Coud\'e Spectropgraph), Keck-I Observatory (HIRES)

\clearpage

\end{document}